\pdfoutput=1
\documentclass[PAPER,cernpreprint,atlasdraft=true,texlive=2016,UKenglish,texmf]{atlasdoc}
\LEcontact{Lily Asquith lily.asquith@cern.ch}
 
\usepackage{atlaspackage}
\usepackage{atlasbiblatex}
\usepackage{ifthen}
\usepackage{mathrsfs}
\usepackage{rotating}
\usepackage{booktabs, array}
\usepackage{siunitx}
\usepackage{atlasphysics}
 
\graphicspath{{logos/}{figures/}}

\makeatletter
\newcommand*{\numtablerange}[1][]{
\kernel@ifnextchar[{\numtable@range{#1}}{\numtable@range{#1}[#1]}}
 
\def\numtable@range#1[#2]#3#4{
\numtablerange@boxme{\expandafter\tablenum\expandafter[\numtablerange@lopt, #1]{#3}} $\pm$
\numtablerange@boxme{\expandafter\tablenum\expandafter[\numtablerange@ropt, #2]{#4}}
}
\newcolumntype{T}[3]{>{\def\numtablerange@lopt{#1}\def\numtablerange@ropt{#2}}#3}
 
\newcommand*{\numtablerange@boxme}[1]{{\sbox0{#1}\usebox0}}
\newcommand{\sma}   {\hspace{-0.2cm} ${\small<}$ 0.1}
\makeatletter

\newcommand{\ilumi} {$ 80.5\, \mathrm {fb^{-1}} $}
\newcommand{\CoMEnergy}{13~\TeV}

\newcommand{\CoMEnergyEight}{8~\TeV}

\newcommand{\CoMEnergySeven}{7~\TeV}
\newcommand{\ilumiRunone} {$ 19.2\, \mathrm {fb^{-1}} $}

\newcommand{\ntotaltext} {2\,977\,526}

\newcommand{\Bdfrac} {\ensuremath{ (4.3 \pm 0.5)\%}}
\newcommand{\Lambdafrac} {\ensuremath{ (2.1 \pm 0.6)\%}}
\newcommand{\nsig} {\ensuremath{ 453\,570 \pm 740  }}
 
\newcommand{\pSfit}     { \ensuremath{-0.081}}
\newcommand{\pSstat}    { 0.041 }
 
\newcommand{\DGfit}     { 0.0607 }
\newcommand{\DGstat}   { 0.0047}
 
\newcommand{\GSfit}    { 0.6687 }
\newcommand{\GSstat}    { 0.0015 }
 
\newcommand{\Apasqfit}    { 0.2213 }
\newcommand{\Apasqstat}  { 0.0019 }
 
\newcommand{\Azesqfit}    { 0.5131 }
\newcommand{\Azesqstat}    { 0.0013 }
 
\newcommand{\ASsqfit}     { 0.0321 }
\newcommand{\ASsqstat}     { 0.0033 }
 
\newcommand{\Deltparafit}    { 2.94 }
\newcommand{\Deltparastat}    { 0.05 }
 
\newcommand{\Deltperpfit}     { 2.91 }
\newcommand{\Deltperpstat}    { 0.11 }

\newcommand{\DeltperpfitSec}  { \ensuremath{3.12} }
\newcommand{\DeltperpstatSec} { \ensuremath{0.11} }

\newcommand{\DeltparafitSec}  { \ensuremath{3.35} }
\newcommand{\DeltparastatSec} { \ensuremath{0.05} }

\newcommand{\DeltPSfitSec}    { \ensuremath{-0.25}}
\newcommand{\DeltPSstatSec}   { \ensuremath{0.05} }

\newcommand{\pAz}    { \ensuremath{-0.003 }}
\newcommand{\pApa}   { \ensuremath{-0.011 }}
\newcommand{\pAs}   { \ensuremath{-0.006 }}
\newcommand{\pG}    { 0.019 }
\newcommand{\pDG}   { \ensuremath{-0.084 }}
\newcommand{\pDpara}    { 0.007 }
\newcommand{\pDperp}   { 0.005 }
\newcommand{\pDpS}  { \ensuremath{-0.006 }}

\newcommand{\DGAz}  { 0.096 }
\newcommand{\DGApa}  { 0.090 }
\newcommand{\DGAs}   { 0.057 }
\newcommand{\DGG}   { \ensuremath{-0.586 }}
\newcommand{\DGDpara}   { \ensuremath{-0.029}}
\newcommand{\DGDperp}  { \ensuremath{ -0.010}}
\newcommand{\DGDpS}   { 0.021 }
 
\newcommand{\GAz}    { \ensuremath{-0.048 }}
\newcommand{\GApa}  { \ensuremath{-0.116 }}
\newcommand{\GAs}   { 0.071 }
\newcommand{\GDpara}    { \ensuremath{0.070 } }
\newcommand{\GDperp}    { \ensuremath{0.017 } }
\newcommand{\GDpS}    { 0.015 }
 
\newcommand{\AzApa}  { \ensuremath{-0.338 }}
\newcommand{\AzAs}  { 0.269 }
\newcommand{\AzDpara}    { \ensuremath{ 0.080 } }
\newcommand{\AzDperp}   { \ensuremath{0.017 } }
\newcommand{\AzDpS}    { 0.070 }
 
\newcommand{\ApaAs}  { \ensuremath{-0.110 }}
\newcommand{\ApaDpara}    { \ensuremath{-0.444}}
\newcommand{\ApaDperp}    { \ensuremath{-0.106}}
\newcommand{\ApaDpS}    { \ensuremath{-0.052 }}
 
\newcommand{\AsDpara}    { \ensuremath{ 0.291 } }
\newcommand{\AsDperp}    { \ensuremath{ 0.060 } }
\newcommand{\AsDpS}  { 0.251 }
 
\newcommand{\DparaDperp}    { 0.235 }
\newcommand{\DparaDpS}    { \ensuremath{0.097 } }
 
\newcommand{\DperpDpS}    { \ensuremath{0.056 } }

\newcommand{\pAzSec}    { \ensuremath{-0.004 }}
\newcommand{\pApaSec}   { \ensuremath{-0.003 }}
\newcommand{\pAsSec}   { \ensuremath{-0.007 }}
\newcommand{\pGSec}    { 0.017 }
\newcommand{\pDGSec}   { \ensuremath{-0.080 }}
\newcommand{\pDparaSec}    { 0.007 }
\newcommand{\pDperpSec}   { 0.004 }
\newcommand{\pDpSSec}  { \ensuremath{-0.007 }}

\newcommand{\DGAzSec}  { 0.095 }
\newcommand{\DGApaSec}  { 0.090 }
\newcommand{\DGAsSec}   { 0.051 }
\newcommand{\DGGSec}   { \ensuremath{-0.586 }}
\newcommand{\DGDparaSec}   { 0.032 }
\newcommand{\DGDperpSec}  { 0.005 }
\newcommand{\DGDpSSec}   { 0.020 }
 
\newcommand{\GAzSec}    { \ensuremath{-0.045 }}
\newcommand{\GApaSec}  { \ensuremath{-0.125 }}
\newcommand{\GAsSec}   { 0.080 }
\newcommand{\GDparaSec}    { \ensuremath{-0.086 }}
\newcommand{\GDperpSec}    { \ensuremath{-0.023 }}
\newcommand{\GDpSSec}    { 0.015 }
 
\newcommand{\AzApaSec}  { \ensuremath{-0.341 }}
\newcommand{\AzAsSec}  { 0.276 }
\newcommand{\AzDparaSec}    { \ensuremath{ -0.103 }}
\newcommand{\AzDperpSec}   { \ensuremath{-0.034 }}
\newcommand{\AzDpSSec}    { 0.070 }
 
\newcommand{\ApaAsSec}  { \ensuremath{-0.172 }}
\newcommand{\ApaDparaSec}    { 0.522 }
\newcommand{\ApaDperpSec}    { 0.133 }
\newcommand{\ApaDpSSec}    { \ensuremath{-0.052 }}
 
\newcommand{\AsDparaSec}    { \ensuremath{ -0.362 }}
\newcommand{\AsDperpSec}    { \ensuremath{ -0.118 }}
\newcommand{\AsDpSSec}  { 0.244 }
 
\newcommand{\DparaDperpSec}    { 0.254 }
\newcommand{\DparaDpSSec}    { \ensuremath{-0.085 }}
 
\newcommand{\DperpDpSSec}    { 0.001 }

\newcommand{\pSsystExp} { 22\phantom{.0}} 
\newcommand{\DGsystExp} { \phantom{1}4.3 }
\newcommand{\GSsystExp}  { \phantom{1}2.2 }
\newcommand{\ApasqsystExp} { \phantom{1}2.3 }
\newcommand{\AzesqsystExp} { \phantom{1}3.8 }
\newcommand{\ASsqsystExp}  { \phantom{1}4.6 }
\newcommand{\DeltperpsystExp} { 55\phantom{.0}  }
\newcommand{\DeltparasystExp} { 88\phantom{.0}  }
\newcommand{\DeltPSsystExp}   { 39\phantom{.0}  }

\newcommand{\pSsyst} { 0.022 } 
\newcommand{\DGsyst} { 0.0043 }
\newcommand{\GSsyst}  { 0.0022 }
\newcommand{\Apasqsyst} { 0.0023}
\newcommand{\Azesqsyst} { 0.0038}
\newcommand{\ASsqsyst}  { 0.0046}
\newcommand{\Deltperpsyst} { 0.06} 
\newcommand{\Deltparasyst} { 0.09} 
\newcommand{\DeltPSsyst}   { 0.04}

\newcommand{\pSfitCombi}        { \ensuremath{-0.087} }
\newcommand{\pSstatCombi}       { 0.036 }
\newcommand{\pSsystCombi}       { 0.021 }
 
\newcommand{\DGfitCombi}        { 0.0657 }
\newcommand{\DGstatCombi}       { 0.0043 }
\newcommand{\DGsystCombi}       { 0.0037 }
 
\newcommand{\GSfitCombi}        { 0.6703 }
\newcommand{\GSstatCombi}       { 0.0014 }
\newcommand{\GSsystCombi}       { 0.0018 }
 
\newcommand{\ApasqfitCombi}     { 0.2220 }
\newcommand{\ApasqstatCombi}    { 0.0017 }
\newcommand{\ApasqsystCombi}    { 0.0021 }
 
\newcommand{\AzesqfitCombi}     { 0.5152 }
\newcommand{\AzesqstatCombi}    { 0.0012 }
\newcommand{\AzesqsystCombi}    { 0.0034 }
 
\newcommand{\ASsqfitCombi}      { 0.0343 }
\newcommand{\ASsqstatCombi}     { 0.0031 }
\newcommand{\ASsqsystCombi}     { 0.0045 }
 
\newcommand{\DeltperpfitCombi}  { 3.22 }
\newcommand{\DeltperpstatCombi} { 0.10 }
\newcommand{\DeltperpsystCombi} { 0.05 }
 
\newcommand{\DeltparafitCombi}  { 3.36 }
\newcommand{\DeltparastatCombi} { 0.05 }
\newcommand{\DeltparasystCombi} { 0.09 }
 
\newcommand{\DeltPSfitCombi}    { \ensuremath{-0.24} }
\newcommand{\DeltPSstatCombi}   { 0.05 }
\newcommand{\DeltPSsystCombi}   { 0.04 }

\newcommand{\pSfitCombiSolB}        { \ensuremath{-0.087} }
\newcommand{\pSstatCombiSolB}       { 0.036 }
\newcommand{\pSsystCombiSolB}       { 0.021 }
 
\newcommand{\DGfitCombiSolB}        { 0.0657 }
\newcommand{\DGstatCombiSolB}       { 0.0043 }
\newcommand{\DGsystCombiSolB}       { 0.0037 }
 
\newcommand{\GSfitCombiSolB}        { 0.6704 }
\newcommand{\GSstatCombiSolB}       { 0.0014 }
\newcommand{\GSsystCombiSolB}       { 0.0018 }
 
\newcommand{\ApasqfitCombiSolB}     { 0.2218 }
\newcommand{\ApasqstatCombiSolB}    { 0.0017 }
\newcommand{\ApasqsystCombiSolB}    { 0.0021 }
 
\newcommand{\AzesqfitCombiSolB}     { 0.5152 }
\newcommand{\AzesqstatCombiSolB}    { 0.0012 }
\newcommand{\AzesqsystCombiSolB}    { 0.0034 }
 
\newcommand{\ASsqfitCombiSolB}      { 0.0348 }
\newcommand{\ASsqstatCombiSolB}     { 0.0031 }
\newcommand{\ASsqsystCombiSolB}     { 0.0045 }
 
\newcommand{\DeltperpfitCombiSolB}  { 3.03 }
\newcommand{\DeltperpstatCombiSolB} { 0.10 }
\newcommand{\DeltperpsystCombiSolB} { 0.05 }
 
\newcommand{\DeltparafitCombiSolB}  { 2.95 }
\newcommand{\DeltparastatCombiSolB} { 0.05 }
\newcommand{\DeltparasystCombiSolB} { 0.09 }
 
\newcommand{\DeltPSfitCombiSolB}    { \ensuremath{-0.24} }
\newcommand{\DeltPSstatCombiSolB}   { 0.05 }
\newcommand{\DeltPSsystCombiSolB}   { 0.04 }

\newcommand{\cutChiJpsi}   {\chisq $< 10$}
 
\newcommand{\cutMassPhi} {$ 1.0085 \; {\mathrm \GeV} < m(K^+ K^-) < 1.0305 \; {\mathrm \GeV} $}
\newcommand{\cutChiBs}   {\chisq $< 3$}
 
\newcommand{\cutMassBsRange}{ 5.150--5.650 \GeV}

\newcommand{\Bst}{\ensuremath{ \Bs \ra \JpsiPhi }\xspace }

\newcommand{\Bsto}{\ensuremath{ \Bs \ra J/\psi (\mumu) \phi(K^+K^-) }}

\newcommand{\BstoKK}{\ensuremath{ \Bs \ra J/\psi K^+K^- (f_{0}) }}
\newcommand{\JpsiPhi}{\ensuremath{ J/\psi\phi }}
\newcommand{\JpsiPhin}{\ensuremath{ J/\psi\phi}}

\newcommand{\Gs}{\ensuremath{ \Gamma_{s}     }}
\newcommand{\DGs}{\ensuremath{ \Delta\Gamma_{s}     }}
 
\newcommand{\phis}{\ensuremath{\phi_{s}  }}
\newcommand{\dms}{\ensuremath{\Delta m_{s}  }}
\newcommand{\BH}{\ensuremath{B_{\mathrm H}  }}
\newcommand{\BL}{\ensuremath{B_{\mathrm L}  }}
\newcommand{\GH}{\ensuremath{\Gamma_{\mathrm H}  }\space}
\newcommand{\GL}{\ensuremath{\Gamma_{\mathrm L}  }}

\newcommand{\kminus} {\Kminus}
\newcommand{\kplus} {\Kplus}

\def \ra{\ensuremath{\rightarrow}}

\newcommand{\Ds}{\ensuremath{D_{s}} \space}
\newcommand{\BsaBs}{\ensuremath{\Bs}--\ensuremath{\bar{B}^{0}_{s}}\space}

\newcommand{\aBs}{\ensuremath{\bar{B}^{0}_{s}}}

\newcommand{\Jpsipa}{\Jpsi \ }

\newcommand{\psubt}{\pt \space}

\newcommand{\pico}{\ipb\   }
\newcommand{\femto}{\ifb\space}

\newcommand{\chisq}{$\mathrm \chi^{2}$/ndof }

\newcommand{\Jpsitomm}{\ensuremath{ \Jpsi \ra \mumu }\space}

\newcommand{\phiKK}{\ensuremath{ \phi \ra \kplus \kminus}\space}

\newcommand{\pb}{\ensuremath{pb^{-1}}}
\newcommand{\fb}{\ensuremath{fb^{-1}}}
\newcommand{\nb}{\ensuremath{nb^{-1}}}

\newcommand{\CP}{\textit{CP}}
 
\AtlasTitle{Measurement of the \textit{CP}-violating phase $\phi_s$ in \Bst decays in ATLAS at 13 TeV}
 
\AtlasAbstract{
A measurement of the \Bst\ decay parameters using \ilumi\ of integrated luminosity collected with the ATLAS detector from \CoMEnergy\ proton--proton collisions at the LHC is presented.
The measured parameters include the \CP-violating phase \phis,  the width difference \DGs \  between the \ensuremath{B_{s}^{0}} \ meson mass eigenstates  and   the average decay width \Gs. The values measured for the physical parameters are  combined with those from \ilumiRunone \ of \CoMEnergySeven \ and \CoMEnergyEight \ data, leading to the following:
 
\begin{eqnarray*}
\phis & = & \pSfitCombi\phantom{0} \pm  \pSstatCombi \phantom{0} ~\mathrm{(stat.)} \pm \pSsystCombi\phantom{0} ~\mathrm{(syst.)~rad}  \\
\DGs  & = & \phantom{-}\DGfitCombi \pm  \DGstatCombi~\mathrm{(stat.)}\pm \DGsystCombi~\mathrm{(syst.)~ps}^{-1} \\
\Gs   & = & \phantom{-}\GSfitCombi \pm \GSstatCombi ~\mathrm{(stat.)}\pm \GSsystCombi ~\mathrm{(syst.)~ps}^{-1}\\
\end{eqnarray*}
 
Results for \phis \ and \DGs \ are also presented as 68\% confidence level contours in the \phis--\DGs\ plane. Furthermore the transversity amplitudes and corresponding strong phases are measured. \phis\ and \DGs\ measurements are in agreement with the Standard Model predictions.
}
 
\author{The ATLAS Collaboration}
 
\AtlasRefCode{BPHY-2018-01}
 
\PreprintIdNumber{CERN-EP-2019-218}

\AtlasJournal{EPJC}
\AtlasJournalRef{Eur. Phys. J. C 81 (2021) 342}
\AtlasDOI{10.1140/epjc/s10052-021-09011-0}

\hypersetup{pdftitle={ATLAS document},pdfauthor={The ATLAS Collaboration}}
 
\begin{document}
 
\maketitle

\section{Introduction}
 
In the presence of new physics (NP) phenomena,  sources of \CP\ violation in  $b$-hadron decays can arise in addition to those  predicted by the  Standard Model (SM)~\cite{BOTELLA20071}.
In the  \Bst\ decay, \CP\  violation occurs due to interference between a direct decay and a decay with \BsaBs mixing. The oscillation frequency of \Bs\ meson mixing is characterised by the mass difference, \dms, of the heavy (\BH) and light (\BL)  mass eigenstates. The \CP-violating phase \phis\ is defined as the weak phase difference between the \BsaBs\ mixing amplitude and the $b \rightarrow c \overline{c} s$ decay amplitude. In the SM the phase \phis\ is small and is  related to the Cabibbo--Kobayashi--Maskawa (CKM) quark mixing matrix elements via the relation $\phis \simeq -2 \beta_{s}$, with $\beta_{s} = \mathrm{arg} [- (V_{ts} V^{*}_{tb})/(V_{cs} V_{cb}^{*}) ]$. By combining beauty and kaon physics observables, and assuming no NP  contributions to \Bs\ mixing and decays, a value of   $-2 \beta_{s} = -0.03696^{+0.00072}_{-0.00082}$ rad was  predicted by the CKMFitter group \cite{PhysRevD.91.073007} and $-2 \beta_{s} = -0.03700 \pm 0.00104$ rad according to the UTfit Collaboration \cite{UTFit2006}.
While large NP enhancements of the mixing amplitude have been excluded by the precise  measurement of the oscillation frequency \cite{Aaij:2013Mix}, the  NP couplings involved in the mixing may  still increase the size of the observed \CP\  violation  by enhancing the mixing phase $\phis $ with respect to the SM value.
 
Other physical quantities involved in \BsaBs\ mixing are the decay width $\Gs=(\GL + \GH)/2$ and the width difference $\DGs = \GL - \GH$, where $\GL$ and $\GH$ are the decay widths of the light and heavy mass eigenstates, respectively. The latest predictions for the width difference in the SM are    $\DGs = 0.091 \pm 0.013$~ps$^{-1}$~\cite{Lenz:2019lvd} and $\DGs = 0.092 \pm 0.014$~ps$^{-1}$~\cite{Davies:2019gnp}.
A potential  NP enhancement of $\phis $ would also decrease the size of \DGs, but it is not expected to be affected as significantly as \phis~\cite{Nierste}. Nevertheless, extracting \DGs\ from the data is an important test of theoretical predictions~\cite{Nierste}.

Theory predictions have been made for the lifetime ratios   $\tau$(\Bs)/$\tau(B_d)$ and $\tau$(\Bs)/$\tau(B^+)$, with the latest update Ref. \cite{LE}. The lifetime $\tau$(\Bs) has not been calculated in theory yet at a precision comparable with those obtained by experiments. The current world combined value of the decay width, $\Gs$,  obtained from experimental results is $\Gs =  0.6600\pm0.0016$~ps$^{-1}$~\cite{Zyla:2020zbs}.

The analysis of the time evolution of the \Bst\  decay provides the most precise determination of $\phis$ and \DGs.  Previous measurements of these quantities have been reported by the D0, CDF,  LHCb, ATLAS and CMS Collaborations \cite{Abazov:2011ry, CDF:2012ie,  LHCb:2014run1, BPHY-2014-03, CMS-BPH-13-012,Aaij:2017zgz, Aaij2019, cmsrun2archive}. Additional improvements in measuring \phis \  from $\Bs \to \psi(2S) \phi$, $\Bs \to \Ds^+\Ds^-$ and $\Bs \to J/\psi \pi^+\pi^-$ decays have been achieved by the LHCb Collaboration \cite{Aaij:2016psitwoS, Aaij:2014Ds,Aaij:2014dka,Aaij:2019mhf}.
 
The analysis presented here introduces a measurement of the \Bst\ decay parameters using \ilumi\ of the LHC proton--proton ($pp$) data collected by the ATLAS detector during 2015--2017, at a centre-of-mass energy, $\sqrt{s}$, equal to \CoMEnergy. The analysis closely follows a previous ATLAS measurement~\cite{BPHY-2014-03} that was performed using  19.2~$\mathrm{fb}^{-1}$ of the data collected at \CoMEnergySeven\ and \CoMEnergyEight, and introduces more precise signal and background models.
 
\section{ATLAS detector and Monte Carlo simulation}
\label{sec:detector}

The ATLAS detector\footnote{ATLAS uses a right-handed coordinate system
with its origin at the nominal interaction point. The $z$-axis is along
the beam pipe, the $x$-axis points to the centre of the LHC ring and
the $y$-axis points upward. Cylindrical coordinates $(r,\phi)$ are used
in the transverse plane, $r$ being the distance from the origin and
$\phi$ being the azimuthal angle around the beam pipe. The pseudorapidity
$\eta$ is defined as $\eta = - \ln [\tan (\theta/2)]$ where
$\theta$ is the polar angle.}
consists of three main components: an inner detector (ID) tracking system
immersed in a $2$~T axial magnetic field, electromagnetic and hadronic calorimeters, and a muon spectrometer (MS).
The inner tracking detector covers the pseudorapidity range \(|\eta| < 2.5\), and consists of silicon pixel, silicon microstrip, and transition radiation tracking detectors.
The ID is surrounded by a high-granularity liquid-argon (LAr) sampling electromagnetic calorimeter.
A steel/scintillator tile calorimeter provides hadronic coverage in the central rapidity range. The endcap and forward regions are equipped with LAr calorimeters for electromagnetic and hadronic measurements.
The MS surrounds the calorimeters and provides a system of tracking chambers and detectors for triggering.
A full description can be found in Refs.~\cite{PERF-2007-01,ATLAS-TDR-2010-19, JINST.13.T05008}.

The data were collected during periods with different  instantaneous luminosity, so several triggers  were used in the analysis. All triggers were based on the identification of    a \Jpsitomm decay, with transverse momentum (\pT)
thresholds of either  $4$~\GeV\ or $6$~\GeV\ for the muons. 
Data quality requirements are imposed on the data, notably on the performance of the MS, ID and calorimeter systems.
The measurement uses \ilumi\ of $pp$ collision data. The uncertainty in the combined 2015--2017 integrated luminosity is 2.0\%~\cite{ATLAS-CONF-2019-021}, obtained using the LUCID-2 detector~\cite{LUCID2} for the primary luminosity measurements.

To study the detector response, estimate backgrounds, and model systematic effects, 100M Monte Carlo (MC) simulated \Bst\ events were generated using  {\normalfont \scshape Pythia} 8.210~\cite{Sjostrand:2014zea} tuned with  ATLAS data, using the A14 set of parameter values~\cite{ATL-PHYS-PUB-2014-021}  together with the CTEQ6L1 set of parton distribution functions~\cite{Pumplin:2002vw}.   The detector response was simulated using the ATLAS simulation framework based on \textsc{Geant4}~\cite{Atlfast, Geant4}.  In order to account for the varying  number of  proton--proton interactions per bunch crossing (pile-up) and trigger configurations during data-taking, the MC events were weighted to reproduce the same pile-up and trigger conditions as in the data.  Additionally, background samples of both the exclusive ($\Bd \ra J/\psi K^{0*}$ and $\Lambda_b \ra J/\psi p K^{-}$) and inclusive
($b\bar{b} \ra J/\psi X$ and $pp \ra J/\psi X$) decays were simulated, using the same simulation tools as in the case of the signal events. For validation studies related to flavour tagging, detailed in Section \ref{sec:tagging},  events with $B^{\pm}\to J/\psi K^{\pm}$ exclusive decays were also simulated.
 
\section{Reconstruction and candidate selection}
\label{sec:selection}
 
The reconstruction and candidate selection for the decay \Bsto\ is described here.
Events must pass the trigger selections described in Section~\ref{sec:detector}. In  addition, each event must contain at least one reconstructed primary vertex, formed from at least four ID tracks, and at least one pair of oppositely charged muon candidates that are reconstructed using information from the MS and the ID.
The muons used in the analysis are required to meet the \textit{Tight}\footnote{Tight muon reconstruction is optimised to maximise the purity of muons at the cost of some efficiency, requiring combined muons with hits in at least two stations of the MS and additional criteria, described in Ref.~\cite{PERF-2015-10}.}
or \textit{Low}-$\mathit{\pt}$\footnote{This working point is optimised to provide good muon reconstruction efficiency down to a $\pT$ of $\approx 3$~\GeV, while controlling the fake-muon rate.
It allows $\geq 1$ ($\geq 2$) MDT station tracks up to $|\eta|<1.3$ ($1.3<|\eta|<1.55$) for candidates reconstructed by algorithms utilising inside-out combined  reconstruction~\cite{PERF-2015-10}. Additional cuts on the number of precision stations and on variables very sensitive to the decays in flight of hadrons are also applied to suppress fake muons.}
working point identification criteria.
The muon track parameters  are determined from the ID measurement alone, since the precision of the measured track parameters is dominated by the ID track reconstruction in the  \pT\ range of interest for this analysis. Pairs of oppositely charged muon tracks are re-fitted to a common vertex and the pair is accepted if the quality of the fit meets the requirement \cutChiJpsi.  In order to account for  varying  mass resolution in different parts of the detector, the \Jpsi\ candidates are divided into three subsets  according to the pseudorapidity $\eta$ of the muons. In the first subset, both muons have $|\eta| <  1.05$, where the values $\eta = \pm 1.05$ correspond to the edges of the barrel part of the MS.  In the second subset, one muon has  $1.05  < |\eta| <  2.5 $ and the other muon  $|\eta| <  1.05$. The third subset contains  candidates where  both muons have $1.05  < |\eta| <  2.5 $.
A maximum likelihood fit is used to extract the $J/\psi$ mass and the corresponding mass resolution for these three subsets, and  in each case the signal region is defined symmetrically around the fitted mass, so as to retain 99.7\% of the \Jpsi\ candidates identified in the fits.

The candidates for  the decay \phiKK are reconstructed from all pairs of oppositely charged tracks, with $\pT > 1$~\GeV \ and $|\eta| < 2.5$, that are not identified as muons. Candidate events for \Bsto\ decays are selected by fitting the tracks for each combination of \Jpsitomm\ and \phiKK to a common vertex. The fit is also constrained by fixing the invariant mass calculated from the two muon tracks to the \Jpsipa mass~\cite{PDG}. A quadruplet of tracks is accepted for further analysis if the vertex fit has  \cutChiBs. For the \phiKK candidate, the invariant mass of the track pairs (using a charged kaon mass hypothesis) must fall within the interval \cutMassPhi. The interval, chosen using MC simulation, is selected  to retain 98\%  of  true \phiKK   decays.   The \Bs \ candidate with the lowest $\chi^2 /$ndof is selected in events where more than one candidate passes all selections.  In total,  \ntotaltext \ \Bs\ candidates are collected within the mass range of \cutMassBsRange. This range is chosen to give enough background events in the sidebands of the mass distributions to allow precise determination of the properties of the background events.
 
The mean number of interactions per bunch crossing is 30, necessitating a choice of the best candidate for the primary vertex at which the \Bs\  meson is produced.
Primary vertex positions are recalculated after removing any tracks used in the \Bs\ meson reconstruction.
The variable used to select the best candidate for the primary vertex is the three-dimensional impact parameter, $a_0$, which is calculated as the minimum distance between each primary vertex candidate and the line extrapolated from the reconstructed  \Bs\  meson vertex in the direction of the  \Bs\  momentum. The chosen primary vertex is the one with the smallest $a_0$. A simulated dataset is used to estimate the fraction of \Bs\
candidates where the incorrect production vertex is selected ($12\%$) and demostrates that the mis-selection of reconstructed primary vertex does not bias the reconstructed proper decay time.
 
For each \Bs\ meson candidate the proper decay time $t$ is estimated using:
\begin{linenomath}
\begin{equation*}
t = \frac{ L_{xy}\ m_{{}_{B}} }     { p_{ \mathrm{T}_{B}  }},
\end{equation*}
\end{linenomath}
where $p_{ \mathrm{T}_{B}  }$ is the  reconstructed transverse momentum of the \Bs\ meson candidate and  $m_{{}_{B}}$ denotes  the  mass  of the \Bs\ meson, taken from Ref.~\cite{PDG}. The transverse decay length, $L_{xy}$, is the displacement in the transverse plane of the  \Bs\ meson  decay vertex relative to the primary vertex, projected onto the direction of the \Bs\ transverse momentum.
\section{Flavour tagging}
\label{sec:tagging}
 
To identify, or \textit{tag}, the flavour of a neutral $B$ meson at the point of production, information is extracted using the decay of the other (or \textit{opposite}) $b$-hadron that is produced from the pair production of $b$ and $\bar{b}$ quarks.
This method is called opposite-side tagging (OST).
 
The OST algorithms each define a discriminating variable, based on charge information, which is sensitive to the flavour (i.e. $b$- or $\bar{b}$-quark) of the opposite-side $b$-hadron. The algorithms thus provide a probability that a signal $B$ meson in a given event is produced in a given flavour.
The calibration of the OST algorithms proceeds using data containing $B^{\pm}\to J/\psi K^{\pm}$ candidate decays,
where the charge of the kaon determines the flavour of the $B$ meson, providing a self-tagging sample of events.
These OST algorithms are calibrated as a function of the discriminating variable, using  yields of signal $B^{\pm}$ mesons extracted from fits to the data. Once calibrated, the OST algorithms are applied to $\Bsto$ candidate events to provide a probability that each candidate was produced in a  $\Bs$ or $\aBs$ meson state, which is used in the maximum likelihood fit (described in Section~\ref{sec:fit}). This approach assumes invariance of the OST algorithm with respect to the specific signal $b$-hadron type (i.e $B^{\pm}$ meson or $\Bs$ meson), which is tested and the difference is considered as a systematic uncertainty.
\subsection{$B^\pm \to J/\psi K^\pm$ event selection}
\label{sec:taggingBplus}
 
Candidate $B^{\pm}\to J/\psi K^{\pm}$ decays are identified in a series of steps.
First, $J/\psi$ candidates are selected from oppositely charged muon pairs forming a good vertex, as described in Section~\ref{sec:selection}.
Each muon is required to have $\pT > 4$~\GeV \ and $|\eta| < 2.5$.
Dimuon candidates with invariant mass $2.8 < m(\mu^+\mu^-) < 3.4$~\GeV, as determined from the re-fitted track parameters of the vertex, are retained for further analysis.
To form the $B^{\pm}$ candidate, an additional track
is required, which is not identified as an electron or muon. The track is assigned the charged-kaon mass hypothesis and combined with the dimuon candidate using
a vertex fit, performed with the mass of the dimuon pair constrained to the $J/\psi$ mass.
Prompt background contributions are suppressed by a requirement on the proper decay time of the $B^{\pm}$ candidate of $t > 0.2$~ps.

\begin{figure}[t]
\begin{center}
{\includegraphics[width=0.60\textwidth]{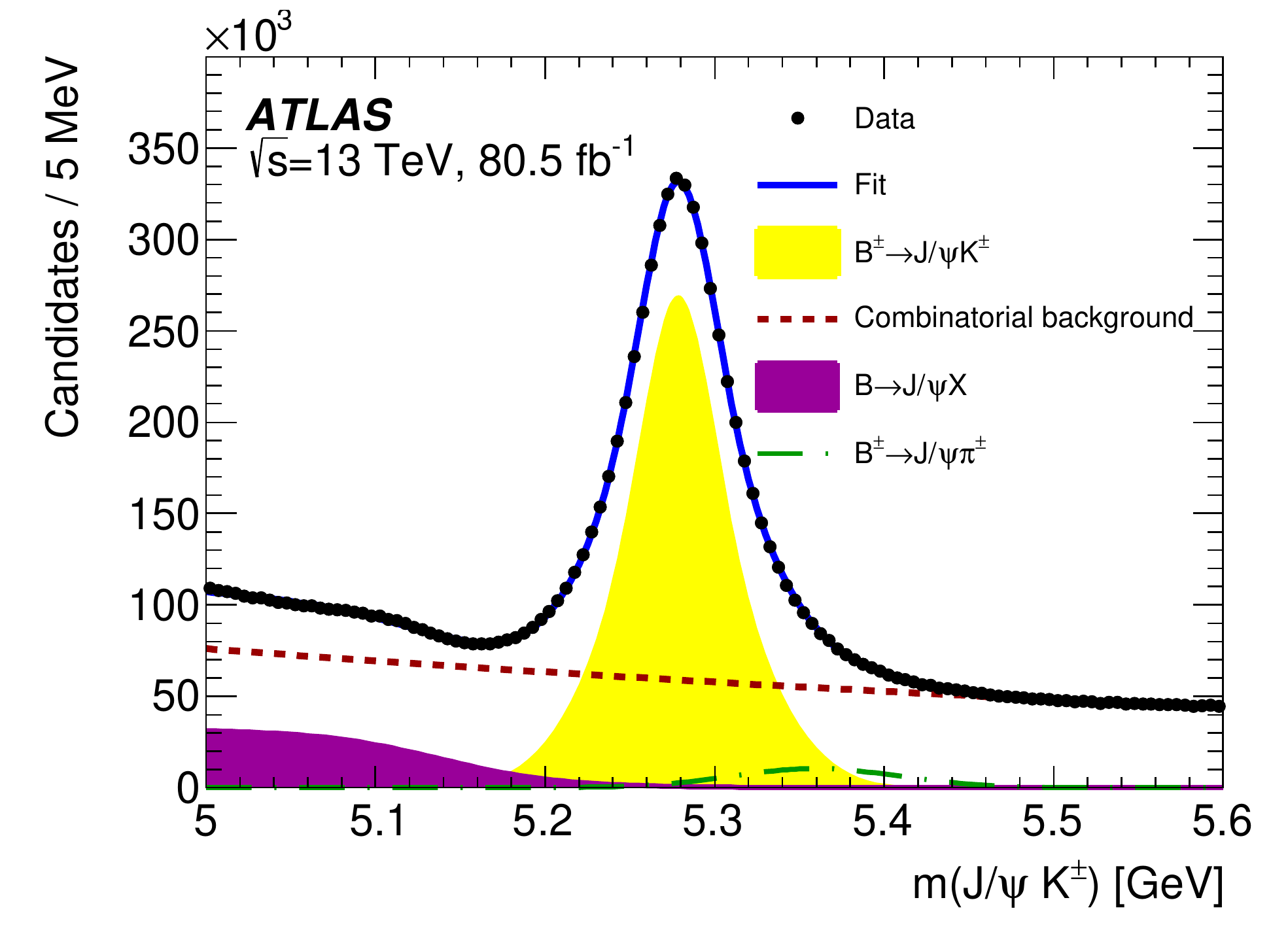}}
\caption{The invariant mass distribution for selected $B^{\pm}\to\jpsi K^{\pm}$ candidates.
Data are shown as points, and the overall result of the fit is given by the blue curve.
The contributions from the combinatorial background component are indicated by the red dotted line,  partially reconstructed $b$-hadron decays by the purple shaded area, and decays of $B^{\pm}\to J/\psi \pi^{\pm}$, where the pion is misassigned as a kaon, by the green dashed line.
\label{fig:bplusmass}}
\end{center}
\end{figure}

The tagging probabilities are determined from $B^{+}$ and $B^{-}$ signal events. These signal yields are derived from  fits to the invariant mass distribution, $m(J/\psi K^{\pm})$, and performed in intervals of the discriminating variables.
To describe the $B^{\pm}\to J/\psi K^{\pm}$ signal, two Gaussian functions with a common mean are used. An exponential function is used to describe the combinatorial background and a hyperbolic tangent function to parameterise the low-mass contribution from incorrectly or partially reconstructed $b$-hadron decays.
A Gaussian function is used to describe the $B^{\pm}\to J/\psi \pi^{\pm}$ contribution, with fixed parameters taken from simulation except for the normalisation, which is a free parameter.
A fit to the overall mass distribution is used to define the shapes of signal and backgrounds.
Subsequent fits are performed in the intervals of the tagging discriminating variables, separately for $B^{+}$ and $B^{-}$ candidate events, with the normalisations and also the slope of the exponential function left free.
The $B^{+}$ and $B^{-}$ signal yields are extracted from these fits.
Figure~\ref{fig:bplusmass} shows the invariant mass distribution of $B^{\pm}$ candidates overlaid with a fit to all selected candidates, and including the individual fit components for the signal and backgrounds.

\subsection{Flavour tagging methods}
\label{sec:taggingMethods}
 
The flavour of the signal $B$ meson at the point of production is inferred using several methods, which differ in their efficiency and discrimination power.
The measured charge of a lepton (electron or muon) from the semileptonic decay of a $B$ meson provides strong discrimination;
however, the ATLAS sensitivity to $b\to \ell$ transitions are diluted through processes that can change the charge of the observed lepton, such as
through neutral $B$ meson oscillations, or through the cascade decays $b\to c\to \ell$.
The separation power of lepton tagging is enhanced by considering a weighted sum of the charge of the tracks in a cone around the lepton.
If no lepton is present, a weighted sum of the charge of the tracks in a jet associated with the opposite-side $b$-hadron decay is used to provide discrimination.
This weighted sum, or \textit {cone charge}, is defined as:
 
\vspace*{-0.3cm}
\begin{equation}\label{eq:conecharge}
Q_{x}  =  \frac{ \sum^{N\;\mathrm{tracks}}_{i} q_{i} \cdot (p_{\mathrm{T} i })^{\kappa}} {  \sum^{N\;\mathrm {tracks}}_{i}(p_{\mathrm{T} i })^{\kappa}},
\end{equation}
 
where $x=\{\mu,e,\mathrm{jet}\}$ refers to muon, electron, or jet charge, respectively, and the summation is made using the charge of the track, $q_{i}$, and its $p_{\mathrm{T} i }$, over a selected set of tracks, including the lepton, in a cone of size $\Delta R = \sqrt{(\Delta\phi)^2+(\Delta\eta)^2}$, around the lepton or jet direction.
The value of the parameter $\kappa$ is optimised on each OST method,  by determining the value of $\kappa$  that maximises the  \textit{tagging power} (defined in Section~\ref{sec:tagperf}).
The requirements on the tracks and $\Delta R$ are described below, dependent on the OST method.

Two subcategories of $Q_{x}$ are considered: the first \textit{discrete} category is used in the case where the cone charge is formed either from only one track or from more than one track of the same charge; this results in a cone charge of $Q_{x}=\pm1$. The second \textit{continuous} category is used when more than one track is considered, and the sum contains tracks of both negative and positive charge. In the continuous case, $Q_{x}$ is divided into intervals within the range $-1<Q_{x}<1$ for each OST algorithm.
 
A probability $P(B|Q_{x})$ is constructed, which is defined as the probability that a $B$ meson is produced in a state
containing a ${\bar b}$-quark, given the value of the cone charge $Q_{x}$.
Since $Q_{x}$ is evaluated on the opposite side, a large, negative value of $Q_{x}$ tends to correspond to a higher value of $P(B|Q_{x})$.
An equivalent probability for the $b$-quark case is defined as $P({\bar B}|Q_{x})$.
Using the $B^{\pm}$ calibration samples, $P(Q_{x}|B^{\pm})$ for each tagging method used can be defined.
The probability to tag a $\Bs$ meson as containing a ${\bar b}$-quark is therefore given as $P(B|Q_{x}) = P(Q_{x}|B^{+}) / (P(Q_{x}|B^{+}) + P(Q_{x}|B^{-}))$, and correspondingly  $P({\bar B}|Q_{x}) = 1-P(B|Q_{x})$.
If there is no OST information available for a given $\Bs$ meson, a probability of $0.5$ is assigned to that candidate.
 
\subsubsection*{Muon tagging}
For muon-based tagging, at least one additional muon is required in the event, with $\pT>2.5$~\GeV,
$|\eta|<2.5$ and $|\Delta z|<5$~mm, where $|\Delta z|$ is the difference in $z$ between the primary vertex and the longitudinal impact parameter of the ID track associated with the muon.
Muons are classified and kept if their identification quality selection working point is either Tight or Low-$\pt$;
these categories are subsequently treated as distinct flavour tagging methods.
For muons with $\pT>4$~\GeV, Tight muons are the dominant category, with the Low-$\mathit{\pt}$ requirement typically identifying muons of $\pt<4$~\GeV.
In the case of multiple muons satisfying selection criteria in one event, Tight muons are chosen over Low-$\pt$ muons. Within the same muon category, the muon with the highest $\pT$ that passes the selections is used.
 
\begin{figure}[htb]
\begin{center}
\includegraphics[width=0.48\textwidth]{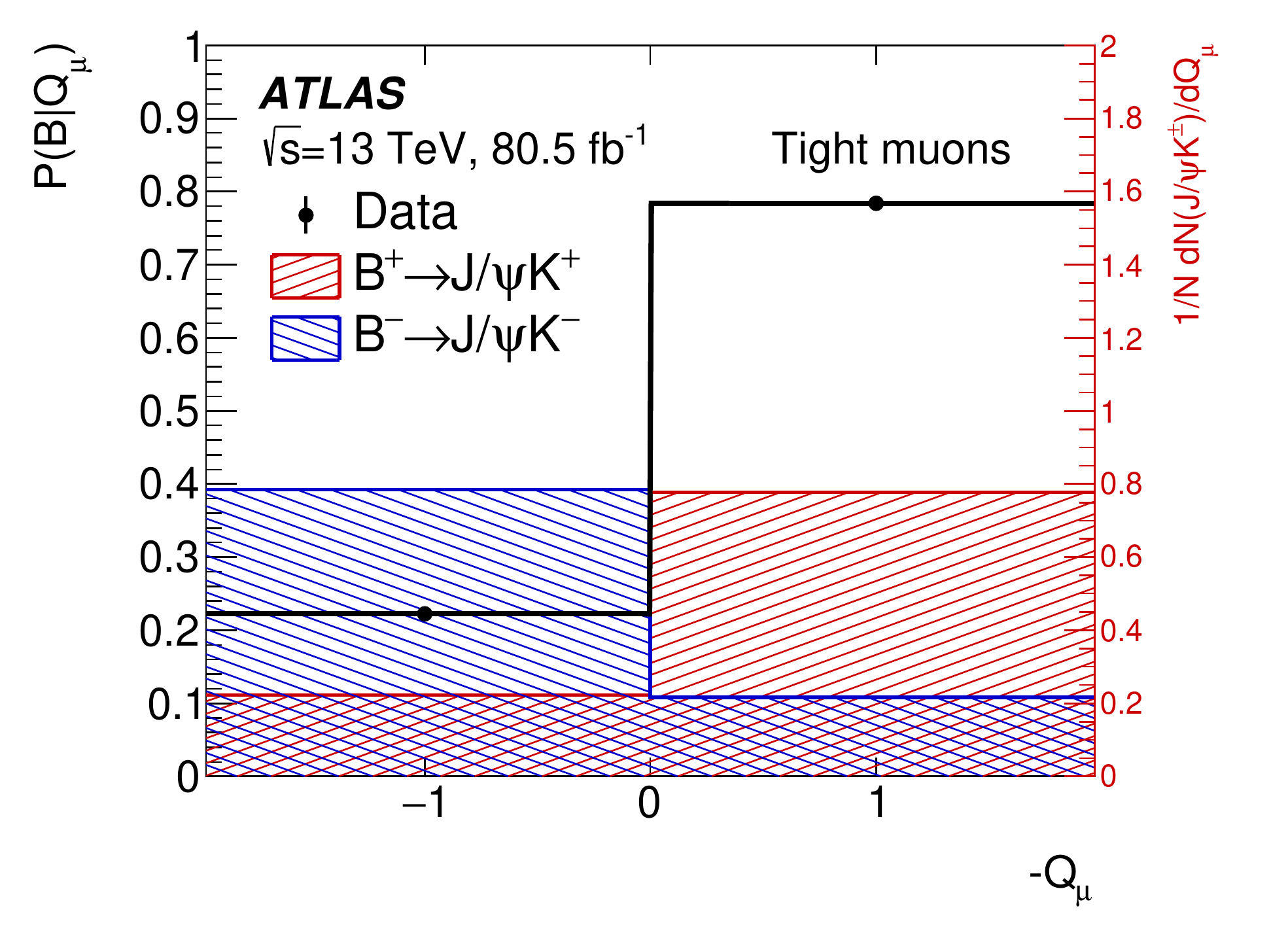} \hfill
\includegraphics[width=0.48\textwidth]{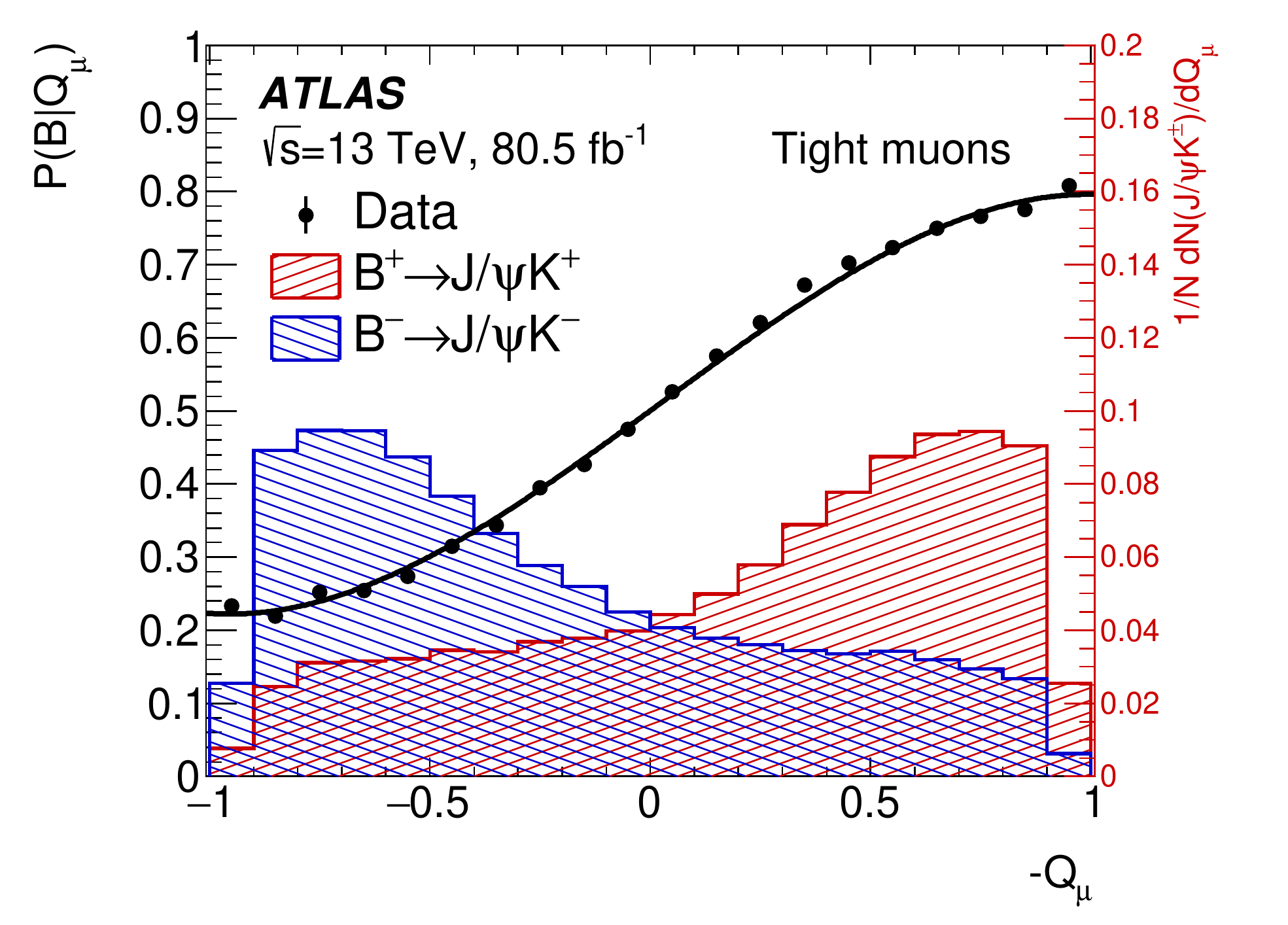} \hfill
\caption{Cone charge distributions, $-Q_{\mu}$, for Tight muons, shown for cases of discrete charge (left), and for the  continuous distribution (right).
For each plot, in red (blue), the normalised $B^{+}$ ($B^{-}$) cone charge distribution is shown (corresponding to the right axis scale).
A $B^{+}$ ($B^{-}$) candidate is more likely to have a large negative (positive) value of $Q_{\mu}$.
Superimposed is the distribution of the tagging probability, $P(B|Q_{\mu})$, as a function of the cone charge, derived from a data sample of $B^{\pm}\to J/\psi K^{\pm}$ decays, and defined as the probability to have a $B^{+}$ meson (on the signal-side) given a particular cone charge $Q_{\mu}$. The fitted parameterisation, shown in black, is used as the calibration curve to infer the probability to have a $\Bs$ or $\aBs$ meson at production in the decays to $\JpsiPhin$.\label{fig:chargedist:tightmu}}
\end{center}
\end{figure}

\begin{figure}[htb]
\begin{center}
\includegraphics[width=0.48\textwidth]{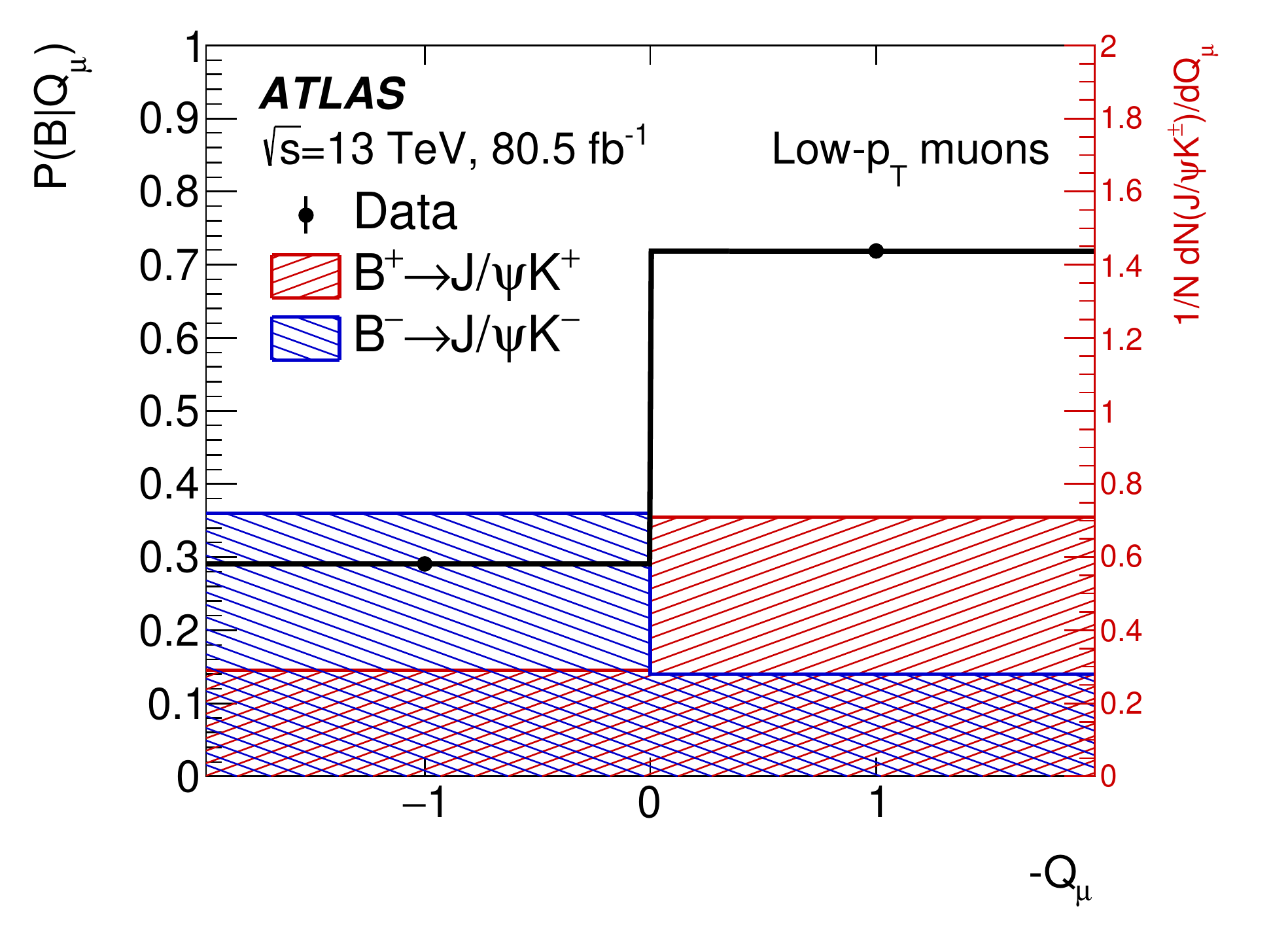} \hfill
\includegraphics[width=0.48\textwidth]{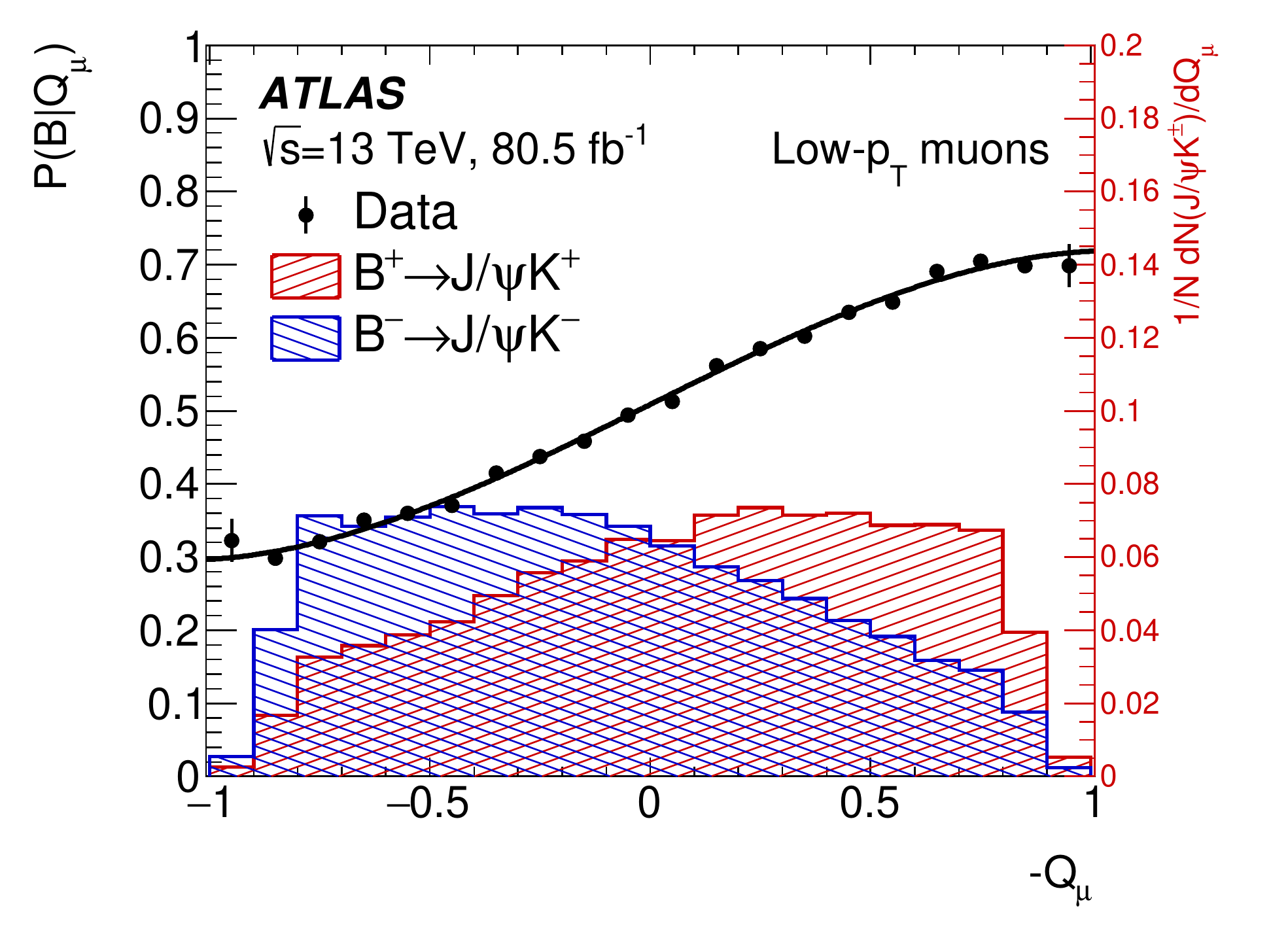} \hfill
\caption{Normalised cone charge distributions (shown against the right axis scale), $-Q_{\mu}$, for $B^{+}$ ($B^{-}$) events shown in red (blue)  for Low-$\pt$ muons, for cases of discrete charge (left), and for the  continuous distribution (right).
Superimposed is the distribution of the tagging probability, $P(B|Q_{\mu})$.\label{fig:chargedist:lowptmu}}
\end{center}
\end{figure}

A muon cone charge variable, $Q_{\mu}$, is constructed according to Eq.~(\ref{eq:conecharge}), with $\kappa=1.1$ and the sum over the reconstructed ID tracks within a cone of size $\Delta R=0.5$ around the muon direction. These tracks must have $\pt>0.5$~\GeV, $|\eta|<2.5$, and $|\Delta z|<5$~mm.
Tracks associated with the decay of a $B$ meson signal candidate are excluded from the sum.
In each interval of $Q_{\mu}$, a fit to the $\jpsi K^{\pm}$ invariant mass spectrum is performed and the number of signal events extracted. The fit model used is described in Section~\ref{sec:taggingBplus}.
Figure~\ref{fig:chargedist:tightmu} shows the distributions of the muon cone charge using $B^{\pm}$ signal candidates for Tight muons, and includes the tagging probability as a function of the cone charge variable. The corresponding distributions for Low-$\pt$ muons are shown in Figure~\ref{fig:chargedist:lowptmu}.
 
 
\subsubsection*{Electron tagging}
\begin{figure}[ht]
\begin{center}
\includegraphics[width=0.48\textwidth]{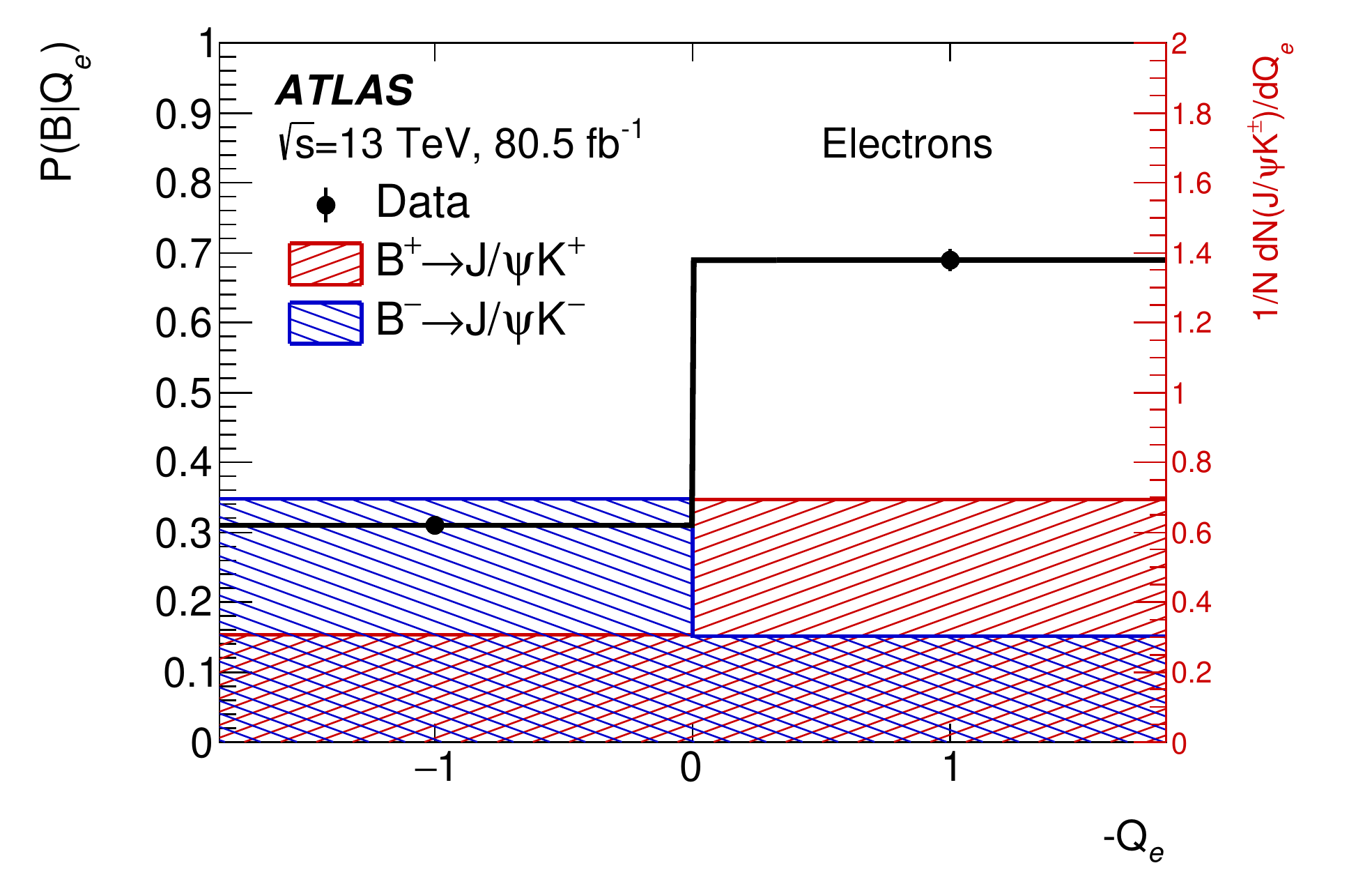} \hfill
\includegraphics[width=0.48\textwidth]{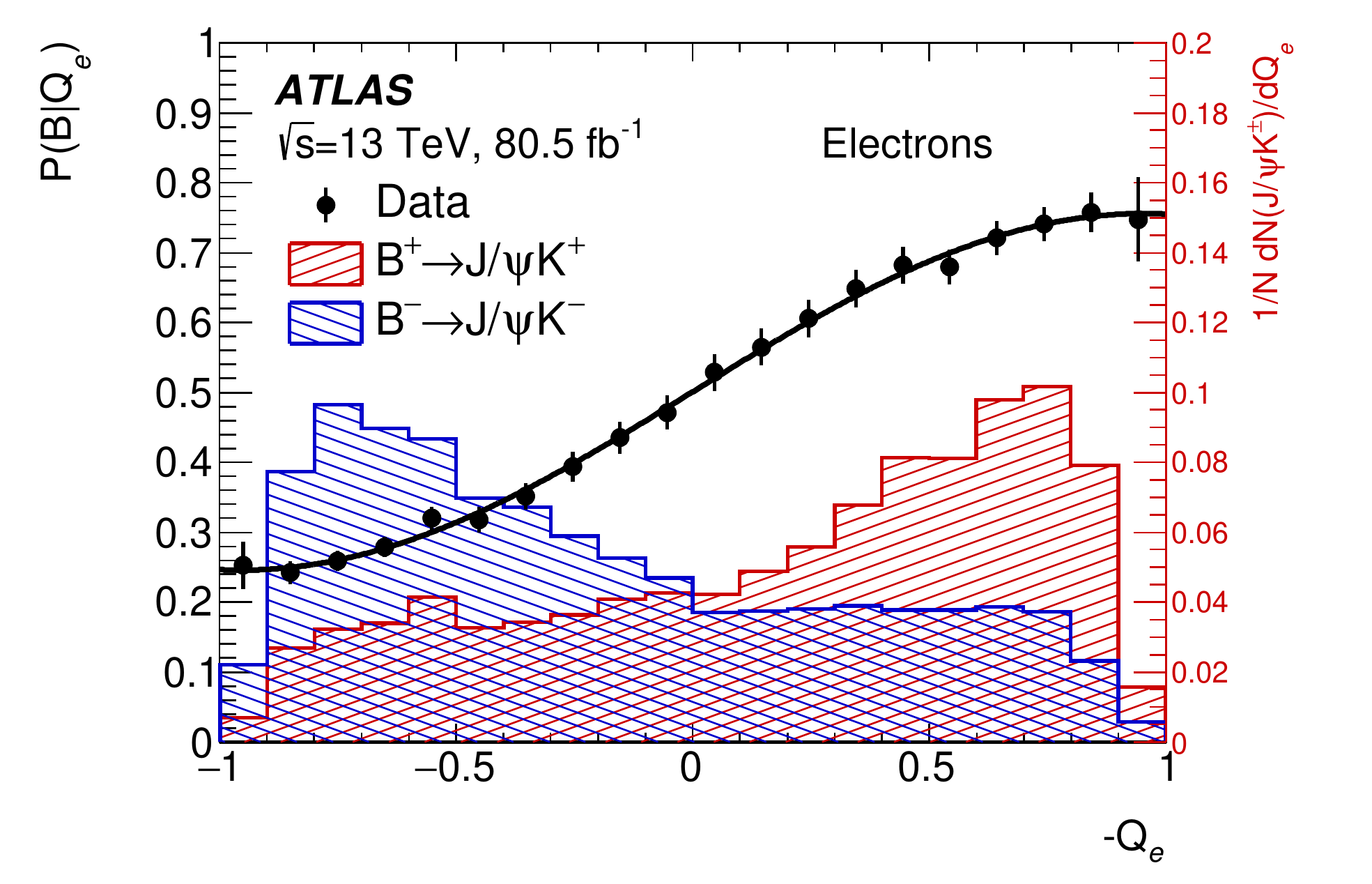} \hfill
\caption{Normalised cone charge distributions (shown against the right axis scale), $-Q_{e}$, for $B^{+}$ ($B^{-}$) events shown in red (blue)  for electrons, for cases of discrete charge (left), and the  continuous distribution (right).
Superimposed is the distribution of the tagging probabilities, $P(B|Q_{e})$.\label{fig:chargedist:el}}
\end{center}
\end{figure}

Electrons are identified using ID and calorimeter information, and must satisfy the \textit{Medium} electron quality criteria~\cite{EGAM-2018-01}. The ID track associated with the electron is required to have $\pT>0.5$ \GeV, $|\eta|<2.5$, and $|\Delta z|<5$~mm.
To reject electrons from the signal-side of the decay, electrons with $\cos(\zeta_{b})>0.93$, where $\zeta_{b}$ is the
opening angle between the momentum of the signal $B$ meson candidate and the electron momentum, are not considered.
In the case of more than one electron passing the selection, the electron with the highest $\pT$ is chosen.
Charged particle tracks within a cone of size $\Delta R = 0.5$ are used to form the electron cone charge $Q_{e}$, constructed according to Eq.~(\ref{eq:conecharge}), with $\kappa=1.0$.
The resulting electron cone charge distributions are shown in Figure~\ref{fig:chargedist:el},
together with the corresponding tagging probability.

\subsubsection*{Jet tagging}
In the absence of a muon or electron, a jet identified as containing a $b$-hadron is required.
Jets are reconstructed from calorimetric information~\cite{PERF-2014-07} using the anti-${k_t}$ algorithm~\cite{Cacciari:2008gp,Fastjet} with a radius parameter $R = 0.4$.
The identification of a $b$-tagged jet uses a multivariate algorithm~\textit {MV2c10}~\cite{FTAG-2018-01}, utilising boosted decision trees (BDT), which output a classifier value. Jets are selected if this value exceeds $0.56$. This value is chosen to maximise the tagging power of the calibration sample. In the case of multiple selected jets, the jet with the highest value of the BDT output classifier is used. Jets associated with the signal decay are not considered in this selection.
 
\begin{figure}[ht]
\begin{center}
\includegraphics[width=0.48\textwidth]{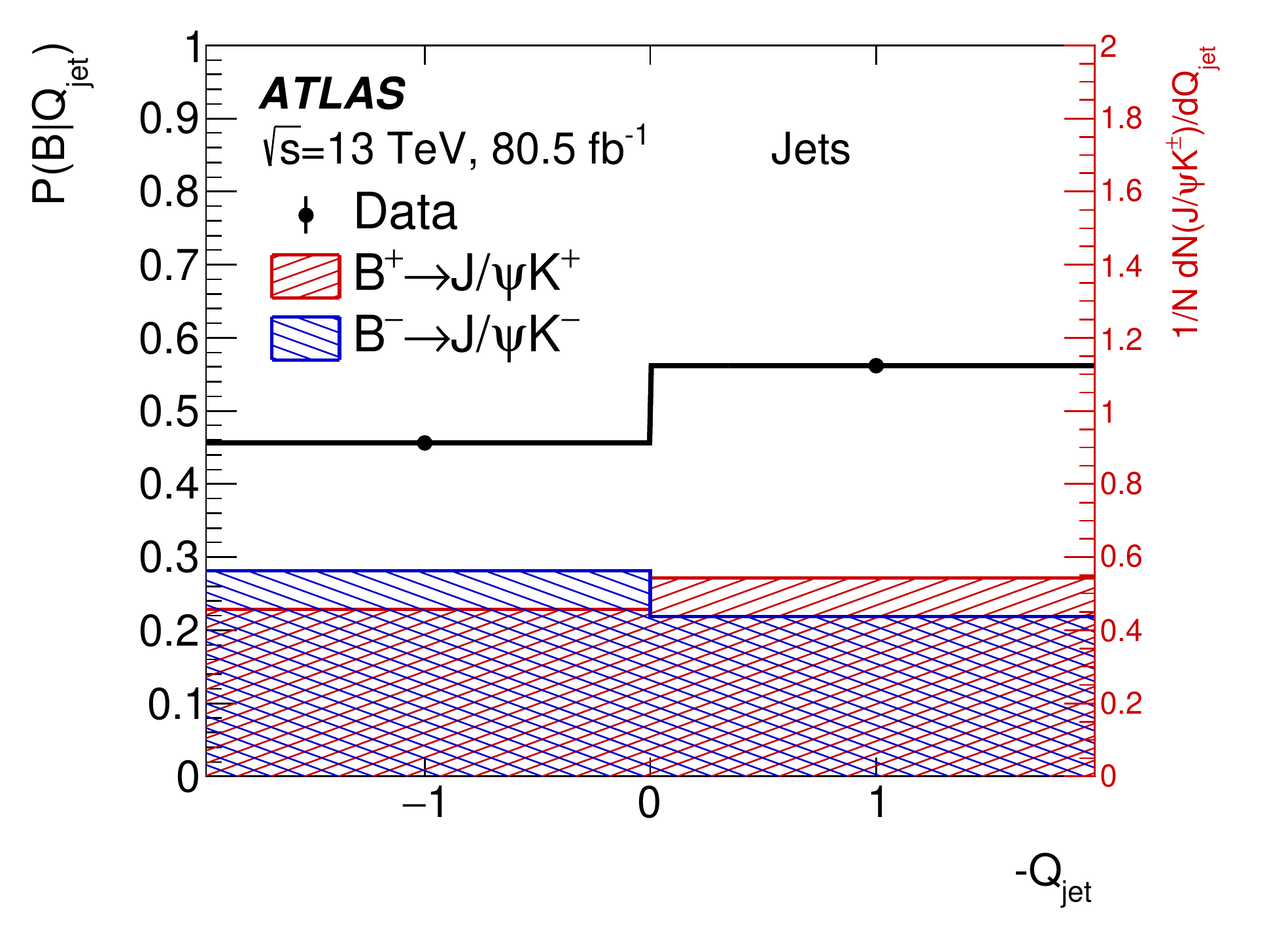} \hfill
\includegraphics[width=0.48\textwidth]{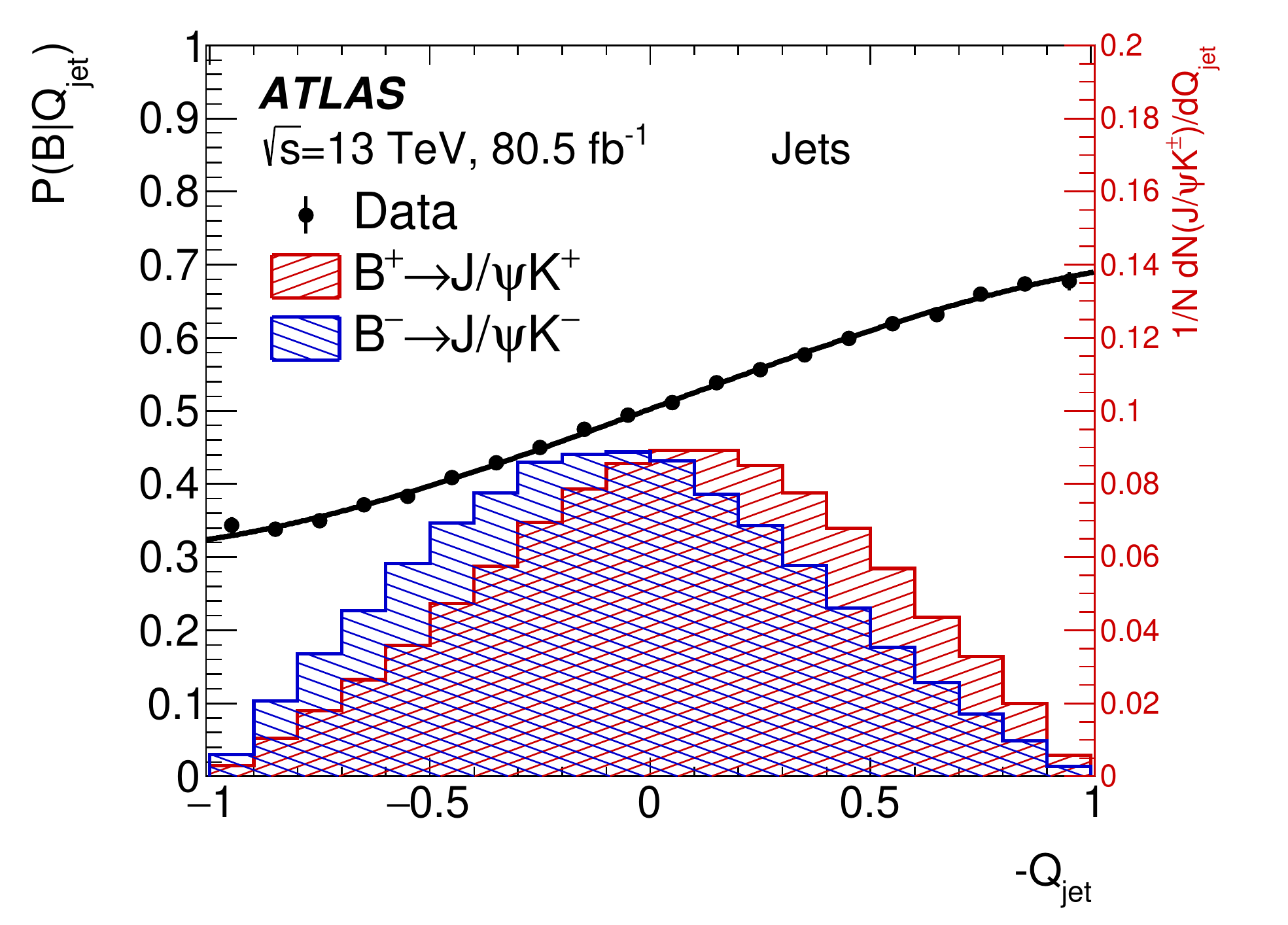} \hfill
\caption{Normalised cone charge distributions (shown against the right axis scale), $-Q_\mathrm{jet}$, for $B^{+}$ ($B^{-}$) events shown in red (blue)  for jets, for cases of discrete charge (left), and the  continuous distribution (right).
Superimposed is the distribution of the tag probability, $P(B|Q_\mathrm{jet})$.\label{fig:chargedist:jet}}
\end{center}
\end{figure}

Tracks within a cone of size $\Delta R = 0.5$ around the jet axis are used to define a jet cone charge, $Q_\mathrm{jet}$, constructed according to Eq.~(\ref{eq:conecharge}),
where $\kappa=1.1$ and the sum is over the tracks associated with the jet, with
$|\Delta z|<5$~mm, and excluding tracks from the decay of the signal $B$ meson candidate.
Figure~\ref{fig:chargedist:jet} shows the distribution of the opposite-side jet cone charge for  $B^{\pm}$ signal candidates.

\subsection{Flavour tagging performance\label{sec:tagperf}}
In order to quantify and compare the performance of the various tagging methods, three figure-of-merit terms are constructed, which describe: the fraction of events used by a given tagging method, the purity of the method, and the overall power of the tagging method in the sample.
The efficiency, $\epsilon_{x}$, of an individual tagging method is defined as the number of signal events tagged by that method divided by the total number of signal events in the sample.
The purity of a particular flavour tagging method, called the dilution, is defined as  ${\mathcal D}(Q_{x}) = 2 P({B}|Q_{x}) - 1$.
The tagging power of a particular tagging method is then defined as $T_{x}= \sum_{i} \epsilon_{x\,i} \cdot {\mathcal D}^2(Q_{x\,i})$, where the sum is over the probability distribution in intervals of the cone charge variable.
An effective dilution, $D_{x}=\sqrt{T_{x}/\epsilon_{x}}$, is calculated from the measured tagging power and efficiency.

By definition, there is no overlap between lepton-tagged and jet-charge-tagged events. The overlap between events with a muon (either Tight or Low-$\pt$) and events with an electron corresponds to around 0.6\% of all tagged events. In the case of multiply tagged events, the OST method is selected in the following order: Tight muon, electron, Low-$\pt$ muon, jet. However, the ordering of muon- and electron-tagged events is shown to have negligible impact on the final results.
A summary of the tagging performance for each method and the overall performance on the $B^{\pm}$ sample is given in Table~\ref{tab:taggingresults}.
 
\begin{table}[t]
\begin{center}
\caption{Summary of tagging performances for the different flavour tagging methods  on the sample of $B^{\pm}$ signal candidates, as described in the text. Uncertainties shown are statistical only.  The efficiency ($\epsilon_{x}$) and tagging power ($T_{x}$) are each determined by summing over the individual bins of the cone charge distribution. The effective dilution ($D_{x}$) is obtained from the measured efficiency and tagging power. For the efficiency, effective dilution, and tagging power, the corresponding uncertainty is determined by combining the appropriate uncertainties in the individual bins of each charge distribution.
\label{tab:taggingresults}}
\begin{tabular}{l|c|c|c}
\hline
Tag method &  $\epsilon_{x}$ $[\%]$ & $D_{x}$ $[\%]$ & $T_{x}$ $[\%]$ \\
\hline\hline
Tight muon         & $\phantom{0}4.50 \pm 0.01 $ & $ 43.8  \pm 0.2 $  & $ 0.862 \pm 0.009$ \\
Electron           & $\phantom{0}1.57 \pm 0.01 $ & $ 41.8  \pm 0.2 $  & $ 0.274 \pm 0.004$ \\
Low-$\pt$ muon     & $\phantom{0}3.12 \pm 0.01 $ & $ 29.9  \pm 0.2 $  & $ 0.278 \pm 0.006$ \\
Jet                & $12.04 \pm 0.02 $ & $ 16.6  \pm 0.1 $  & $ 0.334 \pm 0.006$ \\
\hline
Total &             $ 21.23 \pm 0.03 $ & $ 28.7 \pm 0.1 $ & $ 1.75\phantom{0} \pm 0.01\phantom{0} $ \\
\hline
\end{tabular}
\end{center}
\end{table}
\subsection{Using tag information in the $B^0_s$ fit}
\label{sec:taggingInFit}
 
For the maximum likelihood fit performed on the $\Bs$ data, and described in detail in Section~\ref{sec:fit}, the per-candidate probability, $P(B|Q_{x})$, that the $B$ meson candidate was produced in a state $\Bs$ (versus a $\aBs$)  is provided by the calibrations derived from the $B^{\pm}\to J/\psi K^{\pm}$ sample, described above, and shown in Figures~\ref{fig:chargedist:tightmu}--\ref{fig:chargedist:jet}.
Since the distributions of $P(B|Q_{x})$  from signal $\Bs$ mesons and backgrounds can be expected to be different, separate probability density functions (PDFs) are necessary to describe these distributions in the likelihood function.
These PDFs are defined as $P_\textrm{s}(P(B|Q_{x}))$ and $P_\textrm{b}(P(B|Q_{x}))$, describing the probability distributions for signal and background, respectively, and are derived from the sample of $\Bs$ candidates.
For the exclusive decays $B_d \ra J/\psi K^{0*}$ and $\Lambda_b \ra J/\psi p K^{-}$  that are present in the sample of $\Bs$ candidates, $P_\textrm{s}(P(B|Q_{x}))$  is used to model the  probability distributions  for these contributions (described further in Section~\ref{sec:fitBackground}).
The PDFs consist of the fraction of events that are tagged with a particular method (or are untagged), the fractions of those events categorised as discrete or continuous, and for those that are continuous, a PDF of the corresponding probability distribution.

\subsubsection*{Continuous PDF}
The parameterisations of the continuous PDF components of $P_\textrm{s,b}(P(B|Q_{x}))$ for each OST method are defined as follows.
In the sideband regions, $5.150 < m(J/\psi KK) <5.317$~\GeV\ and $5.417 < m(J/\psi KK) < 5.650$~\GeV, unbinned maximum likelihood fits to the $P(B|Q_x)$ distributions are performed to extract the background (continuous category) PDFs for $P_\textrm{b}(P(B|Q_{x}))$.
For the Tight muon and electron methods, the parameterisation has the form of the sum of a second-order polynomial and two exponential functions.  A Gaussian function is used for the Low-$p_\mathrm{T}$ muons. For the jet  tagging algorithm an eighth-order polynomial is used.
 
For the signal, fits are performed to the $P(B|Q_{x})$ distributions, using all events in the
$m(J/\psi KK)$ distributions to extract the signal (continuous category) PDFs for
$P_\textrm{s}(P(B|Q_{x}))$.
In these fits, the parameters describing the background PDFs are fixed to their previously extracted values, as is the relative normalisation of signal and background, extracted from a fit to the $m(J/\psi KK)$ distribution.
For the signal PDFs,  the Tight muon tagging method uses the sum of two exponential functions and a constant function to describe the signal.
For the electron tagging method, the signal function has the form of the sum of a second-order polynomial and two exponential functions, and for the Low-$p_\mathrm{T}$ muon and jet tagging methods a Gaussian function is used.
 
\subsubsection*{Discrete PDF}
In the case where the cone charge is discrete, the  fractions of events $f_{+1}$ ($f_{-1}$)  with  cone charges $+1$ ($-1$)
are determined separately for signal and background using events from the signal and sideband regions of the \Bs\ mass distribution (as defined in Section~\ref{sec:selection}).
The remaining fraction of events, $1 - f_{+1} - f_{-1}$, corresponds to the continuous parts of the distribution.
Positive and negative charges are equally probable for background candidates formed from a random combination of a
$J/\psi$  and a pair of tracks,  but this is not necessarily the case for background candidates formed from  a partially reconstructed $b$-hadron.
Table~\ref{tab:TagPunziSpkes} summarises the fractions $f_{+1}$ and $f_{-1}$ obtained from each tagging method for signal and background events.
\begin{table}[!th]
\begin{center}
\caption{Fractions $f_{+1}$  and $f_{-1}$ of events with cone charges of $+1$ and $-1$, respectively, for signal and background events and for the different tagging methods. Only statistical uncertainties are given.\label{tab:TagPunziSpkes} }
\begin{tabular}{l|cc|cc}
\hline
Tag method & \multicolumn{2}{c|}{Signal} & \multicolumn{2}{c}{Background} \\
& $f_{+1}$ [\%]& $f_{-1}$ [\%] & $f_{+1}$ [\%] &$f_{-1}$ [\%]\\
\hline \hline
Tight muon     & $\phantom{0}6.9\phantom{0} \pm 0.3\phantom{0}$ & $\phantom{0}7.5\phantom{0} \pm 0.3\phantom{0}$ & $\phantom{0}4.7\phantom{0} \pm 0.1\phantom{0}$ &$\phantom{0}4.9\phantom{0}  \pm 0.1\phantom{0}$  \\
Electron       & $20\phantom{.00}  \pm 1\phantom{.00}$          & $19\phantom{.00}  \pm 1\phantom{.00}$          & $16.8\phantom{0}\pm 0.2\phantom{0}$            &$17.3\phantom{0} \pm 0.2\phantom{0}$  \\
Low-$\pt$ muon & $10.9\phantom{0}\pm 0.5\phantom{0}$          & $11.6\phantom{0}\pm 0.5\phantom{0}$            & $\phantom{0}7.0\phantom{0} \pm 0.1\phantom{0}$ &$\phantom{0}7.5\phantom{0}  \pm 0.1\phantom{0}$  \\
Jet            & $\phantom{0}3.60\pm 0.15$                      & $\phantom{0}3.54\pm 0.15$                      & $\phantom{0}3.05\pm 0.03$                      &$\phantom{0}3.17 \pm 0.03$ \\
\hline
\end{tabular}
 
\end{center}
\end{table}
 
The fractions of signal and background events tagged using the different OST methods
are found using a similar sideband-subtraction method, and are summarised in
Table~\ref{tab:TagPunziRelat}.
 
\begin{table}[!th]
\begin{center}
\caption{Fractions of signal and background events tagged
using the different methods. The efficiencies include both the continuous and discrete contributions.
Only statistical uncertainties are quoted. \label{tab:TagPunziRelat}}
\begin{tabular}{l|c|c}
\hline
Tag method & Signal efficiency [\%] & Background efficiency [\%] \\
\hline\hline
Tight muon & $\phantom{0}4.06\pm 0.06$            & $\phantom{0}3.21 \pm 0.01$  \\
Electron    & $\phantom{0}1.86\pm 0.04$            & $\phantom{0}1.48 \pm 0.01$  \\
Low-$\pt$ muon & $\phantom{0}2.95\pm 0.05$            & $\phantom{0}2.70 \pm 0.01$  \\
Jet         & $12.1\phantom{0}\pm 0.1\phantom{0}$  & $\phantom{0}9.41\pm 0.02$  \\
Untagged    & $79.1\phantom{0}\pm 0.3\phantom{0}$  & $83.20\pm 0.05$  \\
\hline
\end{tabular}
\end{center}
\end{table}
 
To account for possible deviations of the data from the selected fit models, variations of the procedure described here are used to determine systematic uncertainties, as described in Section~\ref{sec:systematics}.
\section{Maximum likelihood fit\label{sec:fit}}
 
An unbinned maximum likelihood fit is performed on the selected events to extract the parameter values of the \Bsto\ decay.
The fit uses information about the reconstructed mass, $m$, the measured proper decay time, $t$,
the measured mass uncertainty, $\sigma_m$, the measured proper decay time uncertainty, $\sigma_t$, the measured transverse momentum, \pT,  the tagging probability, $P(B|Q_{x})$, and the transversity angles, $\Omega$, of each \Bst\ decay candidate.
The measured value of the proper decay time uncertainty, $\sigma_t$, is calculated from the covariance matrix associated with the vertex fit for each candidate event.
The transversity angles $\Omega = (\theta_T,\psi_T,\phi_T)$ are defined in Section \ref{sec:fitSignal}.
The likelihood function is defined as a combination of the signal and background PDFs as follows:
 
\vspace*{-0.2cm} 
{\small
\begin{alignat}{3} \label{eq:likelihood}
\ln~{\cal L}  =  \sum_{i=1}^N    w_i \cdot  \mathrm{ln} [  f_{\textrm{s}} \cdot   {\cal F}_{\textrm{s}}(  m_{i}, t_{i},  \sigma_{m_{i}}, \sigma_{t_{i}}, \Omega_i , P_{i}(B|Q_{x}), {\mathit p_{\mathrm{T}_{i}}} )
+  f_{\textrm{s}} \cdot f_{B^0} \cdot {\cal F}_{B^0}  (  m_{i}, t_{i}, \sigma_{m_{i}}, \sigma_{t_{i}}, \Omega_i , P_{i}(B|Q_{x}), {\mathit p_{\mathrm{T}_{i}}} ) \nonumber \\
+  f_{\textrm{s}} \cdot f_{\Lambda_b} \cdot {\cal F}_{\Lambda_b}  (  m_{i}, t_{i}, \sigma_{m_{i}}, \sigma_{t_{i}}, \Omega_i , P_{i}(B|Q_{x}), {\mathit p_{\mathrm{T}_{i}}} ) \nonumber \\
\vphantom{\sum_{i=1}^N} + ( 1 -  f_{\textrm{s}} \cdot (1 + f_{B^0} + f_{\Lambda_b})    )  {\cal F}_{\textrm{bkg}} (  m_{i}, t_{i}, \sigma_{m_{i}}, \sigma_{t_{i}}, \Omega_i, P_{i}(B|Q_{x}), {\mathit p_{\mathrm{T}_{i}}} )    ],   \nonumber \\
\end{alignat}
}
\noindent  where $N$ is the number of selected candidates,  $w_i$ is a weighting factor to account for the trigger efficiency (described in Section~\ref{sec:fitEfficiency}). The terms ${ \cal F}_{\textrm{s} }$, ${\cal F}_{B^0}$, ${\cal F}_{\Lambda_b}$ and  ${\cal F}_{\textrm{bkg} }$  are the PDFs modelling the signal, $B^0$  background, $\Lambda_b$ background, and the other background distributions, respectively. The term
$f_\textrm{s}$ is the fraction of signal candidates and $f_{B^0}$ and $f_{\Lambda_b}$ are the background  fractions of $B^0$ mesons and $\Lambda_b$ baryons misidentified as \Bs\ candidates, calculated relative to the number of signal events.
These background fractions are fixed to their expectation from the MC simulation, and variations are applied as part of the evaluation of the effects of systematic uncertainties. The  mass $m_{i}$, the proper decay time $t_{i}$ and the decay angles $\Omega_i $ are the values measured from the data for each event $i$.
A detailed description of the signal PDF terms in Eq.~(\ref{eq:likelihood}) is given in Section \ref{sec:fitSignal}.  The three background functions are described in Section \ref{sec:fitBackground}.

\subsection{Signal PDF}
\label{sec:fitSignal}
The PDF used to describe  the signal events, ${\cal F}_{\textrm{s}}$, has the following composition:
\begin{eqnarray*}
{\cal F}_{\textrm{s}}(  {\mathit m_{i}, t_{i},} \sigma_{m_{i}}, \sigma_{t_{i}}, \Omega_i, P_{i}(B|Q_{x}), {\mathit p_{\mathrm{T}_{i}}} ) & =  & P_\textrm{s}(m_{i}|\sigma_{m_{i}})  \cdot P_\textrm{s}(\sigma_{m_{i}}|{\mathit p_{\mathrm{T}_{i}}}) \cdot P_\textrm{s}(t_{i},\Omega_i|\sigma_{t_{i}},P_{i}(B|Q_{x})) \nonumber \\
& & \cdot P_\textrm{s}(\sigma_{t_{i}}|{\mathit p_{\mathrm{T}_{i}}}) \cdot P_\textrm{s}(P_{i}(B|Q_{x})) \cdot A(\Omega_i,{\mathit p_{\mathrm{T}_{i}}})  \cdot P_\textrm{s}({\mathit p_{\mathrm{T}_{i}}}).
\end{eqnarray*}
The mass  term $P_\textrm{s}(m_{i}|\sigma_{m_{i}})$ is modelled  in the following way:
\begin{equation}
{ P}_{\mathrm {s}}  (\mathit{m}_{i}|\sigma_{m_{i}} )  \equiv  \frac{1}{  \sqrt{2\pi}  S_m   \sigma_{m_{i}}     } \cdot
\mathrm{e}^{   \frac{   -(m_i -m_{B_s})^{2}   }     {2( S_m \sigma_{m_{i}}    )^{2}  }  }.
\label{pdfsig}
\end{equation}
The   term $P_\textrm{s}(m_{i}|\sigma_{m_{i}})$  uses per-candidate mass errors, $\sigma_{m_{i}}$,  calculated for each  \JpsiPhi  \  candidate from the covariance matrix associated with the four-track vertex fit. Each measured candidate mass is convolved with a Gaussian function with a width equal to $\sigma_{m_{i}}$ multiplied by a scale factor  $S_m$, introduced to account for any mismeasurements. Both  $S_m$ and  the mean value $m_{B_s} $, which is the   \Bs \ meson  mass,  are free parameters determined in the fit.
 
The  PDF term $P_\textrm{s}(t_{i}, \Omega_i|\sigma_{t_{i}}, P_{i}(B|Q_{x}))$ takes into account the lifetime resolution, so each time element in Table \ref{tab:SigPDF} is convolved  with a Gaussian function defined as:
 
\begin{equation}
{R}(t^{'}-t_{i}, \sigma_{t_{i}}) \equiv \frac{1}{\sqrt{2\pi}~S_{t}~\sigma_{t_{i}}} \cdot
\mathrm{e}^{\frac{-(t^{'}_i-t_{i})^{2}} {2(S_{t} \sigma_{t_{i}})^{2}}}.
\label{Gausst}
\end{equation}

\begin{figure}[ht]
\centering
\includegraphics[width=0.5\textwidth]{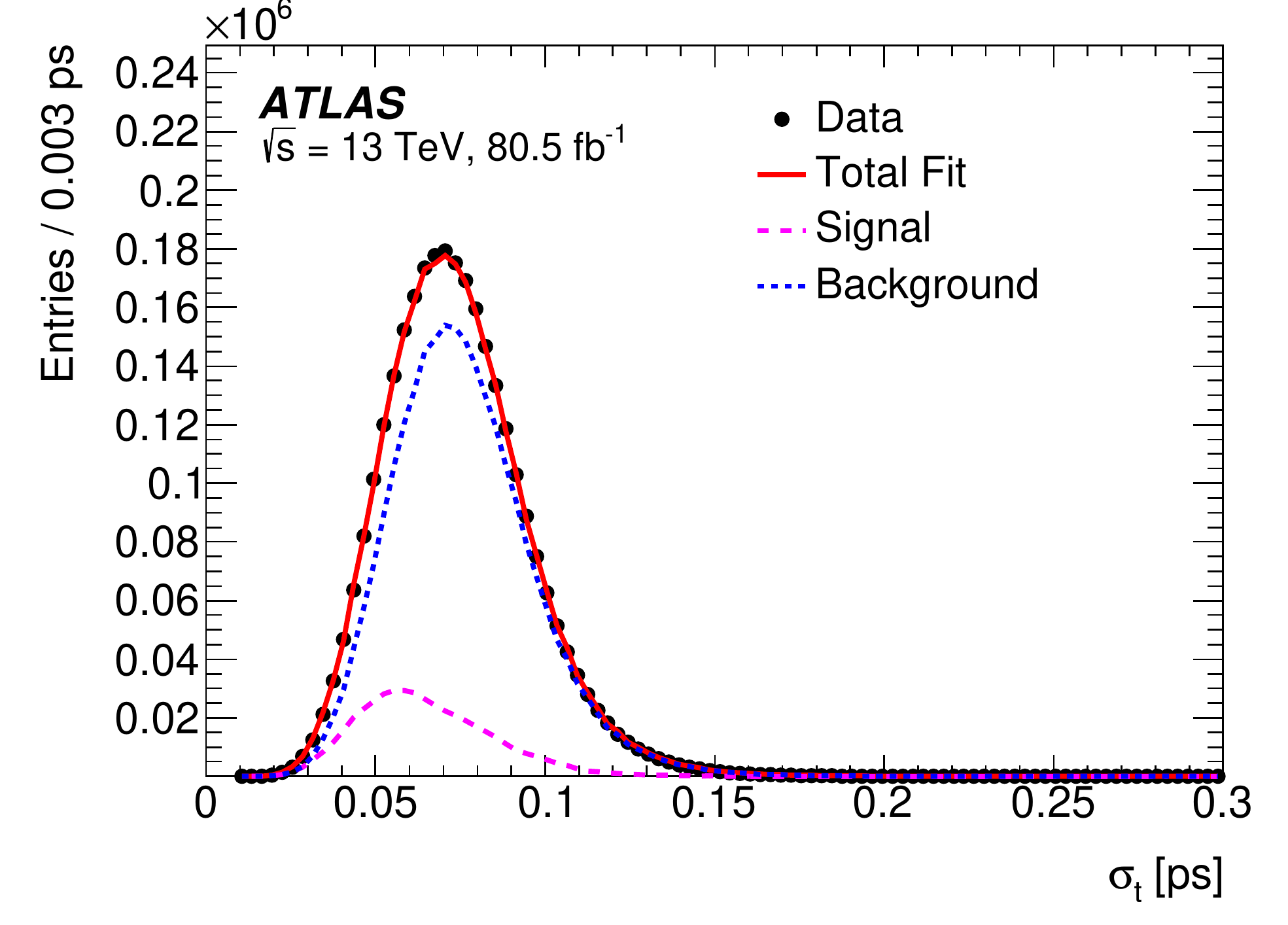}
\caption{The proper decay time uncertainty distribution for data (black), and the fits to the background (blue) and the signal (purple) contributions. The total fit is shown as a red curve.}
\label{fig:timeErr}
\end{figure}

$S_{t}$ is a scale factor (a parameter of the fit) and $\sigma_{t_{i}}$ is the per-candidate uncertainty on proper decay time $t_{i}$. This convolution is performed numerically on an event-by-event basis and the value  $\sigma_{t_{i}}$ is measured for each \Bs candidate, based on the tracking error matrix of the four final state particles. The probability term $P_\textrm{s}(\sigma_{t_{i}}|{\mathit p_{\mathrm{T}_{i}}})$ is   introduced  to account for differences between signal and background events  for the values of  the per-candidate time errors.  Distributions of this variable for signal and background described by gamma functions are shown in Figure \ref{fig:timeErr}. The average value of  the  time error for signal events is 69 fs.
 
The same approach was applied for the probability terms $P_\textrm{s}(\sigma_{m_{i}}|{\mathit p_{\mathrm{T}_{i}}})$ and $P_\textrm{s}({\mathit p_{\mathrm{T}i}})$ accounting for differences between signal and background events  for the values of  the per-candidate mass error and  ${\mathit p_{\mathrm{T}_{i}}}$  values, respectively. The tagging probability term for signal $P_\textrm{s}(P_{i}(B|Q_{x}))$ is described in Section~\ref{sec:taggingInFit}.

The term $P_\textrm{s}(t_{i}, \Omega_i|\sigma_{t_{i}}, P_{i}(B|Q_{x})) $ is a joint PDF for the decay time $t$ and the transversity angles $\Omega$ for the \Bsto\ decay.  Ignoring detector effects, the distribution for the time $t$ and the angles $\Omega$  is given by the differential decay rate \cite{CDFAngles}:
 
\begin{equation*}
\frac{\mathrm{d}^4 \Gamma}{\mathrm{d}t\ \mathrm{d}\Omega}= \sum_{k=1}^{10} {\cal O}^{(k)}(t) g^{(k)}(\theta_T,\psi_T,\phi_T),
\end{equation*}
where  ${\cal O}^{(k)}(t)$ are the time-dependent functions corresponding to the contributions of the four different amplitudes ($A_0$, $A_{||}$, $A_\perp$, and $A_S$) and their interference terms, and $g^{(k)}(\theta_T,\psi_T,\phi_T)$ are the angular functions.   Table \ref{tab:SigPDF} shows the time-dependent  and the angular functions of the transversity angles.  The formulae for the time-dependent functions have the same structure for \Bs\ and \aBs\ but with a sign reversal in the terms containing $\Delta m_s$,
which is a fixed parameter of the fit (using Ref.~\cite{PDG}).
The formalism used throughout this analysis assumes no direct \CP\ violation, i.e. $\lambda = (q/p)(\bar{A}/A) = 1$.
In Table \ref{tab:SigPDF}, the parameter $A_\perp(t)$ is the time-dependent amplitude for the \CP-odd final-state configuration while $A_0(t)$ and $A_\parallel(t)$ correspond to \CP-even final-state configurations.  The amplitude $A_S(t)$ gives the contribution from the \CP-odd non-resonant $B_s^0\rightarrow J/\psi K^+K^-$ $S$-wave state~\cite{Stone:2008ak} (which includes the $f_0$ meson). The corresponding functions are given in the last four lines of Table \ref{tab:SigPDF}  ($k=7$--$10$). The amplitudes are parameterised by $|A_i|$e$^{ i \delta_i }$, where $i = \{ 0, ||, \perp, S \}$, $\delta_0 = 0$ and are normalised such that $|A_0(0)|^2 + |A_\perp(0)|^2 +  |A_{\|}(0)|^2=1$. The amplitude $|A_\perp(0)|$ is determined according to this condition, while the remaining three amplitudes are parameters of the fit.  The value $|A_S|^2$ is the ratio of the $S$-wave yield to the $\phi \rightarrow K^+K^-$  yield in the interval of $m(K^+K^-)$ used in the analysis.
In the sum over the mass interval,  the  interference terms (lines 8--10 in Table~\ref{tab:SigPDF}) are corrected by a factor $\alpha$
that takes into account the mass-dependent differences in absolute amplitude and phase between the $\phi \rightarrow K^+K^-$ and the $S$-wave amplitudes.
The correction is based on the Breit–Wigner description of the $\phi $
and on model assumptions for the shape and the phase variations of the $S$-wave amplitude.
The phase $\delta_S$ is the phase difference between $A_0(0)$ and the $S$-wave amplitudes at the $\phi \rightarrow K^+K^- $ peak.
The values of $\alpha$ and the related systematic uncertainty are discussed in Section \ref{sec:systematics}.

The angles ($\theta_T,\psi_T,\phi_T$), are defined in the rest frames of the final-state particles.  The $x$-axis is determined by the direction of the $\phi$ meson in the $J/\psi$ rest frame, and the $K^{+}K^{-}$  system defines the $x$--$y$ plane, where $p_{y}(K^{+})>0$. The three angles are defined as:
\begin{itemize}
\item$\theta_T$, the angle between $\vec{p}(\mu^{+})$ and the normal to the $x$--$y$ plane, in the $J/\psi$ meson rest frame,
\item $\phi_T$, the angle between the $x$-axis and $\vec{p}_{xy}(\mu^{+})$, the projection of the $\mu^+$ momentum in the $x$--$y$ plane, in the $J/\psi$ meson rest frame,
\item $\psi_T$, the angle between $\vec{p}(K^{+})$ and $-\vec{p}(J/\psi)$ in the $\phi$ meson rest frame.
\end{itemize}
 
The angular acceptance of the detector and the kinematic cuts on the angular distributions are included in the likelihood function through $A(\Omega_i, {\mathit p_{\text{T}i}})$. This is calculated using a four-dimensional binned acceptance method, applying an event-by-event efficiency according to the transversity angles ($\theta_T,\psi_T,\phi_T$) and the \pT\ of the candidate. The \pT\  binning is necessary, because the angular acceptance is influenced by the \pT\  of the \Bs\ candidate.  The acceptance is calculated from the \Bst\ MC events with additional weighting for \pT\ and $\eta$ distributions. 
In the likelihood function, the acceptance is treated as an angular acceptance PDF, which is multiplied with the time- and angle-dependent PDF describing the \Bsto\ decays.  As both the acceptance and time- and angle-dependent decay PDFs depend on the transversity angles they must be normalised together. This normalisation is done numerically during the likelihood fit.
The PDF is normalised over the entire \Bs\  mass range, \cutMassBsRange.
 
\begin{sidewaystable}[hp]
\caption{The ten time-dependent functions, ${\cal O}^{(k)}(t)$ and the functions of the transversity angles $g^{(k)}(\theta_T,\psi_T,\phi_T)$. The amplitudes $|A_0(0)|^2$ and $|A_\parallel(0)|^2$ are for the \CP-even components of the \Bst\ decay,  $|A_{\perp}(0)|^2$ is the \CP-odd amplitude; they have corresponding strong phases $\delta_{0}$, $\delta_{\parallel}$ and $\delta_{\perp}$. By convention, $\delta_0$ is set to be zero.  The $S$-wave amplitude $|A_S(0)|^2$ gives the fraction of \BstoKK\ and has a related strong phase $\delta_S$. The factor $\alpha$ is described in the text of Section \ref{sec:fitSignal}.  The $\pm$ and $\mp$ terms denote two cases: the upper sign describes the decay of a meson that was initially a \Bs\ meson, while the lower sign describes the decays of a meson that was initially \aBs. \label{tab:SigPDF}}
\begin{center}
\renewcommand{\arraystretch}{2}
\scriptsize
\begin{tabular}{|c| l | l |}
\hline
$k$ & ${\cal O}^{(k)}(t)$                                                                                                                                                                                                                                                                                         & $g^{(k)}(\theta_T,\psi_T,\phi_T)$ \\
\hline \hline
1   & $\frac{1}{2} |A_0(0)|^2 \left[\left(1+\cos\phis\right) \text{e}^{-\Gamma_{\mathrm L}^{(s)} t} + \left(1-\cos\phis\right) \text{e}^{-\Gamma_{\mathrm H}^{(s)} t} \pm 2 \text{e}^{-\Gamma_st} \sin (\Delta m_s t)  \sin \phi_s \right]$                                                                                            & $2 \cos^2\psi_T(1 - \sin^2\theta_T\cos^2\phi_T)$ \\
2   & $\frac{1}{2}|A_{\|}(0)|^2 \left[\left(1+\cos\phis\right) \text{e}^{-\Gamma_{\mathrm L}^{(s)} t} + \left(1-\cos\phis\right) \text{e}^{-\Gamma_{\mathrm H}^{(s)} t} \pm 2 \text{e}^{-\Gamma_st} \sin (\Delta m_s t)  \sin \phi_s \right]$                                                                                          & $\sin^2\psi_T(1 - \sin^2\theta_T\sin^2\phi_T)$ \\
3   & $\frac{1}{2} |A_{\perp}(0)|^2 \left[\left(1- \cos\phis\right)\text{e}^{-\Gamma_{\mathrm L}^{(s)} t} + \left(1+\cos\phis\right)\text{e}^{-\Gamma_{\mathrm H}^{(s)} t} \mp 2 \text{e}^{-\Gamma_st} \sin (\Delta m_s t)  \sin \phi_s \right]$                                                                                       & $\sin^2\psi_T\sin^2\theta_T$ \\
4   & $\frac{1}{2}|A_0(0)||A_{\|}(0)|\cos\delta_{||} \left[\left(1+\cos\phis\right)\text{e}^{-\Gamma_{\mathrm L}^{(s)} t} + \left(1-\cos\phis\right)\text{e}^{-\Gamma_{\mathrm H}^{(s)} t} \pm 2 \text{e}^{-\Gamma_st} \sin (\Delta m_s t)  \sin \phi_s \right]$                                                                       & $\frac{1}{\sqrt{2}}\sin2\psi_T\sin^2\theta_T\sin2\phi_T $ \\
5   & $|A_{\|}(0)||A_{\perp}(0)| \left[  \frac{1}{2}(\text{e}^{-\Gamma_{\mathrm L}^{(s)} t}-\text{e}^{-\Gamma_{\mathrm H}^{(s)} t})\cos(\delta_{\perp} - \delta_{||})\sin\phis \pm \text{e}^{-\Gamma_st}(\sin (\delta_\perp -\delta_\|) \cos(\Delta m_s t) - \cos (\delta_\perp - \delta_\|)\cos \phi_s \sin (\Delta m_st) ) \right]$  & $-\sin^2\psi_T\sin2\theta_T\sin\phi_T$ \\
6   & $|A_{0}(0)||A_{\perp}(0)| \left[\frac{1}{2}(\text{e}^{-\Gamma_{\mathrm L}^{(s)} t}-\text{e}^{-\Gamma_{\mathrm H}^{(s)} t})\cos \delta_{\perp} \sin\phis \pm \text{e}^{-\Gamma_st}(\sin \delta_\perp \cos(\Delta m_s t) - \cos \delta_\perp \cos \phi_s \sin (\Delta m_st) ) \right]$                                             & $\frac{1}{\sqrt{2}} \sin2\psi_T\sin2\theta_T\cos\phi_T$ \\
7   & $\frac{1}{2} |A_{S}(0)|^2 \left[\left(1- \cos\phis\right)\text{e}^{-\Gamma_{\mathrm L}^{(s)} t} + \left(1+\cos\phis\right)\text{e}^{-\Gamma_{\mathrm H}^{(s)} t} \mp 2 \text{e}^{-\Gamma_st} \sin (\Delta m_s t)  \sin \phi_s \right]$                                                                                           & $\frac{2}{3}\left(1-\sin^2\theta_T\cos^2\phi_T\right) $ \\
8   & $ \alpha |A_{S}(0)||A_{\parallel}(0)| \left[\frac{1}{2}(\text{e}^{-\Gamma_{\mathrm L}^{(s)} t}-\text{e}^{-\Gamma_{\mathrm H}^{(s)} t})\sin (\delta_\| - \delta_S) \sin\phis \pm \text{e}^{-\Gamma_st}(\cos ( \delta_\| - \delta_S ) \cos(\Delta m_s t) - \sin ( \delta_\| - \delta_S ) \cos \phi_s \sin (\Delta m_st) ) \right]$ & $\frac{1}{3} \sqrt{6} \sin\psi_T\sin^2\theta_T\sin 2\phi_T$ \\
9   & $\frac{1}{2} \alpha|A_{S}(0)| |A_{\perp}(0)|\sin(\delta_{\perp}-\delta_{S}) \left[\left(1- \cos\phis\right)\text{e}^{-\Gamma_{\mathrm L}^{(s)} t} + \left(1+\cos\phis\right)\text{e}^{-\Gamma_{\mathrm H}^{(s)} t} \mp 2 \text{e}^{-\Gamma_st} \sin (\Delta m_s t)  \sin \phi_s \right]$                                         & $\frac{1}{3} \sqrt{6} \sin\psi_T\sin 2\theta_T\cos\phi_T$ \\
10  & $ \alpha |A_{0}(0)||A_{S}(0)| \left[\frac{1}{2}(\text{e}^{-\Gamma_{\mathrm H}^{(s)} t}-\text{e}^{-\Gamma_{\mathrm L}^{(s)} t})\sin  \delta_S \sin\phis \pm \text{e}^{-\Gamma_st}(\cos \delta_S \cos(\Delta m_s t) + \sin \delta_S \cos \phi_s \sin (\Delta m_st) ) \right]$                                                      & $\frac{4}{3}\sqrt{3}\cos\psi_T\left(1-\sin^2\theta_T\cos^2\phi_T\right)$ \\
\hline
\end{tabular}
\end{center}
\end{sidewaystable}

\subsection{Background PDF}
\label{sec:fitBackground}
 
The background PDF has the following composition:
\begin{eqnarray*}
{\cal F}_{\textrm{bkg}}(  {\mathit m_{i}, t_{i}}, \sigma_{t_{i}}, \Omega_i, P_{i}(B|Q_{x}), { \mathit p_{\mathrm{T}_{i}}} ) & = & P_\textrm{b}(m_{i})\cdot  P_\textrm{b}(t_{i}|\sigma_{t_{i}}) \cdot P_\textrm{b}(P_{i}(B|Q_{x}))  \nonumber \\
&   & \cdot P_\textrm{b}(\Omega_i)\cdot P_\textrm{b}(\sigma_{m_{i}}|{\mathit p_{\mathrm{T}_{i}}})  \cdot P_\textrm{b}(\sigma_{t_{i}}|{\mathit p_{\mathrm{T}_{i}}})  \cdot P_\textrm{b}({\mathit p_{\mathrm{T}_{i}}}).
\end{eqnarray*}
 
The proper decay time function $P_\textrm{b}(t_{i}|\sigma_{t_{i}})$ is parameterised as a peak modelled by a Gaussian distribution, two positive-time exponential functions and two negative-time exponential functions. These functions are convolved with the same resolution function, defined in Eq.~(\ref{Gausst}) as the signal decay time-dependence. The prompt peak models the combinatorial background events, which are expected to have reconstructed lifetimes distributed around zero. The two positive-time exponential functions represent a  fraction of longer-lived backgrounds with non-prompt $J/\psi$, combined with hadrons from the primary vertex or  from a  $ B/D$ meson in the same event.  The two negative-time exponential functions take into account events with poor vertex resolution. The probability terms $P_\textrm{b}(\sigma_{m_{i}}|{\mathit p_{\mathrm{T}_{i}}})$, $P_\textrm{b}(\sigma_{t_{i}}|{\mathit p_{\mathrm{T}_{i}}})$ and $P_\textrm{b}({\mathit p_{\mathrm{T}_{i}}})$ are described by gamma functions. They are unchanged from the  analysis described in  Ref.~\cite{BPHY-2011-05}  and explained in detail there.  The tagging probability term for background events $P_\textrm{b}(P_{i}(B|Q_{x}))$ is described in Section~\ref{sec:taggingInFit}.
 
The shape of the background angular distribution, $P_\textrm{b}(\Omega_i)$ arises primarily from detector and kinematic acceptance effects.  The best description is achieved by Legendre polynomial functions:
\begin{eqnarray} \label{eq:likelihoodBkgAng}
Y^m_l (\theta_T) & = &  \sqrt{(2l+1)/(4\pi)}\sqrt{(l-m)!/(l+m)!}\ P^{|m|}_l(\cos\theta_T),  \nonumber  \\
P_k(x)  & =  & \frac{1}{2^k k!} \frac{\mathrm{d}^k}{\mathrm{d}x^k}(x^2-1)^k, \nonumber  \\
P^{|m|}_l(x) & = & (-1)^{|m|}(1-x^2)^{|m|/2} \frac{\mathrm{d}^{|m|}}{\mathrm{d}x^{|m|}}(P_l(x)), \nonumber \\
P_{\mathrm{b}}(\theta_T,\psi_T,\phi_T)  & = &  \sum_{k=0}^{k_{max}}\sum_{l=0}^{l_{max}}\sum_{m=-l}^{l}
\begin{cases}
a_{k,l,m}\sqrt{2} Y^m_l(\theta_T) \cos(m\phi_T) P_k(\cos\psi_T)  & \textrm{where } m > 0 \nonumber \\
a_{k,l,m}\sqrt{2} Y^{-m}_l(\theta_T) \sin(m\phi_T) P_k(\cos\psi_T)  & \textrm{where } m < 0 \nonumber\\
a_{k,l,m}\sqrt{2} Y^{0}_l(\theta_T)  P_k(\cos\psi_T)  & \textrm{where } m = 0 \nonumber \\
\end{cases}
\end{eqnarray}
where $l_{max}=14$, $k_{max}=14$ and the coefficients $a_{k,l,m}$  are adjusted to give the best fit to the angular distributions for events in the sidebands of the \Bs\ mass distribution.  These parameters are then fixed in the main fit,  defined by Eq.~(\ref{eq:likelihood}).
The \Bs \ mass interval used for the background fit is between 5.150  and  5.650 \GeV\ excluding the signal mass region $|(m(B_s^0) -5.366|< 0.110$ \GeV.
Higher-order Legendre polynomial functions were tested as a systematic check, described  in
Section~\ref{sec:systematics}.
 
The background mass model, $P_\textrm{b}({\mathit m_{i}})$ is the sum of an exponential and a constant, with the exponential slope and the relative normalisation left as free parameters of the fit.
 
Contamination from $B_d \ra J/\psi K^{0*}$ and $\Lambda_b \ra J/\psi p K^{-}$ events misreconstructed as \Bst\ is accounted for in the fit through the ${\cal F}_{B^0}$ and ${\cal F}_{\Lambda_b}$ terms in the PDF described in Eq.~(\ref{eq:likelihood}). The PDFs  are determined using MC simulation of these decays. The fractions of these contributions, $f_{B^0} = \Bdfrac$ and $f_{\Lambda_b} = \Lambdafrac$, are defined relative to the number of the \Bst \ signal events and are evaluated  from  MC simulation using production cross sections and branching fractions from  Refs.~\cite{PDG,LHCb-CONF-2013-011,BABAR,LHCbLambdabProduction,1674-1137-40-1-011001,PhysRevLett.115.072001}.
 
MC simulated events are also used to determine the shape of the mass and transversity angle distributions. The 3D angular distributions of $B_d^0\rightarrow J/\psi K^{0*}$ and of the conjugate decay are modelled using input from  Ref.~\cite{LHCbK0Star}, while angular distributions for $\Lambda_b \ra J/\psi p K^{-}$ and the conjugate decay are considered flat. These distributions are sculpted for detector acceptance effects and then described by Legendre polynomial functions with $l_{max}=10$ and $k_{max}=10$, Eq.~(\ref{eq:likelihoodBkgAng}). These shapes are  used as templates in the fit. The $B_d$ and $\Lambda_b$ lifetimes are accounted for in the fit by adding additional exponential terms, scaled by the  ratio of $B_d$/$B_s^0$ or $\Lambda_b$/$B_s^0$ masses as appropriate, where the lifetimes and masses are taken from Ref.~\cite{PDG}.
The PDF terms that describe each of the tagging,  mass,  decay time and $\pT$ probability distributions are taken from the same PDFs used to describe the \Bst\  signal (described in Sections~\ref{sec:taggingInFit} and~\ref{sec:fitSignal}) to account for the fact that these dedicated background events are fully reconstructed $b$-hadron decays.
Systematic uncertainties due to the background from $B_d \ra J/\psi K^{0*}$ and $\Lambda_b \ra J/\psi p K^{-}$ decays are described in Section \ref{sec:systematics}. The contribution of the $S$-wave $B_d \ra J/\psi K\pi$ decays as well as their interference with the $P$-wave $B_d \ra J/\psi K^{0*}$ decays  are included in the PDF of the fit, using the parameters measured in Ref. \cite{LHCbK0Star}.
 
\subsection{Proper decay time dependence of muon trigger efficiency}
\label{sec:fitEfficiency}
 
In the triggers used in this analysis, there is no minimum cut applied on the transverse impact parameter $d_0$ of muons. On the other side, trigger muons with values of $d_0$ > 10 mm are not accepted.  This results in inefficiency at large values of the proper decay time. This inefficiency is  estimated using MC simulated events,  by comparing  the \Bs\ proper decay time distribution  obtained before and after applying the trigger selection.   To account for this inefficiency in the fit, the events are reweighted by a factor $w$, inversely proportional to the trigger efficiency:
\begin{equation}
1/w =     p_0 \cdot [ 1 - p_1 \cdot (\textrm{Erf} ((t-p_3)/p_2) + 1)],
\label{eq:LifeTimeWeight}
\end{equation}
where Erf denotes the error function and  $p_0, p_1, p_2$ and $p_3$ are parameters determined in the fit to MC events.
For more than 99\% of the \Bs\ candidates the inefficiency is below 2\%. For the most affected candidates the inefficiency reaches up to 50\% at high decay time.
No significant bias or inefficiency due to offline track reconstruction, vertex reconstruction, or track quality selection criteria is observed.
\subsection{Summary of the fit parameters}

The joint PDF of proper decay time and decay angles includes the main physics
parameters of interest:
\begin{itemize}
\item
the \CP-violating phase \phis,
\item
the average decay width \Gs\ and the decay width difference \DGs,
\item
the size of the \CP-state amplitudes at $t=0$: $|A_{\parallel}(0)|^2$,
$|A_{0}(0)|^2$ and their corresponding strong phases $\delta_{\perp}$ and
$\delta_{\parallel}$
\item
and the size of the $S$-wave amplitude at $t=0$: $|A_{S}(0)|^2$ and
corresponding strong phase $\delta_{S}$.
\end{itemize}
The size of the remaining amplitude $|A_{\perp}(0)|^2$ is constrained by the
normalisation condition, phase $\delta_{0}$ is set to zero and $\Delta m_s$ is
fixed as mentioned above.

The likelihood function also includes other parameters referred to as ``nuisance parameters'' such as: the \Bs\ signal fraction $f_s$, parameters describing the invariant mass and decay time-angular distributions of combinatorial background events and scale factors of mass and decay time uncertainties.  In addition, there are also other nuisance parameters describing: acceptance functions, parametrisations of the angles of dedicated backgrounds $B_d\rightarrow J/\psi K^{0*}$ and $\Lambda_b\rightarrow J/\psi p K^{-}$ and their fractions $f_{B^0}$ and $f_{\Lambda_b}$, the probability density functions of time error distributions $P(\sigma_{t_{i}}|p_{\mathrm{T}_{i}})$, mass error distributions $P(\sigma_{m_{i}}|p_{\mathrm{T}_{i}})$, $p_\mathrm{T}$ distributions $P(p_{\mathrm{T}_{i}})$ and tagging parameters and calibrations. These parameter values are mainly fixed in the fit to the values extracted from the \Bs\ mass signal and sideband regions or from MC simulations.
 
\section{Systematic uncertainties}
\label{sec:systematics}
 
Systematic uncertainties are evaluated for effects that are described below.
\begin{itemize}
 
\item {\bfseries Flavour tagging:} The effects on the main physics parameters from the fit, due to uncertainties introduced by the flavour tagging procedure, are assessed as follows: The statistical uncertainty due to the size of the sample of $B^\pm \ra J/\psi K^\pm$ decays is included in the overall statistical uncertainty.
The systematic uncertainty arising from the precision of the OST calibration, described in Section~\ref{sec:taggingMethods}, is estimated by changing the models used to parameterise the probability distribution, $P(B|Q_{x})$, as a function of the cone charge from the function used by default (a third-order polynomial for muons and a sinusoid for electrons) to one of several alternative functions: a linear function; a fifth-order polynomial; or two third-order polynomials that describe the positive and negative regions and have common constant and linear terms, but independent quadratic and cubic terms. The \Bs\ fit is repeated using the alternative models and the largest deviation from the nominal fit is assigned as the systematic uncertainty.
To validate the calibration procedure, calibration curves are derived from simulated samples of $B^{\pm}$ and $\Bs$ signals.
The variations between the curves from these two samples are propagated to the calibration curves derived from data.
The differences in the parameter values between the nominal fit and that with the varied calibration curves are included in the systematic uncertainty.
 
An additional systematic uncertainty is assigned to account for potential dependencies on the pile-up distribution. The calibration data are split into subsets of approximately equal size, separated according to the estimated pile-up of the event, and separate calibrations are made for each subset. For the \Bs\ fit, the fit is repeated using the calibrations corresponding to the estimated  pile-up of that event. Differences between the nominal and the modified fit for the parameters of interest are taken as the systematic uncertainty. For the terms $P_{\mathrm{b}}(P(B|Q_{x}))$ and $P_{\mathrm{s}}(P(B|Q_{x}))$, variations of the parameterisation are considered (including using histograms in place of a parameterisation). The resulting changes in the parameter values of the $\Bs$ fit are similarly included in the systematic uncertainties.

\item {\bfseries ID alignment:}  The changes of the fit parameters due to residual misalignments of the ID    were studied and the observed deviations are included in the systematic uncertainties.
 
\item {\bfseries Angular acceptance method:} The angular acceptance of the detector and the kinematic cuts, $A(\Omega_i,{\mathit p_{\mathrm{T}_{i}}})$, described in Section~\ref{sec:fitSignal}, is calculated from a binned fit to MC simulated data.  In order to estimate the systematic uncertainty introduced by the choice of binning, different acceptance functions are calculated using different numbers of \pt bins as well as different widths and central values of the bins.

\item {\bfseries Time efficiency:}  To correct for the proper decay time dependence of trigger inefficiencies, the events are reweighted according to Eq.~(\ref{eq:LifeTimeWeight}).
To estimate systematic uncertainties connected with this procedure, several alternative fits are performed. Firstly, using different sets of \pT\ binning in the MC sample used to determine the efficiency.  Secondly, to  assess the effects of mis-modelling of the \Bs\ vertex $\chi^2 /$ndof in simulated data, an alternative fit is done with MC $\chi^2 /$ndof  reweighted  according to data.
Finally, the groups of similar triggers used to determine the efficiencies were subdivided further by their common features. These smaller groups allowed to address even more details of trigger varieties in time-efficiency determination.
Deviations from the default fit result are included in the systematic uncertainties of the measurement.
 
\item {\bfseries Best candidate selection:} After applying all selection cuts for \Bs \ signal events, approximately $5\%$ of the events  are found to contain multiple candidates. In the default fit, the   \Bs \ candidate with the lowest $\chi^2 /$ndof is selected. To assess the systematic uncertainty due to this selection, an equivalent sample is created where all candidates in the event are retained, each weighted by a factor of $1/N_{\mathrm{cand}}$, where $N_{\mathrm{cand}}$ is number of \Bs \ candidates in the event. Deviations from the default fit are included in the systematic uncertainties of the measurement.

\item  {\bfseries Background angles model:}
The shape of the background angular distribution, $P_\textrm{b}(\theta_T, \varphi_T, \psi_T)$, is described by the fourteenth-order Legendre polynomial functions, given in  Eq.~(\ref{eq:likelihoodBkgAng}).  Alternatively, higher-order Legendre polynomial functions with $l_{max}=16$ and $k_{max}=16$ were tested, and the changes in the fit parameter values relative to the default fit are taken as systematic uncertainties.
 
The shapes are primarily determined by detector and kinematic acceptance effects and are sensitive to the \psubt \ of the \Bs \ meson candidate. For this reason, the parameterisation using the Legendre polynomial functions is performed in  six \psubt \ intervals: 10--15~\GeV, 15--20~\GeV, 20--25~\GeV, 25--30~\GeV, 30--35~\GeV \ and
$>$35~\GeV.
The systematic uncertainties due to the choice of  \psubt intervals are estimated by repeating the fit,  with  these  intervals enlarged and reduced by 1~\GeV \ and by 2~\GeV. The largest changes in the fit results are taken to represent the systematic uncertainties.
 
The sensitivity of the fit to the choice of the invariant mass window is tested by varying its size.  The difference to the default fit results is assigned as a systematic uncertainty.
 
The parameters of the Legendre polynomial functions given in  Eq.~(\ref{eq:likelihoodBkgAng}) are adjusted to give the best fit to the angular distributions for events in the \Bs\ mass sidebands.  To test the sensitivity of the fit results to the choice of sideband regions, the fit is repeated with alternative  choices for the excluded signal mass regions: $|(m(B_s^0) - 5.366 \hspace{0.1 cm} \GeV|> 0.085$ \GeV\ and $|(m(B_s^0) - 5.366\hspace{0.1 cm} \GeV|> 0.160$ \GeV\ (instead of the default $|(m(B_s^0) - 5.366 \hspace{0.1 cm} \GeV|> 0.110$ \GeV). The changes in the fit results are assigned as systematic uncertainties.
 
\relpenalty=10000
\item  {\bfseries \boldmath{$B_d$}  contribution:} The contamination from {$B_d~\ra~J/\psi~K^{0*}$} events misreconstructed as \relpenalty=10000 {\Bst}\ is accounted for in the final fit. Studies are performed to evaluate the systematic uncertainties due to trigger effects, the $B_d \ra J/\psi K^{0*}$  fraction, and the distributions of the mass, transversity angles, and lifetime PDFs.  In the MC events the angular distribution of the $B_d \ra J/\psi K^{0*}$ decay is modelled using parameters taken from  Ref.~\cite{LHCbK0Star}. The contribution of the $S$-wave $B_d \ra J/\psi K\pi$ decays as well as  its interference with the $P$-wave $B_d \ra J/\psi K^{0*}$ decays  are also included in the PDF of the fit, following the parameters measured  in  Ref.~\cite{LHCbK0Star}.
The uncertainties of these parameters are taken into account in the estimation of the systematic uncertainty.  After  applying the \Bs \ signal selection cuts,   the angular distributions are fitted  using Legendre polynomial functions. The  uncertainties of this fit are included in the  systematic uncertainty.
 
\item  {\bfseries \boldmath{$\Lambda_b$}  contribution:} The contamination from $\Lambda_b \ra J/\psi p K^{-}$ events misreconstructed as \relpenalty=10000 \Bst\ is accounted for in the final fit. Studies are performed to evaluate the effect of the uncertainties in the $\Lambda_b \ra J/\psi p K^{-}$ fraction $f_{\Lambda_b}$, and the shapes of the distributions of the mass, transversity angles, and lifetime. Additional studies are performed to determine the effect of the uncertainties in the $\Lambda_b \ra J/\psi \Lambda^{*}$ branching ratios used to reweight the generated MC sample.
 
\item {\bfseries Alternate $\Delta m_{s}$}:  The systematics due to fixing the parameter $\Delta m_{s}$ to the PDG value were estimated by running an alternative fit where the default model was altered by releasing $\Delta m_{s}$ within a Gaussian constraint of which the width was taken from uncertainties assigned to $\Delta m_{s}$ in \cite{PDG}. The resulting changes to  all the fit parameter values are found to be negligible except for $\delta_{\perp}$. The effects on the other variables are small for the parameters \phis\ and $\delta_{\parallel}$.
 
\item {\bfseries Fit model mass and lifetime:}
To estimate the systematic uncertainties due to the signal \Bs mass model,  the default model was altered by adding a second Gaussian function in Eq.~(\ref{pdfsig}),
which has the same structure as the first Gaussian function but a different scale factor, $S^1_m$, which is an additional free parameter of the fit.
The resulting changes to  other fit parameter values  are found to be negligible.
 
To test the sensitivity of the part of the  fit  model describing the lifetime, two systematic tests are performed.
The determination of signal and background lifetime errors is  sensitive to the choice  of   \pt   bins,  in which  the relative contributions of these two components are evaluated. To estimate the systematic uncertainty,  the fit is  repeated  varying the intervals of the default \pt binning.
The determination of signal and background lifetime errors  is also sensitive to the determination of the signal fraction. The fit is repeated by varying this fraction within one  standard deviation of its uncertainty and differences are included in the systematic uncertainty.

\item {\bfseries Fit model $S$-wave phase}: As explained in Section \ref{sec:fitSignal}, the model for the interference between
$B_s^0 \rightarrow J/\psi \phi(K^+K^-)$ and $S$-wave $B_s^0 \rightarrow J/\psi K^+K^-$ is corrected by a factor
$\alpha$ to account for the mass-dependent variations in absolute amplitude and phase of the two amplitudes within the interval 1.0085--1.0305 GeV in $m(K^+K^-)$. The value of $\alpha$ is 0.51$\pm$ 0.08. The central value is obtained under the hypothesis of uniformity of the $S$-wave amplitude. The uncertainty is systematic and it is due to: detector mass resolution and mass scale uncertainties, uncertainties in the description of the $\phi$ resonance (mass, width, and shape descriptions as relativistic Breit-Wigner or  Flatt\'e parameterisation ~\cite{Flatte:1976xu}), and to uncertainties in the description of the shape and phase variation of the $S$-wave amplitude. For the last effect, which is dominant, the $S$-wave amplitude is assumed to be due to the $f_0$(980) resonance and it is described as in Ref. ~\cite{Aaij:2014emv}, similarly to what was done in Ref. ~\cite{Aaij2019}.  The variation from the value of $\alpha$ obtained with the default assumption is taken as a systematic uncertainty. This procedure uses descriptions of the $f_0$(980) based on other measurements \cite{PDG}.
To account for the uncertainty in $\alpha $, the fit was repeated with $\alpha = 0.51 + 0.08$ and $\alpha =  0.51 - 0.08$ values.
The variations of the parameter values relative to those from the default fit using the central value of $\alpha$ are included in the systematic uncertainties.

\item {\bfseries Fit bias:} Due to its complexity, the fit model  can be  sensitive to some   nuisance parameters. This limited sensitivity  could potentially lead to a bias in the measured physics parameters, even when the model describes the fitted data well. To test the stability of the results obtained from the chosen default fit model, a set of pseudo-experiments is conducted using the default model in both the generation and fit. The systematic uncertainties are determined from the mean of the pull distributions of the pseudo-experiments scaled according to the statistical uncertainty of that parameter in the fit to data.  The observed deviations are included in the systematic uncertainties.

\end{itemize}
The systematic uncertainties are listed in Table~\ref{tab:syst_totals}. For each parameter, the total systematic uncertainty is obtained by adding  all of the contributions in quadrature.

\begin{table}[ht]
\footnotesize
\begin{center}
\caption{Summary of systematic uncertainties assigned to the physical parameters of interest.}
\label{tab:syst_totals}
\tabcolsep=0.11cm
\scriptsize
\begin{tabular}{l c c c c c c c c c }
\hline \hline
& \phis            & \DGs             & \Gs            & $|A_{\parallel}(0)|^2$ & $|A_{0}(0)|^2$        & $ |A_{S}(0)|^2 $      & $\delta_{\perp}$       & $\delta_{\parallel}$  & $\delta_{\perp} -\delta_{S} $\\
& [$10^{-3}$ rad]  & [$10^{-3}$ ps$^{-1}$]           & [$10^{-3}$ ps$^{-1}$]          &        [$10^{-3}$]                &          [$10^{-3}$]             &   [$10^{-3}$]                    & [$10^{-3}$ rad]                  & [$10^{-3}$ rad]                 & [$10^{-3}$ rad] \\
\hline
&                  &                  &                  &                   &                  &                  &                   &                  &                 \\
Tagging                                     & 19\phantom{.0}   & \phantom{1}0.4   & \phantom{1}0.3   & \phantom{1}0.2    & \phantom{1}0.2   & \phantom{1}1.1   & 17\phantom{.0}    & 19\phantom{.0}   & \phantom{1}2.3  \\
ID alignment                                & \phantom{1}0.8   & \phantom{1}0.2   & \phantom{1}0.5   & \sma              & \sma             & \sma             & 11\phantom{.0}    & \phantom{1}7.2   & \sma            \\
Acceptance                                  & \phantom{1}0.5   & \phantom{1}0.3   & \sma             & \phantom{1}1.0    & \phantom{1}0.9   & \phantom{1}2.9   & 37\phantom{.0}    & 64\phantom{.0}   & \phantom{1}8.6  \\
Time efficiency                             & \phantom{1}0.2   &  \phantom{1}0.2   & \phantom{1}0.5   & \sma              & \sma             & \phantom{1}0.1   &  \phantom{1}3.0   & \phantom{1}5.7   & \phantom{1}0.5  \\
Best candidate selection                    & \phantom{1}0.4   & \phantom{1}1.6   & \phantom{1}1.3   & \phantom{1}0.1    & \phantom{1}1.0   & \phantom{1}0.5   & \phantom{1}2.3    & \phantom{1}7.0   & \phantom{1}7.4  \\
Background angles model:                    &                       &                       &                      &                        &                       &                       &                        &                       & \\
\hspace{0.5cm}Choice of fit function        & \phantom{1}2.5   & \sma              & \phantom{1}0.3  & \phantom{1}1.1    & \sma             & \phantom{1}0.6   & 12\phantom{.0}    & \phantom{1}0.9   & \phantom{1}1.1  \\
\hspace{0.5cm}Choice of \psubt bins         & \phantom{1}1.3   & \phantom{1}0.5    & \sma            & \phantom{1}0.4    & \phantom{1}0.5   & \phantom{1}1.2   & \phantom{1}1.5    & \phantom{1}7.2   & \phantom{1}1.0  \\
\hspace{0.5cm}Choice of mass window         & \phantom{1}9.3   & \phantom{1}3.3    & \phantom{1}0.2  & \phantom{1}0.4    & \phantom{1}0.8   & \phantom{1}0.9   & 17\phantom{.0}     & \phantom{1}8.6   & \phantom{1}6.0  \\
\hspace{0.5cm}Choice of sidebands  intervals& \phantom{1}0.4   & \phantom{1}0.1    & \phantom{1}0.1  & \phantom{1}0.3    & \phantom{1}0.3   & \phantom{1}1.3   & \phantom{1}4.4    & \phantom{1}7.4   & \phantom{1}2.3  \\
Dedicated backgrounds:                      &                  &                   &                  &                   &                  &                  &                   &                  &                 \\
\hspace{0.5cm}\Bd                           & \phantom{1}2.6   & \phantom{1}1.1    & \sma             & \phantom{1}0.2    & \phantom{1}3.1   & \phantom{1}1.5   & 10\phantom{.0}    & 23\phantom{.0}   & \phantom{1}2.1  \\
\hspace{0.5cm} $\Lambda_b$                  & \phantom{1}1.6   & \phantom{1}0.3    & \phantom{1}0.2   & \phantom{1}0.5    & \phantom{1}1.2   & \phantom{1}1.8   & 14\phantom{.0}    & 30\phantom{.0}   & \phantom{1}0.8  \\
Alternate $\Delta m_{s}$                    & \phantom{1}1.0   &  \sma             & \sma             & \sma              & \sma             & \sma             &  15 \phantom{1}   & \phantom{1}4.0   & \sma            \\
Fit model:                                  &                  &                   &                  &                   &                  &                  &                   &                  &                 \\
\hspace{0.5cm} Time res. sig frac           & \phantom{1}1.4   & \phantom{1}1.1    & \phantom{1}0.5   & \phantom{1}0.5    & \phantom{1}0.6   & \phantom{1}0.8   & 12\phantom{.0}    & 30\phantom{.0}   & \phantom{1}0.4  \\
\hspace{0.5cm} Time res. \psubt bins        & \phantom{1}0.7   & \phantom{1}0.5    & \phantom{1}0.8   & \phantom{1}0.1    & \phantom{1}0.1   & \phantom{1}0.1   & \phantom{1}2.2    & 14\phantom{.0}   & \phantom{1}0.7  \\
\hspace{0.5cm} $S$-wave phase               & \phantom{1}0.3   & \sma              & \sma             & \sma              & \sma             & \phantom{1}0.2   & \phantom{1}8.0    & 15\phantom{.0}   & 37\phantom{.0}  \\
\hspace{0.5cm} Fit bias                     & \phantom{1}5.7   & \phantom{1}1.3    & \phantom{1}1.2   & \phantom{1}1.3    & \phantom{1}0.4   & \phantom{1}1.1   & \phantom{1}3.3    & 19\phantom{.0}   & \phantom{1}0.3  \\
\hline
&                       &                       &                      &                        &                       &                       &                        &                       & \\
\bfseries Total                             & \pSsystExp            & \DGsystExp            & \GSsystExp           & \ApasqsystExp          & \AzesqsystExp         & \ASsqsystExp          & \DeltperpsystExp       & \DeltparasystExp      & \DeltPSsystExp \\
\hline \hline
\end{tabular}
\end{center}
\end{table}
 
\FloatBarrier
\newpage
\section{Results}
\subsection{Fit results}
\label{sec:results}
 
\begin{figure}[htbp]
\begin{center}
\includegraphics[width=0.45\textwidth]{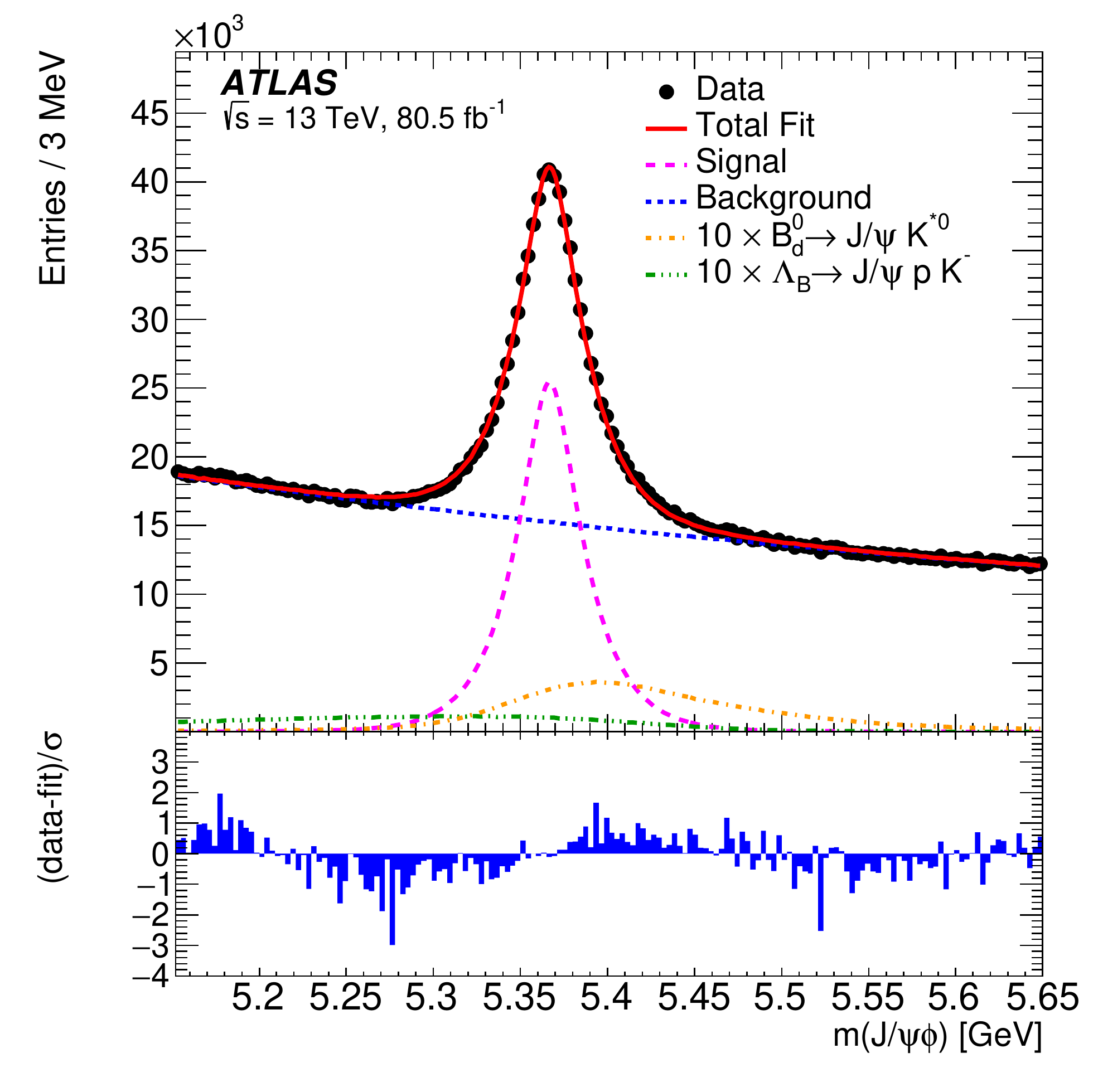}
\includegraphics[width=0.45\textwidth]{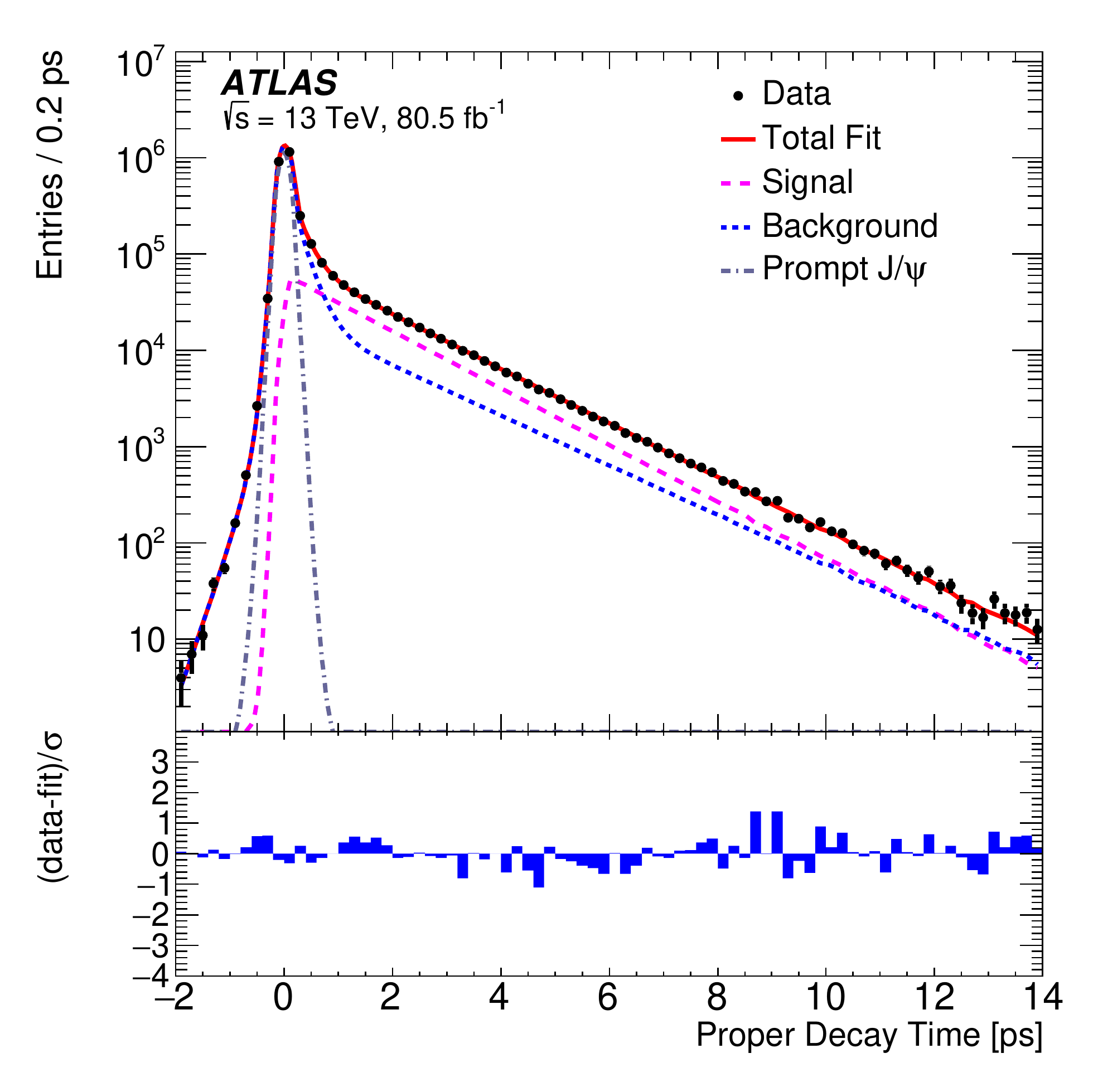}
 
\caption{(Left)   Mass fit projection for the \Bst \ sample.  The red line shows the total fit, the short-dashed magenta line shows the \Bst \  signal component, the combinatorial background is shown as a blue dotted line, the orange dash-dotted line shows the $\Bd \ra J/\psi K^{0*}$ component, and the green dash-dot-dot line shows the contribution from $\Lambda_b \ra J/\psi pK^{-}$ events. (Right) Proper decay time fit projection for the \Bst \ sample. The red line shows the total fit while the short-dashed magenta line shows the total signal.
The total background is shown as a blue dotted line, and a long-dashed grey line shows the prompt $J/\psi$ background component. Below each figure is a ratio plot that shows the difference between each data point and the total fit line divided by the statistical and systematic uncertainties summed in quadrature ($\sigma$) of that point.}
\label{fig:simulFit}
\end{center}
\end{figure}

\begin{figure}[htbp]
\begin{center}
\includegraphics[width=0.45\textwidth]{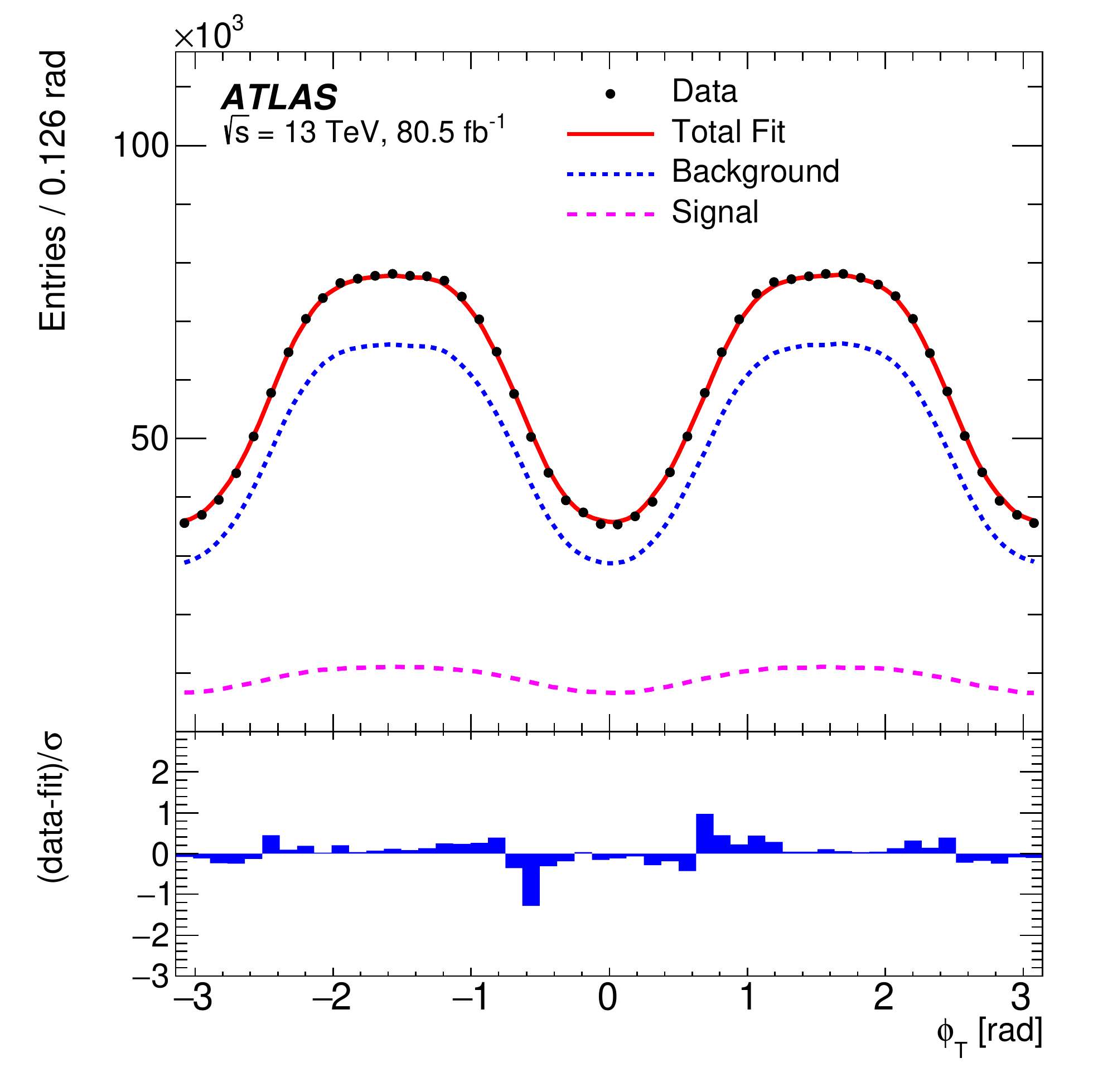}
\includegraphics[width=0.45\textwidth]{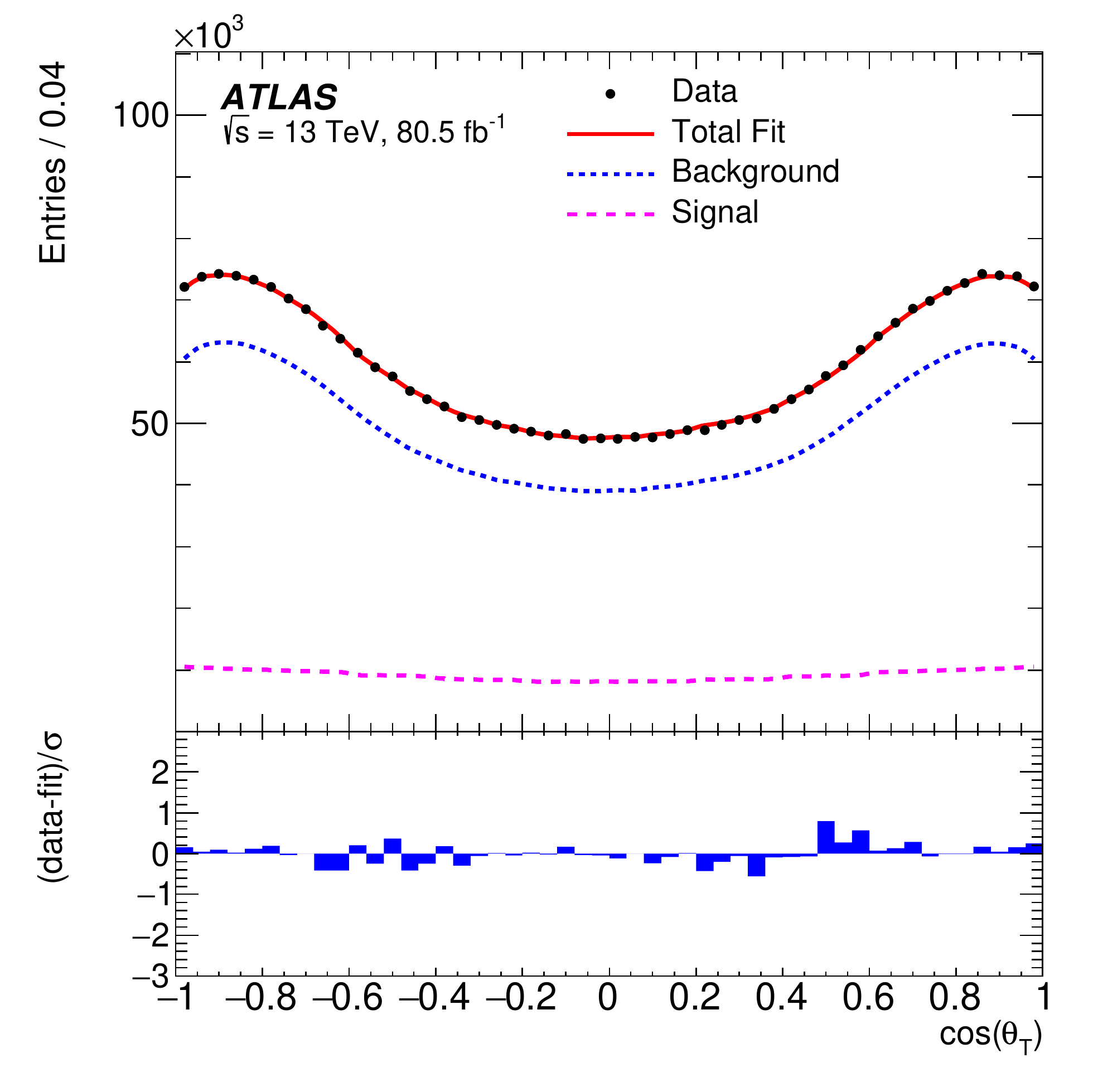}
\includegraphics[width=0.45\textwidth]{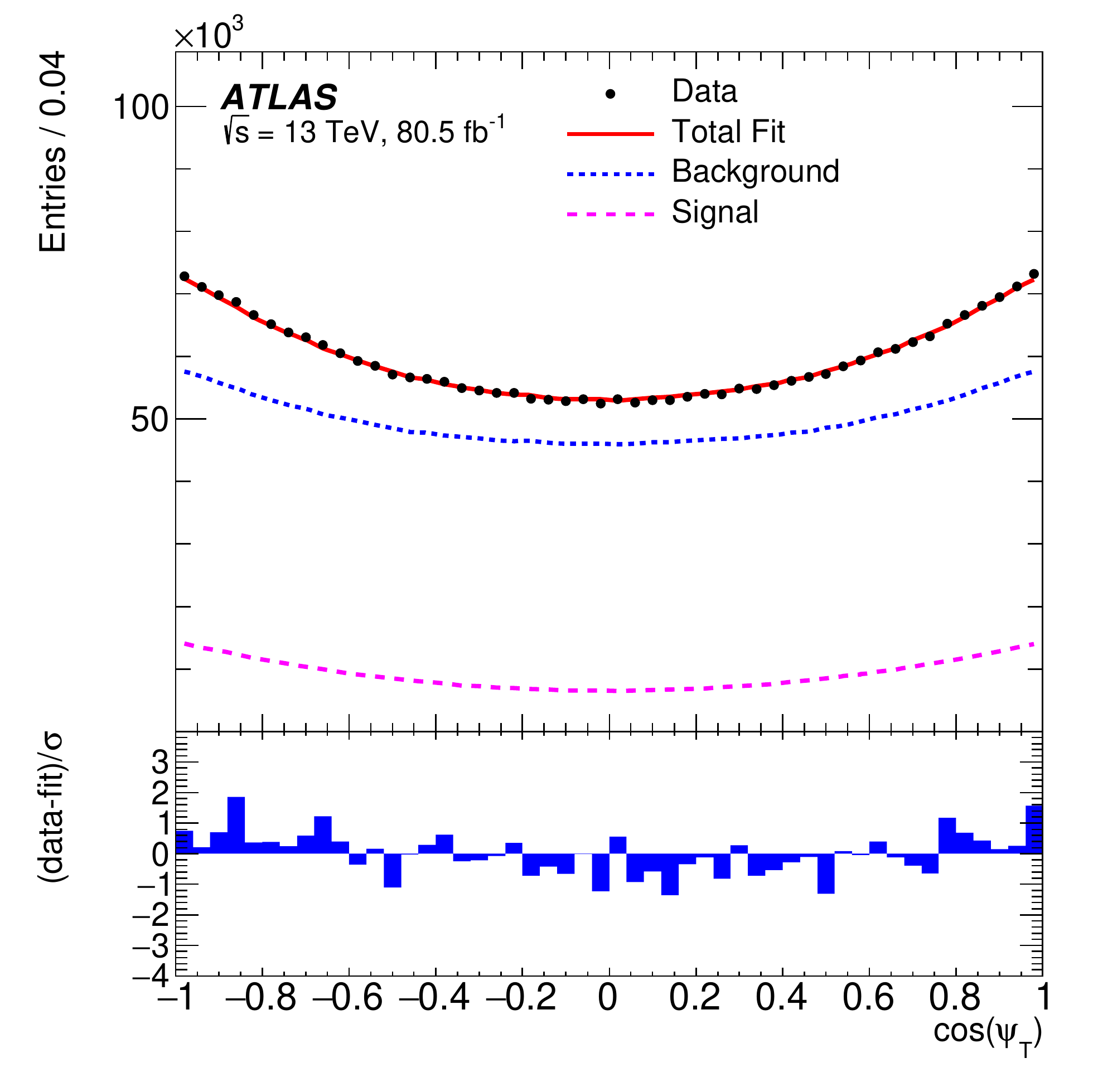}

\caption{Fit projections for the transversity angles
$\phi_T$ (top left), $\cos(\theta_T)$ (top right), and $\cos(\psi_T)$ (bottom).  In all three plots the red solid line shows the total fit, the  \Bst \  signal component is shown by the  magenta dashed line  and the  blue dotted line shows the contribution of all background components. 
Below each figure is a ratio plot that shows the difference between each data point and the total fit line divided by the statistical and systematic uncertainties summed in quadrature ($\sigma$) of that point.
}
\label{fig:AngularProjections}
\end{center}
\end{figure}

The results of the likelihood fit  are shown in Table~\ref{tab:FitResults}. The total number of \Bs\ meson candidates is \nsig.
The fitted value of the \Bs\ mass agrees well with the world average value~\cite{PDG}. Fit projections, including ratio plots, are shown in Figure~\ref{fig:simulFit} for the mass and proper decay time and in Figure~\ref{fig:AngularProjections} for the angles. The ratio plots show the difference between each data point and the total fit line divided by the statistical and systematic uncertainties summed in quadrature ($\sigma$) for that point. The deviations of ratio plots are within $2\sigma$, which shows that the total uncertainties cover any discrepancy between data and fit model.
 
While for most of the physics parameters, including \phis, \DGs \phantom{} and \Gs, the fit determines a single solution with Gaussian behaviour of the projection of the log-likelihood (see Figure \ref{fig:phiSDeltaGLikelihoodScansNew} in Section \ref{sec:twominima}),  for the strong-phases $\delta_{\parallel}$  and $\delta_\perp$ two well separated local maxima of the likelihood are found, and shown as solution (a) and (b) in  Table~\ref{tab:FitResults}. The difference in
-~2$\Delta${ln($L$)}   between the two solutions is 0.03. As discussed in Section \ref{sec:twominima}, the two-fold behaviour of the likelihood in the strong phases is the result of an approximate symmetry of the signal PDF. The effect is completely negligible for all other variables, for which the fit values and uncertainty ranges overlap accurately.

The correlation between statistical uncertainties of the parameters are also shown in Table \ref{tab:FitResultsCorA} for solution (a)   and in Table \ref{tab:FitResultsCorB} for solution (b).  The correlations do not change significantly between the  physics variables which remain stable between the solutions (a) and (b). The correlations with the two strong phases: $\delta_{\parallel}$  and $\delta_\perp$  flip the sign, as expected, except for the correlations that are smaller than ~0.01. The correlations between $\delta_{\parallel}$ and $\delta_\perp$  themselves keep the same sign in both solutions (a) and (b).
 
Releasing the direct \CP\ related $\lambda$ parameter (see
\Sect{\ref{sec:fitSignal}}) in the fit results in a $\lambda$ value compatible
within the statistical precision with unity as well as with results of other
experiments. The effect on the values of the other parameters of interest is
negligible, due to small correlations with $\lambda$.

\begin{table}[htbp]
\begin{center}
\caption{Fitted values for the physical parameters of interest with their statistical and systematic uncertainties. For variables $\delta_\perp$ and $\delta_{\parallel}$  the values are given for the two solutions (a) and (b). The difference in   -~2$\Delta${ln($L$)} between solutions (b) and (a) is 0.03.  For the rest of the variables the values for the two minima are consistent. The same is true for the statistical and systematic uncertainties of all the variables. \label{tab:FitResults} }
\begin{tabular}{c|c|c|c}
\hline
Parameter & Value & Statistical & Systematic \\
& & uncertainty & uncertainty \\
\hline\hline
\phis\ [rad]                         & \pSfit\phantom{0 }      & \pSstat\phantom{0}         &  \phantom{}\pSsyst\phantom{0}        \\
\DGs\ [ps$^{-1}$]                    & \phantom{\ensuremath{-}}\DGfit\phantom{ }       & \DGstat\phantom{-}        &  \phantom{}\DGsyst\phantom{{}}       \\
\Gs\ [ps$^{-1}$]                     & \phantom{\ensuremath{-}}\GSfit\phantom{ }     & \GSstat    \phantom{{}}    & \phantom{}\GSsyst\phantom{{}}       \\
$|A_{\parallel}(0)|^2$               & \phantom{\ensuremath{-}}\Apasqfit\phantom{ }   & \Apasqstat \phantom{{}}    & \Apasqsyst \phantom{}   \\
$|A_{0}(0)|^2$                       & \phantom{\ensuremath{-}}\Azesqfit\phantom{ }   & \Azesqstat \phantom{{}}    & \Azesqsyst \phantom{}   \\
$|A_{S}(0)|^2$                       & \phantom{\ensuremath{-}}\ASsqfit\phantom{ }   & \ASsqstat  \phantom{{}}    & \ASsqsyst \phantom{}    \\
$\delta_{\perp} -\delta_{S}$ [rad]   & \DeltPSfitSec\phantom{00 }             & \DeltPSstatSec \phantom{00}   & \DeltPSsyst\phantom{00}  \\
\hline
\multicolumn{3}{c}{ Solution (a) }  \\

$\delta_\perp$ [rad]    &     \phantom{\ensuremath{-}}\DeltperpfitSec\phantom{00 } & \DeltperpstatSec \phantom{00} &  \Deltperpsyst\phantom{00}\\
$\delta_{\parallel}$ [rad]  & \phantom{\ensuremath{-}}\DeltparafitSec\phantom{00 } & \DeltparastatSec \phantom{00} & \Deltparasyst\phantom{00}\\

\hline
\multicolumn{3}{c}{ Solution (b) }  \\
 
$\delta_\perp$ [rad]        & \phantom{\ensuremath{-}}\Deltperpfit \phantom{00 } & \Deltperpstat \phantom{00} & \Deltperpsyst\phantom{00}\\
$\delta_{\parallel}$ [rad]  & \phantom{\ensuremath{-}}\Deltparafit \phantom{00 } & \Deltparastat \phantom{00} & \Deltparasyst\phantom{00}\\

\hline
\end{tabular}
\end{center}
\end{table}

\begin{table}[htp]
\begin{center}
\caption{Fit correlations between the physical parameters of interest, obtained from the fit for solution (a).   \label{tab:FitResultsCorA}}
\begin{tabular}{c|c|c|c|c|c|c|c|c}

\hline
& $\Delta\Gamma$             & $\Gamma_{s}$    & $|A_{||}(0)|^2$            & $|A_0(0)|^2$               & $|A_S(0)|^2$               & $\delta_{\parallel}$       & $\delta_{\perp}$           & $\delta_{\perp} -\delta_{S} $ \\
\hline
\hline
$\phi_{s}$                     & \pDGSec                    & \phantom{\ensuremath{-}}\pGSec\phantom{ }     & \pApaSec\phantom{ }           				& \pAzSec\phantom{ }                                    & \pAsSec\phantom{ }                          & \phantom{\ensuremath{-}}\pDparaSec\phantom{ }         & \phantom{\ensuremath{-}}\pDperpSec\phantom{ }      & \pDpSSec\phantom{ }  \\
$\Delta\Gamma$                 & \phantom{--}1\phantom{.000}& \DGGSec\phantom{ }                            & \phantom{\ensuremath{-}}\DGApaSec\phantom{ }  & \phantom{\ensuremath{-}}\DGAzSec\phantom{ }           & \phantom{\ensuremath{-}}\DGAsSec\phantom{ } & \phantom{\ensuremath{-}}\DGDparaSec\phantom{ }        & \phantom{\ensuremath{-}}\DGDperpSec\phantom{ }     & \phantom{\ensuremath{-}}\DGDpSSec\phantom{ }  \\
$\Gamma_{s}$                   &                            & \phantom{--}1\phantom{.000}   				& \GApaSec\phantom{ }                           & \GAzSec\phantom{ }                                    & \phantom{\ensuremath{-}}\GAsSec\phantom{ }  & \GDparaSec\phantom{ }                                 & \GDperpSec\phantom{ }                              & \phantom{\ensuremath{-}}\GDpSSec\phantom{ }  \\
$|A_{||}(0)|^2$                &                            &                 								& \phantom{--}1\phantom{.000}                   & \AzApaSec\phantom{ }                                  & \ApaAsSec\phantom{ }                        & \phantom{\ensuremath{-}}\ApaDparaSec\phantom{ }       & \phantom{\ensuremath{-}}\ApaDperpSec\phantom{ }    & \ApaDpSSec\phantom{ }  \\
$|A_0(0)|^2$                   &                            &                 								&                                               & \phantom{--}1\phantom{.000}                           & \phantom{\ensuremath{-}}\AzAsSec\phantom{ } & \AzDparaSec\phantom{ }                                & \AzDperpSec\phantom{ }                             & \phantom{\ensuremath{-}}\AzDpSSec\phantom{ }  \\
$|A_S(0)|^2$                   &                            &                 								&                                               &                                                       & \phantom{--}1\phantom{.000}                 & \AsDparaSec\phantom{ }                                & \AsDperpSec\phantom{ }                             & \phantom{\ensuremath{-}}\AsDpSSec\phantom{ }  \\
$\delta_{\parallel}$           &                            &                 								&                                               &                                                       &                                             & \phantom{--}1\phantom{.000}                           & \phantom{\ensuremath{-}}\DparaDperpSec\phantom{ }  & \DparaDpSSec\phantom{ }  \\
$\delta_{\perp}$               &                            &                 								&                                               &                                                       &                                             &                                                       & \phantom{--}1\phantom{.000}                        & \phantom{\ensuremath{-}}\DperpDpSSec\phantom{ }  \\

\hline
\end{tabular}
\end{center}
\end{table}
 
\begin{table}[htp]
\begin{center}
\caption{Fit correlations between the physical parameters of interest, obtained from the fit for solution (b).   \label{tab:FitResultsCorB}}
\begin{tabular}{c|c|c|c|c|c|c|c|c}
 
\hline
& $\Delta\Gamma$             & $\Gamma_{s}$    & $|A_{||}(0)|^2$            & $|A_0(0)|^2$               & $|A_S(0)|^2$               & $\delta_{\parallel}$       & $\delta_{\perp}$           & $\delta_{\perp} -\delta_{S} $ \\
\hline
\hline
$\phi_{s}$                     & \pDG                        & \phantom{\ensuremath{-}}\pG\phantom{ }     & \pApa\phantom{ }                          & \pAz\phantom{ }                          & \pAs\phantom{ }                          & \phantom{\ensuremath{-}}\pDpara\phantom{ }  & \phantom{\ensuremath{-}}\pDperp\phantom{ }         & \pDpS\phantom{ } \\
$\Delta\Gamma$                 & \phantom{--}1\phantom{.000} & \DGG\phantom{ }                            & \phantom{\ensuremath{-}}\DGApa\phantom{ } & \phantom{\ensuremath{-}}\DGAz\phantom{ } & \phantom{\ensuremath{-}}\DGAs\phantom{ } & \DGDpara\phantom{ }                         & \DGDperp\phantom{ }                                & \phantom{\ensuremath{-}}\DGDpS\phantom{ } \\
$\Gamma_{s}$                   &                             & \phantom{--}1\phantom{.000}                & \GApa\phantom{ }                          & \GAz\phantom{ }                          & \phantom{\ensuremath{-}}\GAs\phantom{ }  & \phantom{\ensuremath{-}}\GDpara\phantom{ }  & \phantom{\ensuremath{-}}\GDperp\phantom{ }         & \phantom{\ensuremath{-}}\GDpS\phantom{ } \\
$|A_{||}(0)|^2$                &                             &                                            & \phantom{--}1\phantom{.000}               & \AzApa\phantom{ }                        & \ApaAs\phantom{ }                        & \ApaDpara\phantom{ }                        & \ApaDperp\phantom{ }                               & \ApaDpS\phantom{ } \\
$|A_0(0)|^2$                   &                             &                                            &                                           & \phantom{--}1\phantom{.000}              & \phantom{\ensuremath{-}}\AzAs\phantom{ } & \phantom{\ensuremath{-}}\AzDpara\phantom{ } & \phantom{\ensuremath{-}}\AzDperp\phantom{ }        & \phantom{\ensuremath{-}}\AzDpS\phantom{ } \\
$|A_S(0)|^2$                   &                             &                                            &                                           &                                          & \phantom{--}1\phantom{.000}              & \phantom{\ensuremath{-}}\AsDpara\phantom{ } & \phantom{\ensuremath{-}}\AsDperp\phantom{ }        & \phantom{\ensuremath{-}}\AsDpS\phantom{ } \\
$\delta_{\parallel}$           &                             &                                            &                                           &                                          &                                          & \phantom{--}1\phantom{.000}                 & \phantom{\ensuremath{-}}\DparaDperp\phantom{ }     & \phantom{\ensuremath{-}}\DparaDpS\phantom{ } \\
$\delta_{\perp}$               &                             &                                            &                                           &                                          &                                          &                                             & \phantom{--}1\phantom{.000}                        & \phantom{\ensuremath{-}}\DperpDpS\phantom{ } \\

\hline
\end{tabular}
\end{center}
\end{table}

\subsection{Fit to strong phases}
\label{sec:twominima}

\begin{figure}[htbp]
\begin{center}
\includegraphics[width=0.8\textwidth]{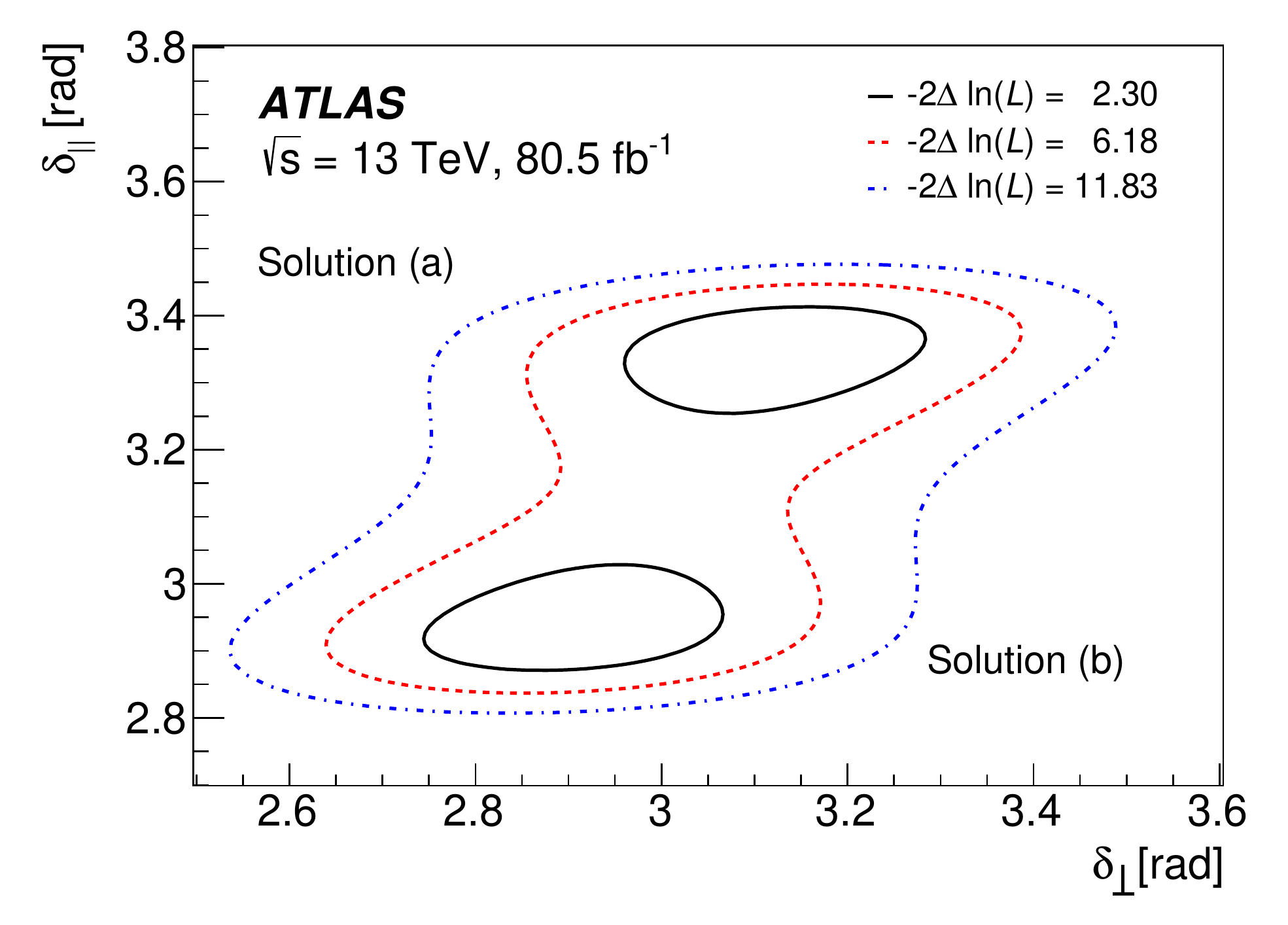}
\caption{
Two-dimensional constraints on the values of $\delta_{||}$ and $\delta_{\perp}$ for solutions (a) and (b) at the level of $-~2 \Delta { \ln(L)} = 2.30, 6.18,$ and $11.83$ respectively created using a full 2D scan. The minimum of the solution (b) is $-~2 \Delta { \ln(L)} =  0.03$ higher than the minimum of the solution (a).
}
\label{fig:2dDeltaScan}
\end{center}
\end{figure}

As shown in Table~\ref{tab:FitResults}, the likelihood fit has determined two solutions with well separated values for the strong phases $\delta_{||}$ and $\delta_{\perp}$.
Figure \ref{fig:2dDeltaScan} shows results of the 2D log-likelihood scan  in the $\delta_{||}$, $\delta_{\perp}$ plane,  revealing two minima,  identified at ($\delta_{||} = 3.35$, $\delta_{\perp}=3.12$)  and  ($\delta_{||} = 2.94$, $\delta_{\perp}=2.91$). These minima are represented by two-dimensional contours at the level of $-~2 \Delta { \ln(L)} =  2.30, 6.18,$ and $11.83$, where $-~2 \Delta { \ln(L)} =  2 {(\ln(L^i) - \ln(L^{a}))} $ is the difference between the likelihood values $ {(L^i)}$  of the fit in which the  two strong phases are fixed to the values shown on the horizontal and vertical axis, and $ L^{a}$ which is the likelihood value for solution (a) of the fit.

An approximate symmetry in the signal PDF is at the origin of this duality.  The strong phases are determined by the six interference terms 4) $-$ 6) and 8) $-$ 10) in Table  \ref{tab:SigPDF}.
Four of them, namely terms 4), 5), 8) and 9), are invariant under the transformation

\begin{equation}
\{ \delta_{\parallel},\phantom{0} \delta_{\perp}, \phantom{0} \delta_{S} \}  \rightarrow
\{ 2 \pi -\delta_{\parallel},  \phantom{0}  \delta_{\perp} + 2 (\pi -\delta_{\parallel}), \phantom{0} \delta_{S} + 2 (\pi - \delta_{\parallel})                           \}.
\label{symmetry}
\end{equation}
This transformation is proportional to $\pi -\delta_{\parallel}$,  which from our data is equal to 0.21.
The two local maxima of the likelihood determined in this analysis satisfy the relation of Eq.~(\ref{symmetry}) very accurately for
$\delta_{||}$ and $(\delta_{\perp}$ - $\delta_{S})$,
and within the statistical uncertainty in the transformation for $\delta_{\perp}$.
The difference in the log-likelihoods,  $-~2 \Delta { \ln(L)}$,  between the two solutions is equal to 0.03, favouring (a) but without ruling out (b).

\begin{figure}[htbp]
\begin{center}
\includegraphics[width=0.32\textwidth]{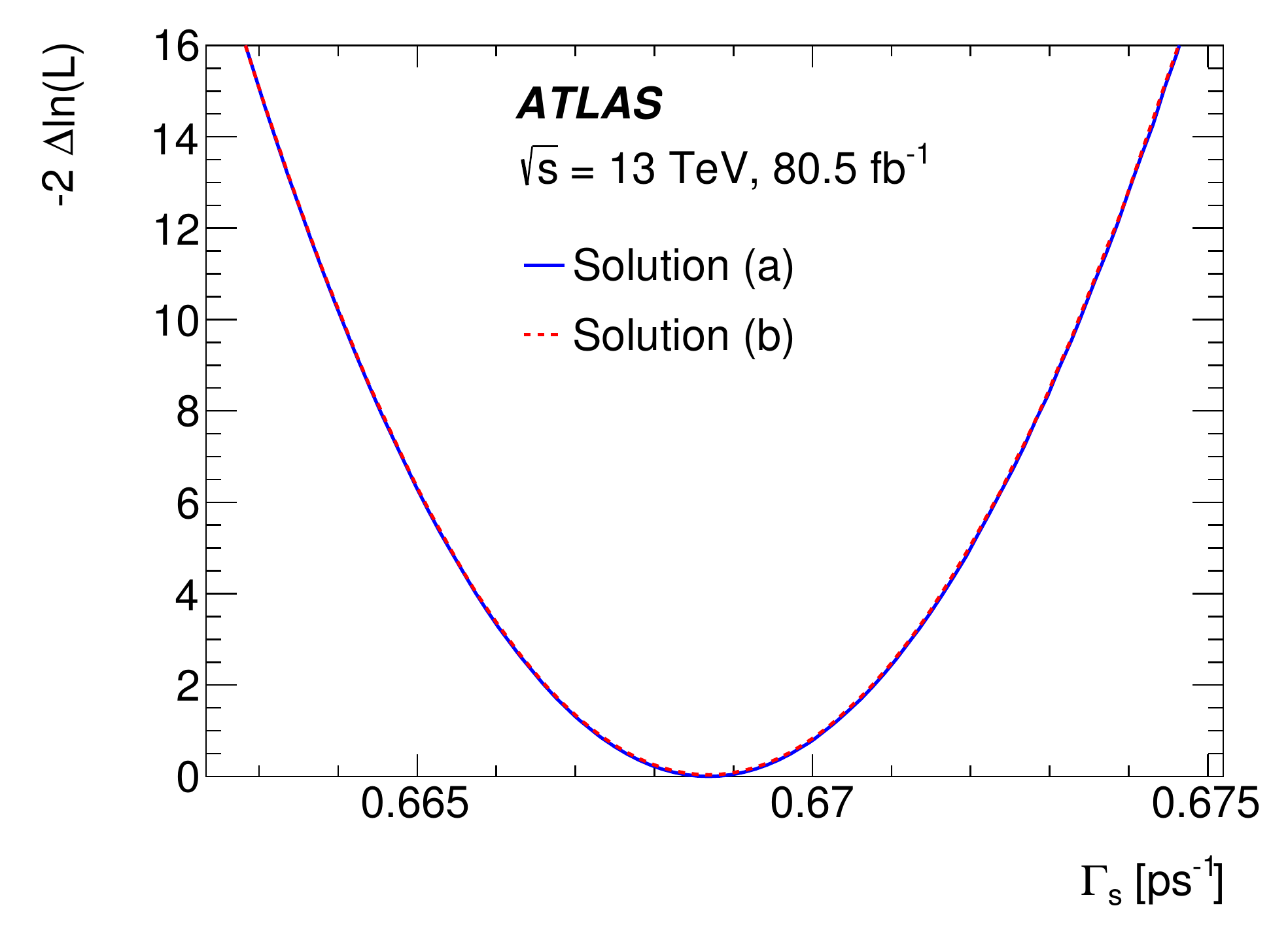}
\includegraphics[width=0.32\textwidth]{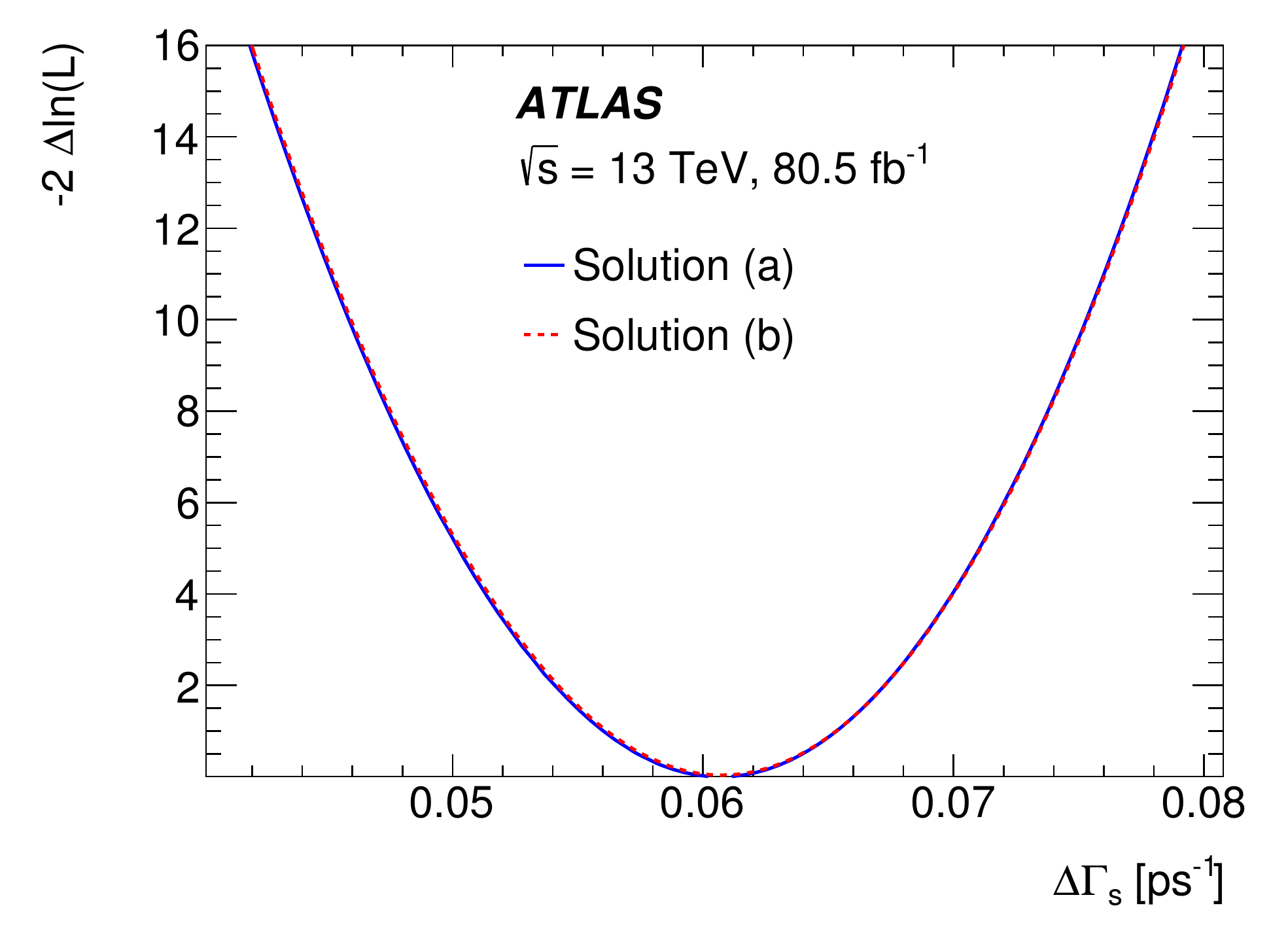}
\includegraphics[width=0.32\textwidth]{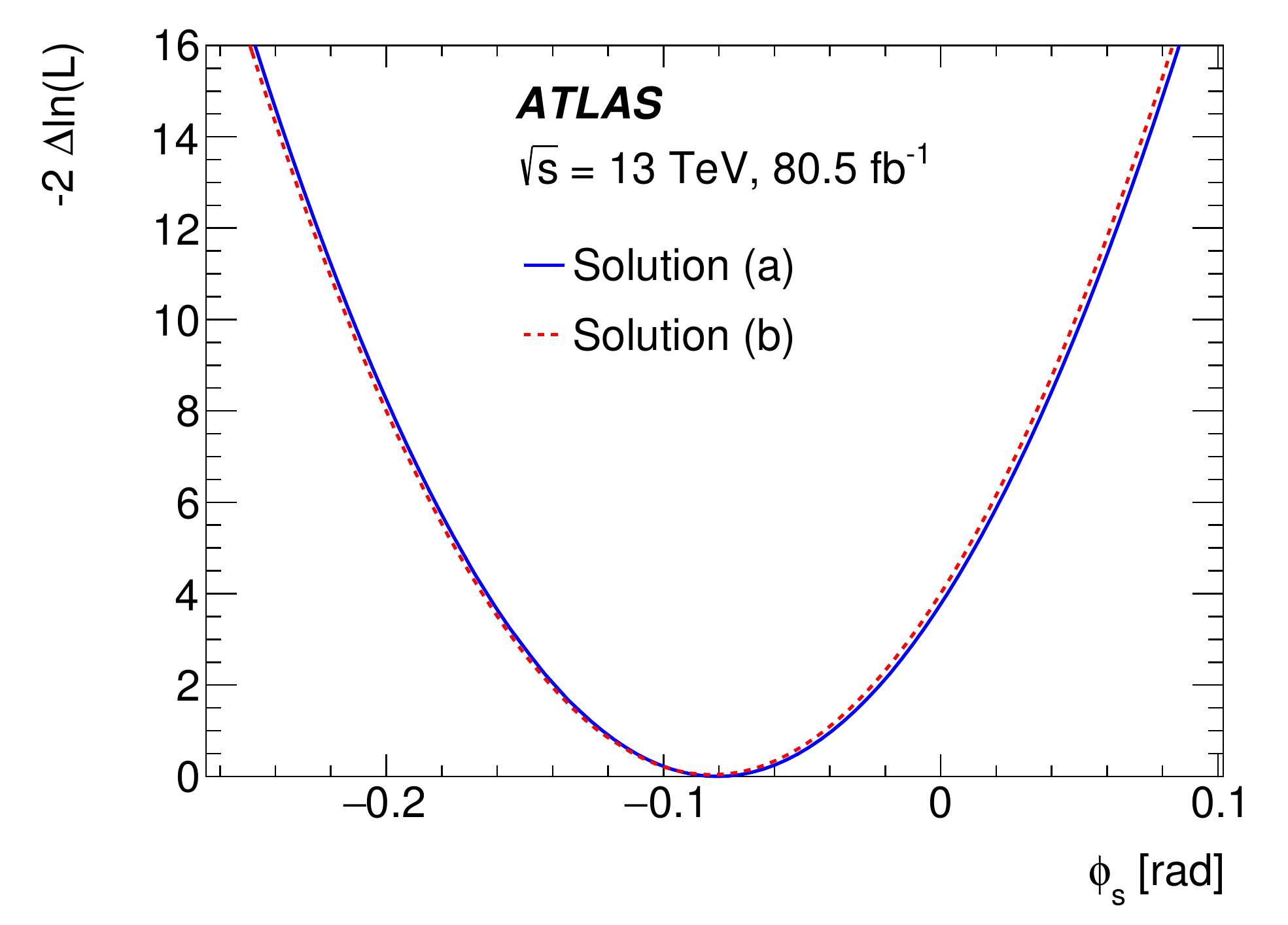}
\includegraphics[width=0.32\textwidth]{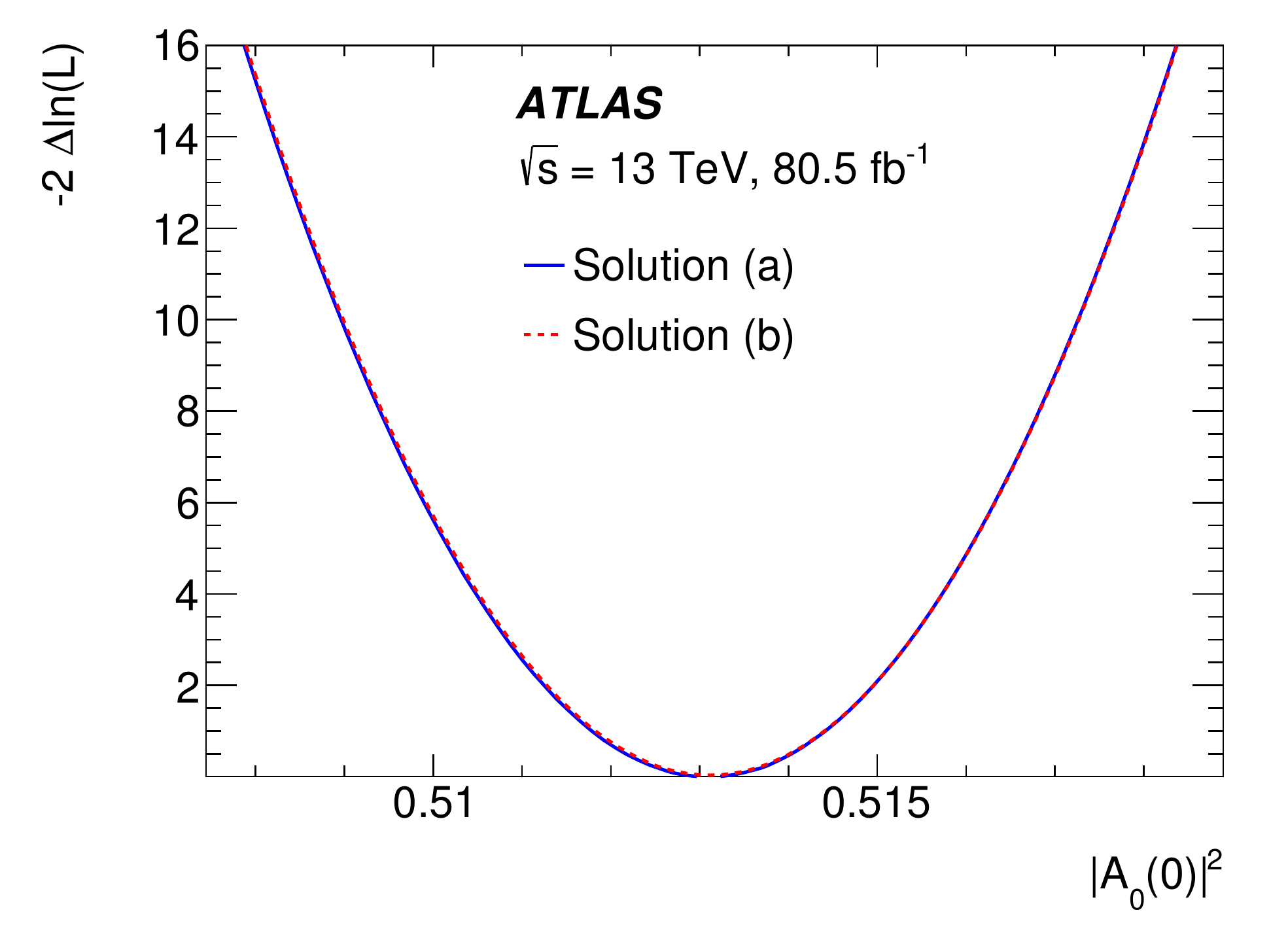}
\includegraphics[width=0.32\textwidth]{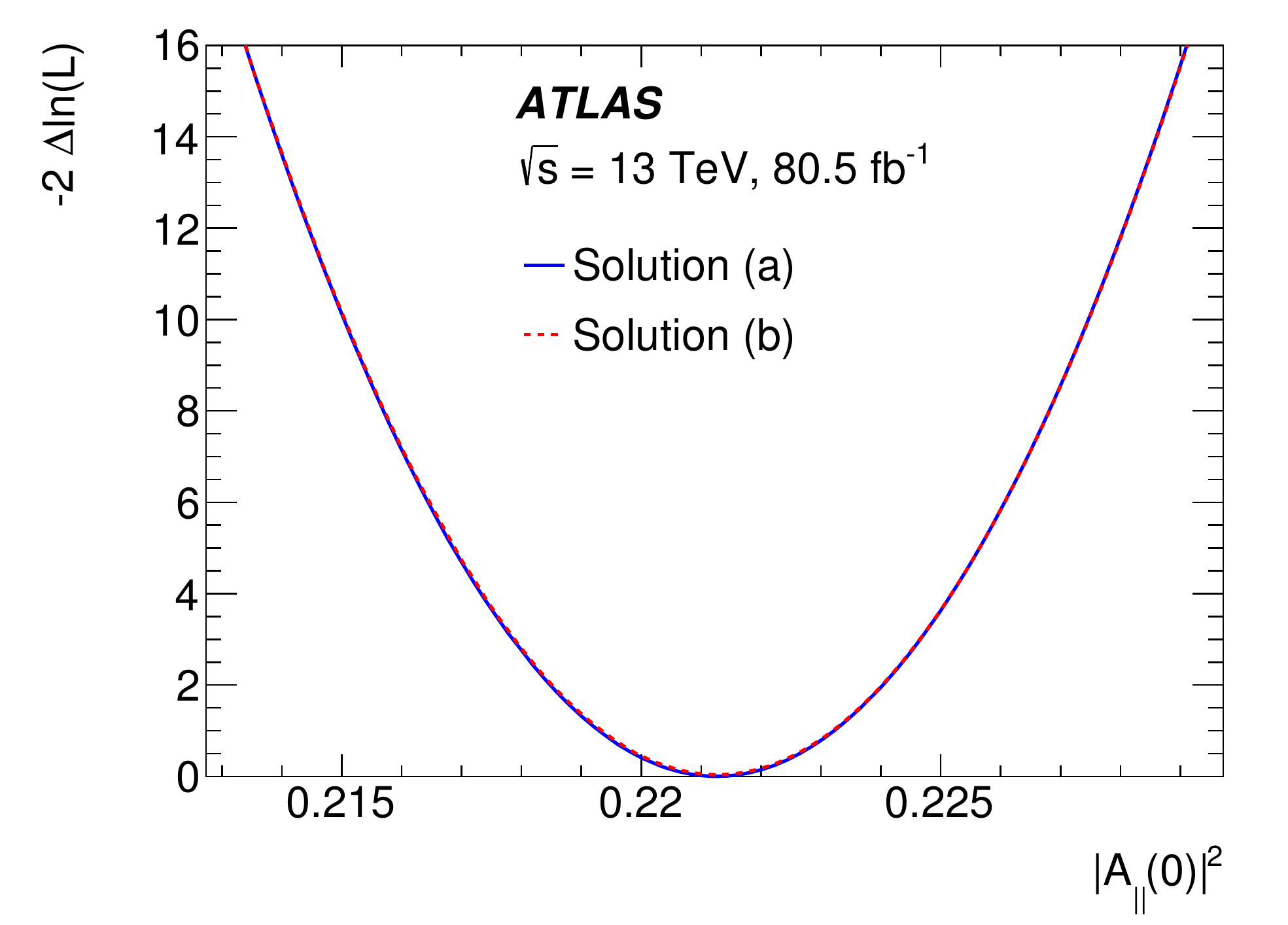}
\includegraphics[width=0.32\textwidth]{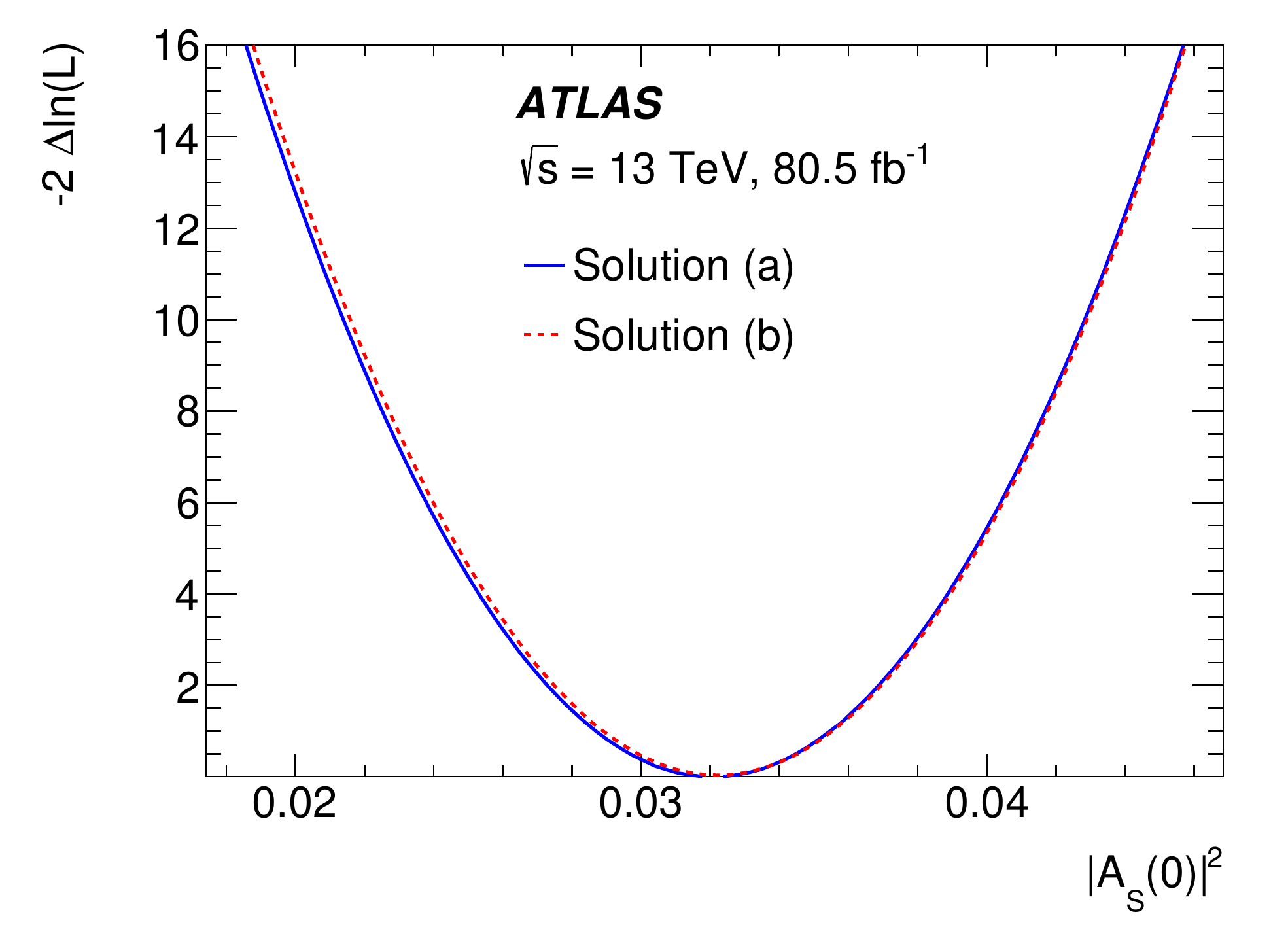}
\includegraphics[width=0.32\textwidth]{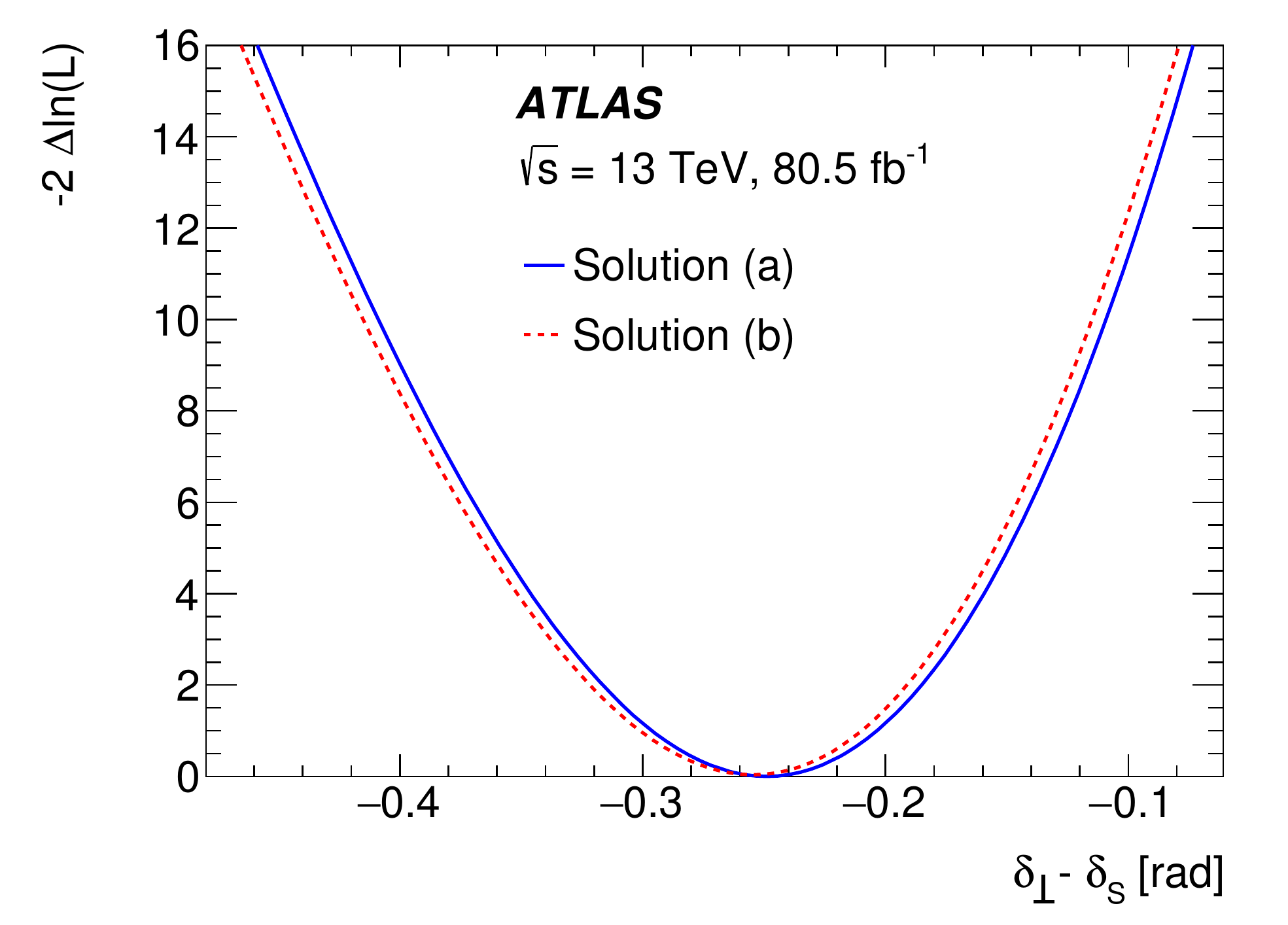}
\caption{1D log-likelihood scans of  all other variables of the fit for  the  solution (a) (blue)  and the solution (b) (dashed red).  The variable on the vertical axis, $-2 \Delta { \ln(L)} =  2 { (\ln(L^{a}) - \ln(L^i))}  $, is the difference between the likelihood values of the default fit, $  {(L^a)}$, and of the fit in which the  physical parameter is fixed to values shown on the horizontal axis.}
\label{fig:phiSDeltaGLikelihoodScansNew}
\end{center}
\end{figure}

As  discussed in Section \ref{sec:results}, the two-fold nature of the likelihood maxima for the strong phases has
a negligible effect on all the other variables. Figure~\ref{fig:phiSDeltaGLikelihoodScansNew} shows the one-dimensional likelihood scans for all other physics parameters, separately for the two solutions (a) and (b).   For each scan, the other physics parameters and all nuisance parameters are optimised in a profile-likelihood fit.
The 1D scans are almost identical for the two solutions.

\section{Combination with \CoMEnergySeven\ and \CoMEnergyEight\ results}
The measured values of  solution (a) and  solution (b) are consistent with those obtained in the previous analysis~\cite{BPHY-2014-03} using \ilumiRunone\ of data collected at $\sqrt{s}$ = \CoMEnergySeven\ and \CoMEnergyEight. A best linear unbiased estimator (BLUE)~\cite{Lyons:1988rp,Valassi:2003mu} is used to combine the current results with those from the previous analysis. The measured values, uncertainties, and correlations are taken from the measurements performed at each centre-of-mass energy. The statistical correlation between these three measurements is zero as the events are different. The correlations of the systematic uncertainties between the three measurements are estimated and tested in several categories depending on whether the given systematic effect changed significantly between the measurements. Solution (a) and solution (b) are treated separately, leading to the two sets of combined results as shown in Table~\ref{tab:FitResultsATLAS}. The correlation matrices of these two combinations are shown in Table~\ref{tab:CombResultsCorA} and Table~\ref{tab:CombResultsCorB}.
 
\begin{figure}[hptb]
\begin{center}
{\includegraphics[width=0.8\textwidth]{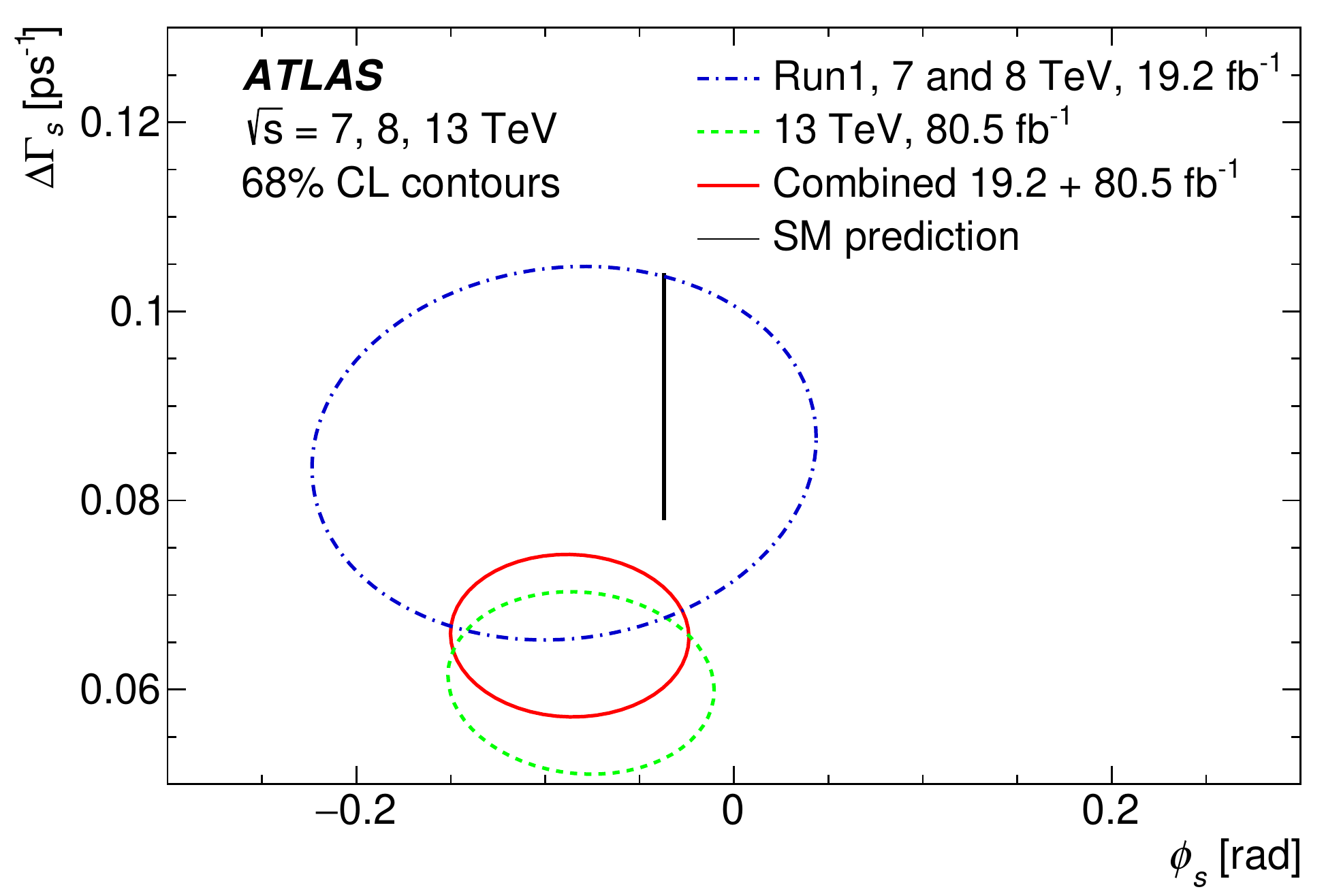}}
\end{center}
\caption{Contours of 68\% confidence level in the \phis--\DGs\ plane, showing ATLAS results for \CoMEnergySeven\ and \CoMEnergyEight\ data (blue dashed-dotted curve), for \CoMEnergy\ data (green dashed curve) and for \CoMEnergy\ data combined  with \CoMEnergySeven\ and \CoMEnergyEight\ (red solid curve) data. The Standard Model prediction \cite{Charles:2015gya, Lenz:2019lvd} is shown as a very thin black rectangle. In  all contours  the statistical and systematic uncertainties are combined in quadrature and correlations are taken into account.}
\label{fig:2DContourA}
\end{figure}

The two-dimensional likelihood contours in the  \phis--\DGs\ plane for the ATLAS result based on \CoMEnergySeven\ and \CoMEnergyEight\ data, the solution (a) of the \CoMEnergy\ measurement, and the combined result for the solution (a) are shown in Figure~\ref{fig:2DContourA}. The statistical and systematic uncertainties are combined in quadrature and correlations are taken into account in the  construction of Gaussian contours. Because there is no significant difference in the \phis\ and \DGs\ values between solution (a) and solution (b), only contours for solution (a) are shown.
 
\begin{table}[thbp]
\begin{center}
\caption{Values of the physical parameters extracted in the combination of solution (a) and solution (b) of \CoMEnergy \ results with those obtained from \CoMEnergySeven\ and \CoMEnergyEight\ data.\label{tab:FitResultsATLAS}}
\begin{tabular}{ c||c|c|c||c|c|c }
 
\hline
& \multicolumn{3}{c||}{ Solution (a) }        & \multicolumn{3}{c}{ Solution (b) } \\
 
Parameter                          & Value                                       & Statistical                    & Systematic                     & Value                                           & Statistical                        & Systematic \\
&                                             & uncertainty                    & uncertainty                    &                                                 & uncertainty                        & uncertainty \\
\hline\hline
\phis\ [rad]                       & \phantom{ }\pSfitCombi\phantom{0}                                   & \pSstatCombi\phantom{  }       & \pSsystCombi\phantom{  }       & \phantom{ }\pSfitCombiSolB\phantom{0}          & \pSstatCombiSolB\phantom{  }       & \pSsystCombiSolB\phantom{  }  \\
\DGs\ [ps$^{-1}$]                  & \phantom{\ensuremath{-}}\DGfitCombi                                 & \DGstatCombi                   & \DGsystCombi                   & \phantom{\ensuremath{-}}\DGfitCombiSolB                                 & \DGstatCombiSolB                   & \DGsystCombiSolB\\
\Gs\ [ps$^{-1}$]                   & \phantom{\ensuremath{-}}\GSfitCombi                                 & \GSstatCombi                   & \GSsystCombi                   & \phantom{\ensuremath{-}}\GSfitCombiSolB                                 & \GSstatCombiSolB                   & \GSsystCombiSolB \\
$|A_{\parallel}(0)|^2$             & \phantom{\ensuremath{-}}\ApasqfitCombi                              & \ApasqstatCombi                & \ApasqsystCombi                & \phantom{\ensuremath{-}}\ApasqfitCombiSolB                              & \ApasqstatCombiSolB                & \ApasqsystCombiSolB  \\
$|A_{0}(0)|^2$                     & \phantom{\ensuremath{-}}\AzesqfitCombi                              & \AzesqstatCombi                & \AzesqsystCombi                & \phantom{\ensuremath{-}}\AzesqfitCombiSolB                              & \AzesqstatCombiSolB                & \AzesqsystCombiSolB \\
$|A_{S}|^2$                        & \phantom{\ensuremath{-}}\ASsqfitCombi                               & \ASsqstatCombi                 & \ASsqsystCombi                 & \phantom{\ensuremath{-}}\ASsqfitCombiSolB                               & \ASsqstatCombiSolB                 & \ASsqsystCombiSolB \\
$\delta_\perp$ [rad]               & \phantom{\ensuremath{-}}\DeltperpfitCombi\phantom{ 0}              & \DeltperpstatCombi\phantom{ 0} & \DeltperpsystCombi\phantom{ 0}  & \phantom{\ensuremath{-}}\DeltperpfitCombiSolB\phantom{0 }               & \DeltperpstatCombiSolB\phantom{0 } & \DeltperpsystCombiSolB\phantom{0 } \\
$\delta_{\parallel}$ [rad]         & \phantom{\ensuremath{-}}\DeltparafitCombi\phantom{ 0}              & \DeltparastatCombi\phantom{ 0} & \DeltparasystCombi\phantom{ 0}  & \phantom{\ensuremath{-}}\DeltparafitCombiSolB\phantom{0 }               & \DeltparastatCombiSolB\phantom{0 } & \DeltparasystCombiSolB\phantom{0 } \\
$\delta_{\perp} -\delta_{S}$ [rad] & \phantom{ }\DeltPSfitCombi\phantom{00}                              & \DeltPSstatCombi\phantom{ 0}   & \DeltPSsystCombi\phantom{ 0}   & \phantom{ }\DeltPSfitCombiSolB\phantom{00}                              & \DeltPSstatCombiSolB\phantom{0 }   & \DeltPSsystCombiSolB\phantom{0 }  \\
\hline
\end{tabular}
\end{center}
\end{table}

\begin{table}[htp]
\begin{center}
\caption{Correlation matrix of the BLUE combination of the \CoMEnergySeven\ and \CoMEnergyEight\ results and the solution (a) of the \CoMEnergy\ result.\label{tab:CombResultsCorA}}
\begin{tabular}{c|c|c|c|c|c|c|c|c}
\hline
& $\Delta\Gamma$              & $\Gamma_{s}$                 & $|A_{\parallel}(0)|^2$             & $|A_0(0)|^2$                & $|A_S(0)|^2$                & $\delta_{\perp}$        & $\delta_{\parallel}$            & $\delta_{\perp} -\delta_{S} $ \\
\hline
\hline
$\phi_{s}$           & \ensuremath{-}0.025         & \phantom{\ensuremath{-}}0.001& \phantom{\ensuremath{-}}0.001            & \phantom{\ensuremath{-}}0.000            & \ensuremath{-}0.001                      & \phantom{\ensuremath{-}}0.003            & \phantom{\ensuremath{-}}0.003            & \ensuremath{-}0.004 \\
$\Delta\Gamma$       & \phantom{--}1\phantom{.000} & \ensuremath{-}0.259          & \phantom{\ensuremath{-}}0.043            & \phantom{\ensuremath{-}}0.027            & \phantom{\ensuremath{-}}0.020            & \phantom{\ensuremath{-}}0.003            & \phantom{\ensuremath{-}}0.011            & \phantom{\ensuremath{-}}0.010 \\
$\Gamma_{s}$         &                             & \phantom{--}1\phantom{.000}  & \ensuremath{-}0.045                      & \ensuremath{-}0.009                      & \phantom{\ensuremath{-}}0.025            & \ensuremath{-}0.010                      & \ensuremath{-}0.021                      & \phantom{\ensuremath{-}}0.006 \\
$|A_{\parallel}(0)|^2$      &                             &                             & \phantom{--}1\phantom{.000} & \ensuremath{-}0.071                      & \ensuremath{-}0.056                      & \phantom{\ensuremath{-}}0.068            & \phantom{\ensuremath{-}}0.155            & \ensuremath{-}0.024 \\
$|A_0(0)|^2$         &                             &                              &                             & \phantom{--}1\phantom{.000} & \phantom{\ensuremath{-}}0.048            & \ensuremath{-}0.009                      & \ensuremath{-}0.015                      & \phantom{\ensuremath{-}}0.016 \\
$|A_S(0)|^2$         &                             &                              &                             &                             & \phantom{--}1\phantom{.000} & \ensuremath{-}0.056                      & \ensuremath{-}0.099                      & \phantom{\ensuremath{-}}0.108 \\
$\delta_{\perp}$ &                             &                              &                             &                             &                             & \phantom{--}1\phantom{.000} & \phantom{\ensuremath{-}}0.106            & \phantom{\ensuremath{-}}0.001 \\
$\delta_{\parallel}$     &                             &                              &                             &                             &                             &                             & \phantom{--}1\phantom{.000} & \ensuremath{-}0.032 \\
\hline
\end{tabular}
\end{center}
\end{table}
 
\begin{table}[htp]
\begin{center}
\caption{Correlation matrix of the BLUE combination of the \CoMEnergySeven\ and \CoMEnergyEight\ results and the solution (b) of the \CoMEnergy\ result.\label{tab:CombResultsCorB}}
\begin{tabular}{c|c|c|c|c|c|c|c|c}
\hline
& $\Delta\Gamma$              & $\Gamma_{s}$                & $|A_{\parallel}(0)|^2$             & $|A_0(0)|^2$                & $|A_S(0)|^2$                & $\delta_{\perp}$        & $\delta_{\parallel}$            & $\delta_{\perp} -\delta_{S} $ \\
\hline
\hline
$\phi_{s}$           & \ensuremath{-}0.027                      & \phantom{\ensuremath{-}}0.002            & \ensuremath{-}0.003                      & \phantom{\ensuremath{-}}0.001            & \ensuremath{-}0.001                      & \phantom{\ensuremath{-}}0.004            & \phantom{\ensuremath{-}}0.002            & \ensuremath{-}0.003 \\
$\Delta\Gamma$       & \phantom{--}1\phantom{.000} & \ensuremath{-}0.259                      & \phantom{\ensuremath{-}}0.043            & \phantom{\ensuremath{-}}0.027            & \phantom{\ensuremath{-}}0.023            & \ensuremath{-}0.006                      & \ensuremath{-}0.010                      & \phantom{\ensuremath{-}}0.011 \\
$\Gamma_{s}$         &                             & \phantom{--}1\phantom{.000} & \ensuremath{-}0.042                      & \ensuremath{-}0.010                      & \phantom{\ensuremath{-}}0.023            & \phantom{\ensuremath{-}}0.007            & \phantom{\ensuremath{-}}0.018            & \phantom{\ensuremath{-}}0.006 \\
$|A_{\parallel}(0)|^2$      &                             &                             & \phantom{--}1\phantom{.000} & \ensuremath{-}0.071                      & \ensuremath{-}0.035                      & \ensuremath{-}0.053                      & \ensuremath{-}0.132                      & \ensuremath{-}0.024 \\
$|A_0(0)|^2$         &                             &                             &                             & \phantom{--}1\phantom{.000} & \phantom{\ensuremath{-}}0.046            & \phantom{\ensuremath{-}}0.004            & \phantom{\ensuremath{-}}0.012            & \phantom{\ensuremath{-}}0.016 \\
$|A_S(0)|^2$         &                             &                             &                             &                             & \phantom{--}1\phantom{.000} & \phantom{\ensuremath{-}}0.027            & \phantom{\ensuremath{-}}0.080            & \phantom{\ensuremath{-}}0.111 \\
$\delta_{\perp}$ &                             &                             &                             &                             &                             & \phantom{--}1\phantom{.000} & \phantom{\ensuremath{-}}0.098            & \phantom{\ensuremath{-}}0.037 \\
$\delta_{\parallel}$     &                             &                             &                             &                             &                             &                             & \phantom{--}1\phantom{.000} & \phantom{\ensuremath{-}}0.036 \\
\hline
\end{tabular}
\end{center}
\end{table}

\begin{figure}[hptb]
\begin{center}
{\includegraphics[width=0.8\textwidth]{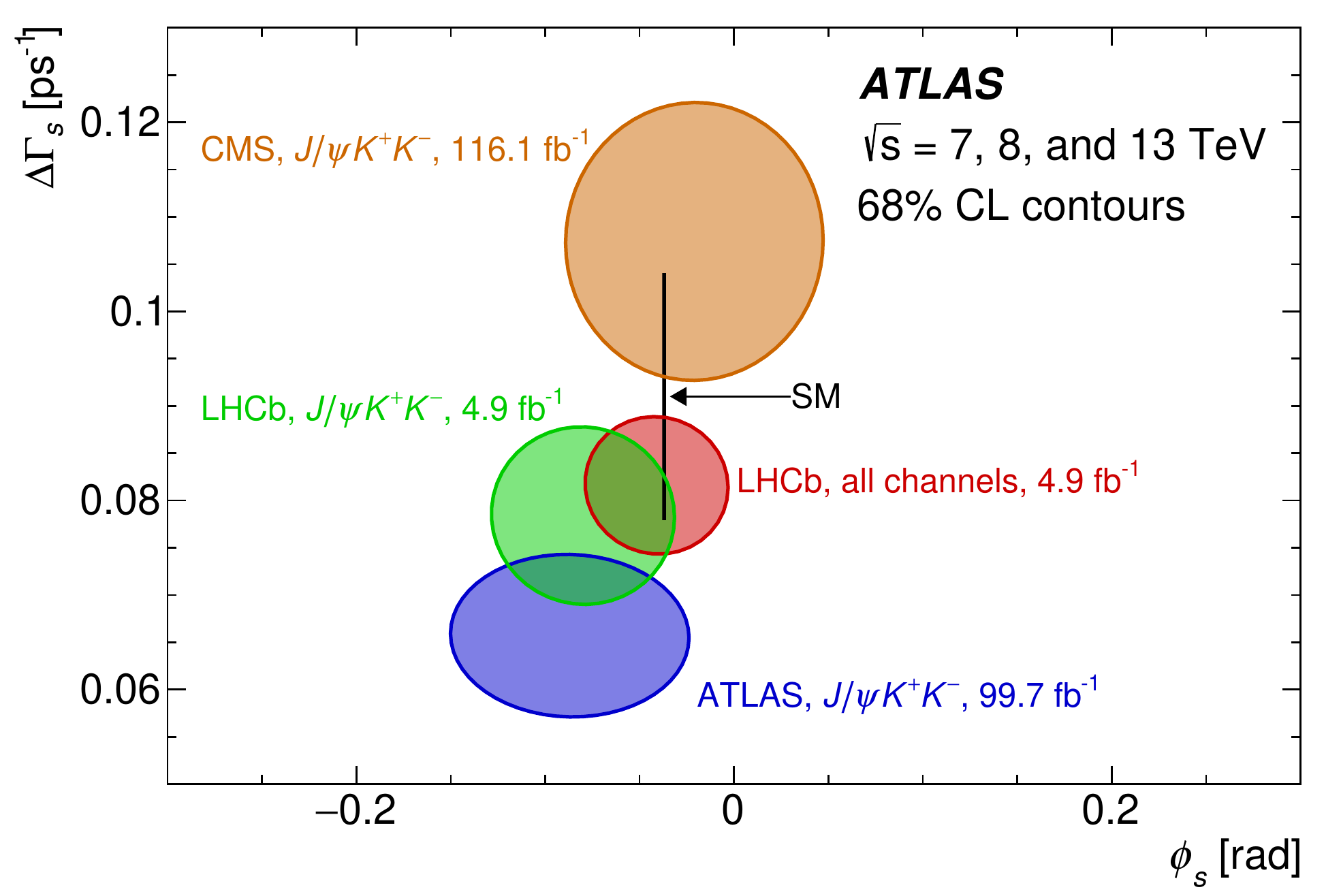}}
\end{center}
\caption{Contours of 68\% confidence level in the \phis--\DGs\ plane, including results from CMS (orange) and LHCb (green) using the $\Bs\to\Jpsi K^+K^-$ decay only and LHCb (red) for all the channels. The  blue contour shows the ATLAS result for  \CoMEnergy\ combined  with \CoMEnergySeven\ and \CoMEnergyEight. The Standard Model prediction \cite{Charles:2015gya, Lenz:2019lvd} is shown as a very thin black rectangle. In  all contours  the statistical and systematic uncertainties are combined in quadrature.}
\label{fig:2DContour}
\end{figure}

Two-dimensional likelihood contours in the  \phis--\DGs \ plane are shown in Figure~\ref{fig:2DContour} for this ATLAS result, the result from CMS~\cite{cmsrun2archive} using the  \Bst \ decay, the result from LHCb~\cite{Aaij2019} using the $\Bs\to\Jpsi K^+K^-$ decay and finally the LHCb result including all \Bs channels \cite{Aaij2019, Aaij:2016psitwoS, Aaij:2014Ds,Aaij:2014dka,Aaij:2019mhf}. The contours are obtained by interpreting each result as a two-dimensional Gaussian probability distribution in the \phis--\DGs \ plane. All results are consistent with each other and with the SM \cite{Charles:2015gya, Lenz:2019lvd}.
 
\FloatBarrier
\section{Summary }
This paper presents a measurement of the time-dependent \CP\ asymmetry parameters in \Bst\ decays from an \ilumi\ data sample of $pp$ collisions collected with the ATLAS detector during the \CoMEnergy\ LHC run. The values from the \CoMEnergy\ analysis are consistent with those obtained in the previous ATLAS analysis using \CoMEnergySeven\ and \CoMEnergyEight\  data. The two measurements are statistically combined.

The \CP-violating phase \phis \  is measured to be \pSfitCombi   $\pm$  \pSstatCombi   (stat.)  $\pm$  \pSsystCombi  (syst.)  rad,
the decay width difference between heavy and light  $B_s^0$ mass   eigenstates, \DGs =   \DGfitCombi  $\pm$  \DGstatCombi    {(stat.)}  $\pm$   \DGsystCombi     {(syst.)}   $\mathrm{ps}^{-1}$ and the measurement for their average decay width,  \Gs  =    \GSfitCombi   $\pm$  \GSstatCombi   {(stat.)}   $\pm$  \GSsystCombi   {(syst.)}  $\mathrm{ps}^{-1}$.  The measurement of the \CP-violating phase \phis\ is consistent with the Standard Model prediction, and it improves on the precision of previous ATLAS measurements,  while for the  average decay width, \Gs, the comparison of the \CoMEnergy\ analysis with the  current  world combined value reveals a tension at the level of $3\sigma$.
\newpage
\section*{Acknowledgements}
 

We thank CERN for the very successful operation of the LHC, as well as the
support staff from our institutions without whom ATLAS could not be
operated efficiently.
 
We acknowledge the support of ANPCyT, Argentina; YerPhI, Armenia; ARC, Australia; BMWFW and FWF, Austria; ANAS, Azerbaijan; SSTC, Belarus; CNPq and FAPESP, Brazil; NSERC, NRC and CFI, Canada; CERN; ANID, Chile; CAS, MOST and NSFC, China; COLCIENCIAS, Colombia; MSMT CR, MPO CR and VSC CR, Czech Republic; DNRF and DNSRC, Denmark; IN2P3-CNRS and CEA-DRF/IRFU, France; SRNSFG, Georgia; BMBF, HGF and MPG, Germany; GSRT, Greece; RGC and Hong Kong SAR, China; ISF and Benoziyo Center, Israel; INFN, Italy; MEXT and JSPS, Japan; CNRST, Morocco; NWO, Netherlands; RCN, Norway; MNiSW and NCN, Poland; FCT, Portugal; MNE/IFA, Romania; JINR; MES of Russia and NRC KI, Russian Federation; MESTD, Serbia; MSSR, Slovakia; ARRS and MIZ\v{S}, Slovenia; DST/NRF, South Africa; MICINN, Spain; SRC and Wallenberg Foundation, Sweden; SERI, SNSF and Cantons of Bern and Geneva, Switzerland; MOST, Taiwan; TAEK, Turkey; STFC, United Kingdom; DOE and NSF, United States of America. In addition, individual groups and members have received support from BCKDF, CANARIE, Compute Canada, CRC and IVADO, Canada; Beijing Municipal Science \& Technology Commission, China; COST, ERC, ERDF, Horizon 2020 and Marie Sk{\l}odowska-Curie Actions, European Union; Investissements d'Avenir Labex, Investissements d'Avenir Idex and ANR, France; DFG and AvH Foundation, Germany; Herakleitos, Thales and Aristeia programmes co-financed by EU-ESF and the Greek NSRF, Greece; BSF-NSF and GIF, Israel; La Caixa Banking Foundation, CERCA Programme Generalitat de Catalunya and PROMETEO and GenT Programmes Generalitat Valenciana, Spain; G\"{o}ran Gustafssons Stiftelse, Sweden; The Royal Society and Leverhulme Trust, United Kingdom.
 
The crucial computing support from all WLCG partners is acknowledged gratefully, in particular from CERN, the ATLAS Tier-1 facilities at TRIUMF (Canada), NDGF (Denmark, Norway, Sweden), CC-IN2P3 (France), KIT/GridKA (Germany), INFN-CNAF (Italy), NL-T1 (Netherlands), PIC (Spain), ASGC (Taiwan), RAL (UK) and BNL (USA), the Tier-2 facilities worldwide and large non-WLCG resource providers. Major contributors of computing resources are listed in Ref.~\cite{ATL-SOFT-PUB-2020-001}.
 
 
\printbibliography

@article{Stone:2008ak,
      author         = "Stone, Sheldon and Zhang, Liming",
      title          = "{S-waves and the Measurement of CP Violating Phases in
                        $B_s$ Decays}",
      journal        = "Phys. Rev.",
      volume         = "D79",
      year           = "2009",
      pages          = "074024",
      doi            = "10.1103/PhysRevD.79.074024",
      eprint         = "0812.2832",
      archivePrefix  = "arXiv",
      primaryClass   = "hep-ph",
      reportNumber   = "SU-HEP-010",
      SLACcitation   = "%%CITATION = ARXIV:0812.2832;%%"
}

@article{LUCID2,
  author={G. Avoni and others},
  title={The new LUCID-2 detector for luminosity measurement and monitoring in ATLAS},
  journal={JINST},
  volume={13},
  number={07},
  pages={P07017},
  doi="10.1088/1748-0221/13/07/P07017",
  year={2018},
}

@Article{Cacciari:2008gp,
     author    = "Cacciari, Matteo and Salam, Gavin P. and Soyez, Gregory",
     title     = "{The anti-$k_t$ jet clustering algorithm}",
     journal   = "JHEP",
     volume    = "04",
     year      = "2008",
     pages     = "063",
     eprint    = "0802.1189",
     archivePrefix = "arXiv",
     primaryClass  =  "hep-ph",
     doi       = "10.1088/1126-6708/2008/04/063",
     SLACcitation  = "%%CITATION = 0802.1189;%%"
}

@Article{Fastjet,
      author         = "Cacciari, Matteo and Salam, Gavin P. and Soyez, Gregory",
      title          = "{FastJet user manual}",
      journal        = "Eur. Phys. J. C",
      volume         = "72",
      year           = "2012",
      pages          = "1896",
      doi            = "10.1140/epjc/s10052-012-1896-2",
      eprint         = "1111.6097",
      archivePrefix  = "arXiv",
      primaryClass   = "hep-ph",
      reportNumber   = "CERN-PH-TH-2011-297",
      SLACcitation   = "%%CITATION = ARXIV:1111.6097;%%"
}

@article{Sjostrand:2014zea,
      author         = "Sj{\"o}strand, Torbjorn and Ask, Stefan and Christiansen,
                        Jesper R. and Corke, Richard and Desai, Nishita and Ilten,
                        Philip and Mrenna, Stephen and Prestel, Stefan and
                        Rasmussen, Christine O. and Skands, Peter Z.",
      title          = "{An Introduction to PYTHIA 8.2}",
      journal        = "Comput. Phys. Commun.",
      volume         = "191",
      year           = "2015",
      pages          = "159",
      doi            = "10.1016/j.cpc.2015.01.024",
      eprint         = "1410.3012",
      archivePrefix  = "arXiv",
      primaryClass   = "hep-ph",
      reportNumber   = "LU-TP-14-36, MCNET-14-22, CERN-PH-TH-2014-190,
                        FERMILAB-PUB-14-316-CD, DESY-14-178, SLAC-PUB-16122,
                        --FERMILAB-PUB-14-316-CD",
      SLACcitation   = "%%CITATION = ARXIV:1410.3012;%%"
}

@article{Aaij2019,
      author         = "{LHCb Collaboration}",
      title          = "{Updated measurement of time-dependent CP-violating
                        observables in $B^{0}_{s}\to J/\psi K^+ K^-$ decays}",
      collaboration  = "LHCb",
      journal        = "Eur. Phys. J.",
      volume         = "C79",
      year           = "2019",
      number         = "8",
      pages          = "706",
      doi            = "10.1140/epjc/s10052-019-7159-8",
      eprint         = "1906.08356",
      archivePrefix  = "arXiv",
      primaryClass   = "hep-ex",
      reportNumber   = "LHCb-PAPER-2019-013, CERN-EP-2019-108",
      SLACcitation   = "%%CITATION = ARXIV:1906.08356;%%",
      related        = "Aaij2019-err",
      relatedstring  = "Erratum:"
}

@article{Aaij:2014Ds,
      author        = "{LHCb Collaboration}",
      title         = "{Measurement of the CP-violating phase $\phi_s$ in
                       $\bar{B}^{0}_{s}\to D_{s}^{+}D_{s}^{-}$ decays}",
      journal       = "Phys. Rev. Lett.",
      collaboration = "LHCb collaboration",
      number        = "LHCB-PAPER-2014-051. LHCB-PAPER-2014-051.
                       CERN-PH-EP-2014-223",
      volume        = "113",
      pages         = "211801,",
      month         = "9",
      year          = "2014",
      reportNumber  = "LHCB-PAPER-2014-051",
%      url           = "https://cds.cern.ch/record/1756036",
      doi           = "10.1103/PhysRevLett.113.211801",
      eprint         = "1409.4619",
      archivePrefix  = "arXiv",
      primaryClass   = "hep-ex",
      SLACcitation   = "%%CITATION = ARXIV:1409.4619;%%"

}

@article{Aaij:2014emv,
	author         = "{LHCb Collaboration}",
	title          = "{Measurement of resonant and CP components in $\bar {B}^0_s\to J/\psi \pi^{+} \pi^{-} $ decays}",
	collaboration  = "LHCb",
	journal        = "Phys.Rev. D",
	volume         = "89",
	pages          = "092006",
	doi            = "10.1103/PhysRevD.89.092006",
	year           = "2014",
	eprint         = "1402.6248",
	archivePrefix  = "arXiv",
	primaryClass   = "hep-ex",
	reportNumber   = "CERN-PH-EP-2014-024, LHCB-PAPER-2013-069",
	SLACcitation   = "%%CITATION = ARXIV:1402.6248;%%",
}

@article{Aaij:2016psitwoS,
      author         = "{LHCb Collaboration}",
      title          = "{First study of the CP -violating phase and decay-width
                        difference in $B_s^0\to\psi(2S)\phi$ decays}",
      collaboration  = "LHCb",
      journal        = "Phys. Lett. B",
      volume         = "762",
      year           = "2016",
      pages          = "253-262",
      doi            = "10.1016/j.physletb.2016.09.028",
      eprint         = "1608.04855",
      archivePrefix  = "arXiv",
      primaryClass   = "hep-ex",
      reportNumber   = "CERN-EP-2016-192, LHCB-PAPER-2016-027",
      SLACcitation   = "%%CITATION = ARXIV:1608.04855;%%"
}

@article{Flatte:1976xu,
      author         = "Flatté, Stanley M.",
      title          = "{Coupled - Channel Analysis of the $\pi \eta$ and $K \bar{K}$
                        Systems Near $K \bar{K}$ Threshold}",
      journal        = "Phys. Lett. B",
      volume         = "63",
      year           = "1976",
      pages          = "224-227",
      doi            = "10.1016/0370-2693(76)90654-7",
      reportNumber   = "CERN-EP-PHYS-76-8",
      SLACcitation   = "%%CITATION = PHLTA,63B,224;%%"
}

@article{Lyons:1988rp,
      author         = "Lyons, Louis and Gibaut, Duncan and Clifford, Peter",
      title          = "{How to Combine Correlated Estimates of a Single Physical
                        Quantity}",
      journal        = "Nucl. Instrum. Meth.",
      volume         = "A270",
      year           = "1988",
      pages          = "110",
      doi            = "10.1016/0168-9002(88)90018-6",
      reportNumber   = "PRINT-88-0048 (OXFORD), OXFORD-NP-5-88",
      SLACcitation   = "%%CITATION = NUIMA,A270,110;%%"
}

@article{Valassi:2003mu,
      author         = "Valassi, A.",
      title          = "{Combining correlated measurements of several different
                        physical quantities}",
      journal        = "Nucl. Instrum. Meth.",
      volume         = "A500",
      year           = "2003",
      pages          = "391-405",
      doi            = "10.1016/S0168-9002(03)00329-2",
      SLACcitation   = "%%CITATION = NUIMA,A500,391;%%"
}

@article{Charles:2015gya,
      author         = "Charles, J. and others",
      title          = "{Current status of the Standard Model CKM fit and
                        constraints on $\Delta F=2$ New Physics}",
      journal        = "Phys. Rev.",
      volume         = "D91",
      year           = "2015",
      number         = "7",
      pages          = "073007",
      doi            = "10.1103/PhysRevD.91.073007",
      eprint         = "1501.05013",
      archivePrefix  = "arXiv",
      primaryClass   = "hep-ph",
      reportNumber   = "LPT-ORSAY-15-04",
      SLACcitation   = "%%CITATION = ARXIV:1501.05013;%%"
}

@article{LE,
    author = "Lenz, Alexander and Piscopo, Maria Laura and Rusov, Aleksey V.",
    title = "{Contribution of the Darwin operator to non-leptonic decays of heavy quarks}",
    eprint = "2004.09527",
    archivePrefix = "arXiv",
    primaryClass = "hep-ph",
    reportNumber = "IPPP/20/12",
    doi = "10.1007/JHEP12(2020)199",
    journal = "JHEP",
    volume = "12",
    pages = "199",
    year = "2020"
}

@article{Aaij:2019mhf,
      author         = "{LHCb Collaboration}",
      title          = "{Measurement of the $CP$-violating phase $\phi_s$ from
                        $B_{s}^{0}\to J/\psi\pi^+\pi^-$ decays in 13 TeV $pp$
                        collisions}",
      collaboration  = "LHCb",
      journal        = "Phys. Lett.",
      volume         = "B797",
      year           = "2019",
      pages          = "134789",
      doi            = "10.1016/j.physletb.2019.07.036",
      eprint         = "1903.05530",
      archivePrefix  = "arXiv",
      primaryClass   = "hep-ex",
      reportNumber   = "LHCb-PAPER-2019-003; CERN-EP-2019-037, CERN-EP-2019-037,
                        LHCb-PAPER-2019-003",
      SLACcitation   = "%%CITATION = ARXIV:1903.05530;%%"
}

@article{Aaij:2014dka,
      author         = "{LHCb Collaboration}",
      title          = "{Measurement of the CP-violating phase $\phi_s$ in
                        $\overline{B}^0_s\rightarrow J/\psi \pi^+\pi^-$ decays}",
      collaboration  = "LHCb",
      journal        = "Phys. Lett.",
      volume         = "B736",
      year           = "2014",
      pages          = "186-195",
      doi            = "10.1016/j.physletb.2014.06.079",
      eprint         = "1405.4140",
      archivePrefix  = "arXiv",
      primaryClass   = "hep-ex",
      reportNumber   = "CERN-PH-EP-2014-086, LHCB-PAPER-2014-019",
      SLACcitation   = "%%CITATION = ARXIV:1405.4140;%%"
}

@article{Aaij:2017zgz,
      author         = "{LHCb Collaboration}",
      title          = "{Resonances and $CP$ violation in $B_s^0$ and
                        $\overline{B}_s^0 \to J/\psi K^+K^-$ decays in the mass
                        region above the $\phi(1020)$}",
      collaboration  = "LHCb",
      journal        = "JHEP",
      volume         = "08",
      year           = "2017",
      pages          = "037",
      doi            = "10.1007/JHEP08(2017)037",
      eprint         = "1704.08217",
      archivePrefix  = "arXiv",
      primaryClass   = "hep-ex",
      reportNumber   = "CERN-EP-2017-062, LHCB-PAPER-2017-008",
      SLACcitation   = "%%CITATION = ARXIV:1704.08217;%%"
}

@article{cmsrun2archive,
    title={Measurement of the $CP$-violating phase $\phi_\mathrm{s}$ in the B$^0_\mathrm{s}\to$ J$/\psi\, \phi$(1020) $\to \mu^+\mu^-$K$^+$K$^-$ channel in proton-proton collisions at $\sqrt{s} =$ 13 TeV},
    collaboration={CMS Collaboration},
    author         = "{CMS Collaboration}",
    institution   = "CERN",
    address       = "Geneva",
    year={2020},
    eprint={2007.02434},
    archivePrefix={arXiv},
    primaryClass={hep-ex},
    reportNumber = "CMS-BPH-20-001, CERN-EP-2020-110",
    doi = "10.1016/j.physletb.2021.136188",
    journal = "Phys. Lett. B",
    volume = "816",
    pages = "136188",
    year = "2021"
}

@article{Lenz:2019lvd,
    author = "Lenz, Alexander and Tetlalmatzi-Xolocotzi, Gilberto",
    title = "{Model-independent bounds on new physics effects in non-leptonic tree-level decays of B-mesons}",
    eprint = "1912.07621",
    archivePrefix = "arXiv",
    primaryClass = "hep-ph",
    reportNumber = "IPPP/19/49, Nikhef-2019-054, SI-HEP-2019, P3H-19-044",
    doi = "10.1007/JHEP07(2020)177",
    journal = "JHEP",
    volume = "07",
    pages = "177",
    year = "2020"
}

@article{Davies:2019gnp,
    author = "Davies, Christine T.H. and Harrison, Judd and Lepage, G. Peter and Monahan, Christopher J. and Shigemitsu, Junko and Wingate, Matthew",
    collaboration = "HPQCD",
    title = "{Lattice QCD matrix elements for the ${B_s^0-\bar{B}_s^0}$ width difference beyond leading order}",
    eprint = "1910.00970",
    archivePrefix = "arXiv",
    primaryClass = "hep-lat",
    reportNumber = "INT-PUB-19-044, IPPP/19/77, JLAB-THY-19-3052",
    doi = "10.1103/PhysRevLett.124.082001",
    journal = "Phys. Rev. Lett.",
    volume = "124",
    number = "8",
    pages = "082001",
    year = "2020"
}

@article{Zyla:2020zbs,
    author = "Zyla, P.A. and others",
    collaboration = "Particle Data Group",
    title = "{Review of Particle Physics}",
    doi = "10.1093/ptep/ptaa104",
    journal = "PTEP",
    volume = "2020",
    number = "8",
    pages = "083C01",
    year = "2020"
}

@Article{PERF-2007-01,
    author         = "{ATLAS Collaboration}",
    title          = "{The ATLAS Experiment at the CERN Large Hadron Collider}",
    journal        = "JINST",
    volume         = "3",
    year           = "2008",
    pages          = "S08003",
    doi            = "10.1088/1748-0221/3/08/S08003",
    primaryClass   = "hep-ex",
}

@Article{BPHY-2011-05,
    author         = "{ATLAS Collaboration}",
    title          = "{Time-dependent angular analysis of the decay \(B_{s}^{0} \rightarrow J/{\psi\phi}\) and extraction of \(\Delta\Gamma_{s}\) and the \(CP\)-violating weak phase \(\phi_s\) by ATLAS}",
    journal        = "JHEP",
    volume         = "12",
    year           = "2012",
    pages          = "072",
    doi            = "10.1007/JHEP12(2012)072",
    reportNumber   = "CERN-PH-EP-2012-182",
    eprint         = "1208.0572",
    archivePrefix  = "arXiv",
    primaryClass   = "hep-ex",
}

@Article{BPHY-2014-03,
    author         = "{ATLAS Collaboration}",
    title          = "{Measurement of the \(CP\)-violating phase \(\phi_s\) and the \(B^0_s\) meson decay width difference with \(B^0_s \rightarrow J/\psi\phi\) decays in ATLAS}",
    journal        = "JHEP",
    volume         = "08",
    year           = "2016",
    pages          = "147",
    doi            = "10.1007/JHEP08(2016)147",
    reportNumber   = "CERN-PH-EP-2015-166",
    eprint         = "1601.03297",
    archivePrefix  = "arXiv",
    primaryClass   = "hep-ex",
}

@Article{PERF-2014-07,
    author         = "{ATLAS Collaboration}",
    title          = "{Topological cell clustering in the ATLAS calorimeters and its performance in LHC Run~1}",
    journal        = "Eur. Phys. J. C",
    volume         = "77",
    year           = "2017",
    pages          = "490",
    doi            = "10.1140/epjc/s10052-017-5004-5",
    reportNumber   = "CERN-PH-EP-2015-304",
    eprint         = "1603.02934",
    archivePrefix  = "arXiv",
    primaryClass   = "hep-ex",
}

@Article{PERF-2015-10,
    author         = "{ATLAS Collaboration}",
    title          = "{Muon reconstruction performance of the ATLAS detector in proton--proton collision data at \(\sqrt{s} = 13\,\text{TeV}\)}",
    journal        = "Eur. Phys. J. C",
    volume         = "76",
    year           = "2016",
    pages          = "292",
    doi            = "10.1140/epjc/s10052-016-4120-y",
    reportNumber   = "CERN-EP-2016-033",
    eprint         = "1603.05598",
    archivePrefix  = "arXiv",
    primaryClass   = "hep-ex",
}

@Article{BPHY-2018-01,
    author         = "{ATLAS Collaboration}",
    title          = "{Measurement of the \(CP\)-violating phase \(\phi_s\) in \(B^0_s \to J/\psi\phi\) decays in ATLAS at \(13\,\text{TeV}\)}",
    year           = "2020",
    reportNumber   = "CERN-EP-2019-218",
    eprint         = "2001.07115",
    archivePrefix  = "arXiv",
    primaryClass   = "hep-ex",
}

@Article{EGAM-2018-01,
    author         = "{ATLAS Collaboration}",
    title          = "{Electron and photon performance measurements with the ATLAS detector using the 2015--2017 LHC proton--proton collision data}",
    journal        = "JINST",
    volume         = "14",
    year           = "2019",
    pages          = "P12006",
    doi            = "10.1088/1748-0221/14/12/P12006",
    reportNumber   = "CERN-EP-2019-145",
    eprint         = "1908.00005",
    archivePrefix  = "arXiv",
    primaryClass   = "hep-ex",
}

@Article{FTAG-2018-01,
    author         = "{ATLAS Collaboration}",
    title          = "{ATLAS \(b\)-jet identification performance and efficiency measurement with \(t\bar{t}\) events in \(pp\) collisions at \(\sqrt{s} = 13\,\text{TeV}\)}",
    journal        = "Eur. Phys. J. C",
    volume         = "79",
    year           = "2019",
    pages          = "970",
    doi            = "10.1140/epjc/s10052-019-7450-8",
    reportNumber   = "CERN-EP-2019-132",
    eprint         = "1907.05120",
    archivePrefix  = "arXiv",
    primaryClass   = "hep-ex",
}

@Booklet{ATL-SOFT-PUB-2020-001,
    author         = "{ATLAS Collaboration}",
    title          = "{ATLAS Computing Acknowledgements}",
    howpublished   = "{ATL-SOFT-PUB-2020-001}",
    url            = "https://cds.cern.ch/record/2717821",
}

@Report{ATLAS-TDR-2010-19,
    author         = "{ATLAS Collaboration}",
    title          = "{ATLAS Insertable B-Layer Technical Design Report}",
    type           = "ATLAS-TDR-19; CERN-LHCC-2010-013",
    year           = "2010",
    url            = "https://cds.cern.ch/record/1291633",
}

@Article{CMS-BPH-13-012,
    author         = "{CMS Collaboration}",
    title          = "{Measurement of the CP-violating weak phase \(\phi_s\) and the decay width difference \(\Delta \Gamma_{s}\) using the \(B^0_s \rightarrow J/\psi \phi(1020)\) decay channel in \(pp\) collisions at \(\sqrt{s} = 8\,\text{TeV}\)}",
    journal        = "Phys. Lett. B",
    volume         = "757",
    year           = "2016",
    pages          = "97",
    doi            = "10.1016/j.physletb.2016.03.046",
    reportNumber   = "CERN-PH-EP-2015-192",
    eprint         = "1507.07527",
    archivePrefix  = "arXiv",
    primaryClass   = "hep-ex",
}

@Booklet{ATLAS-CONF-2019-021,
    author         = "{ATLAS Collaboration}",
    title          = "{Luminosity determination in \(pp\) collisions at \(\sqrt{s} = 13\,\text{TeV}\) using the ATLAS detector at the LHC}",
    howpublished   = "{ATLAS-CONF-2019-021}",
    url            = "https://cds.cern.ch/record/2677054",
    year           = "2019",
}

@Booklet{ATL-PHYS-PUB-2014-021,
    author         = "{ATLAS Collaboration}",
    title          = "{ATLAS Pythia~8 tunes to \(7~\text{TeV}\) data}",
    howpublished   = "{ATL-PHYS-PUB-2014-021}",
    url            = "https://cds.cern.ch/record/1966419",
    year           = "2014",
}

@article{PhysRevD.91.073007,
      author         = "{CKMfitter group, Charles, J. et al.}",
      title          = "{Current status of the Standard Model CKM fit and constraints on $\Delta F$ = 2 New Physics}",
      journal        = "\PRD",
      volume         = "91",
      year           = "2015",
      pages          = "073007",
      doi            = "10.1103/PhysRevD.91.073007",
      eprint         = "1501.05013",
      archivePrefix  = "arXiv",
      primaryClass   = "hep-ph",
      reportNumber   = "LPT-ORSAY-15-04",
      note = {Numbers updated using the results from the 2019 values in \url{https://ckmfitter.in2p3.fr/www/results/plots_summer19/ckm_res_summer19.html}},
      SLACcitation   = "%%CITATION = ARXIV:1501.05013;%%"
}

@article{UTFit2006,
      author         = "{UTfit Collaboration, M. Bona et al.}",
      title          = "{The unitarity triangle fit in the standard model and hadronic parameters from lattice QCD: A reappraisal after the measurements of $\Delta m_s$ and BR($ B \rightarrow \tau \nu_{\tau})$ }",
      collaboration  = "UTfit",
      journal        = "JHEP",
      volume         = "10",
      year           = "2006",
      pages          = "081",
      doi            = "10.1088/1126-6708/2006/10/081",
      eprint         = {hep-ph/0606167},
      archivePrefix  = "arXiv",
      primaryClass   = "hep-ph",
      note  = {Numbers updated to the 2018 results from \url{https://www.utfit.org/UTfit/ResultsSummer2018SM}},
%      reportNumber   = "CERN-PH-EP-2015-153, LHCB-PAPER-2015-029",
      SLACcitation   = "%%CITATION = ARXIV:0606167;%%"
}

@article{PhysRevLett.115.072001,
      author         = "{LHCb Collaboration}",
      title          = "{Observation of $J/\psi p$ Resonances Consistent with
                        Pentaquark States in $\Lambda_b^0 \to J/\psi K^- p$
                        Decays}",
      collaboration  = "LHCb",
      journal        = "Phys. Rev. Lett.",
      volume         = "115",
      year           = "2015",
      pages          = "072001",
      doi            = "10.1103/PhysRevLett.115.072001",
      eprint         = "1507.03414",
      archivePrefix  = "arXiv",
      primaryClass   = "hep-ex",
      reportNumber   = "CERN-PH-EP-2015-153, LHCB-PAPER-2015-029",
      SLACcitation   = "%%CITATION = ARXIV:1507.03414;%%"
}

@article{1674-1137-40-1-011001,
      author         = "{LHCb Collaboration}",
      title          = "{Study of the production of $\Lambda_b^0$ and
                        $\overline{B}^0$ hadrons in $pp$ collisions and first
                        measurement of the $\Lambda_b^0\rightarrow J/\psi pK^-$
                        branching fraction}",
      collaboration  = "LHCb",
      journal        = "Chin. Phys. C",
      volume         = "40",
      year           = "2016",
      number         = "1",
      pages          = "011001",
      doi            = "10.1088/1674-1137/40/1/011001",
      eprint         = "1509.00292",
      archivePrefix  = "arXiv",
      primaryClass   = "hep-ex",
      reportNumber   = "CERN-PH-EP-2015-223, LHCB-PAPER-2015-032",
      SLACcitation   = "%%CITATION = ARXIV:1509.00292;%%"
}

@article{Nierste,
      author         = "Lenz, Alexander and Nierste, Ulrich",
      title          = "{Theoretical update of $B_{s} - \bar{B}_{s}$ mixing}",
      journal        = "JHEP",
      volume         = "06",
      pages          = "072",
      doi            = "10.1088/1126-6708/2007/06/072",
      year           = "2007",
      eprint         = "hep-ph/0612167",
      archivePrefix  = "arXiv",
      %primaryClass   = "hep-ph",
      reportNumber   = "TTP06-31",
      SLACcitation   = "%%CITATION = HEP-PH/0612167;%%",
}

@article{PDG,
  title = {Review of Particle Physics},
  author = {Tanabashi, M. and others},
  collaboration = {Particle Data Group},
  journal = {Phys. Rev. D},
  volume = {98},
  
  pages = {030001},
  numpages = {1898},
  year = {2018},
  month = {8},
  publisher = {American Physical Society},
  doi = {10.1103/PhysRevD.98.030001},
  %url = {https://link.aps.org/doi/10.1103/PhysRevD.98.030001}
}

@article{CDFAngles,
      author         = "Dighe, Amol S. and Dunietz, Isard and Fleischer, Robert",
      title          = "{Extracting CKM phases and $B_s - \bar{B}_s$ mixing
                        parameters from angular distributions of non-leptonic $B$
                       decays}",
      journal        = "\EPJC",
      volume         = "6",
      pages          = "647-662",
      doi            = "10.1007/s100529800954",
      year           = "1999",
      eprint         = "hep-ph/9804253",
      archivePrefix  = "arXiv",
    %  primaryClass   = "hep-ph",
      reportNumber   = "CERN-TH-98-85, FERMILAB-PUB-98-093-T, IC-98-25",
%       SLACcitation   = "%%CITATION = ARXIV:1411.3104;%%",
}

@article{LHCb:2014run1,
      author         = "{LHCb Collaboration}",
      title          = "{Precision Measurement of $CP$ violation in $B_s^0 \to
                        J/\psi K^+K^-$ Decays}",
      collaboration  = "LHCb",
      journal        = "\PRL",
      volume         = "114",
      pages          = "041801",
      doi            = "10.1103/PhysRevLett.114.041801",
      year           = "2015",
      eprint         = "1411.3104",
      archivePrefix  = "arXiv",
      primaryClass   = "hep-ex",
      reportNumber   = "LHCB-PAPER-2014-059, CERN-PH-EP-2014-271,
      LHCB-PAPER-2014-059-AND-CERN-PH-EP-2014-271",
      SLACcitation   = "%%CITATION = ARXIV:1411.3104;%%",
}

@article{CDF:2012ie,
      author         = "{CDF Collaboration}",
      title          = "{Measurement of the Bottom-Strange Meson Mixing Phase in the Full CDF Data Set}",
      collaboration  = "CDF",
      journal        = "\PRL",
      volume         = "109",
      pages          = "171802",
      doi            = "10.1103/PhysRevLett.109.171802",
      year           = "2012",
      eprint         = "1208.2967",
      archivePrefix  = "arXiv",
      primaryClass   = "hep-ex",
      reportNumber   = "FERMILAB-PUB-12-471-PPD",
      SLACcitation   = "%%CITATION = ARXIV:1208.2967;%%",
}

@article{Abazov:2011ry,
      author         = "{D0 Collaboration}",
      title          = "{Measurement of the CP-violating phase $\phi_s^{J/\psi
                        \phi}$ using the flavor-tagged decay $B^0_s \rightarrow
                        J/\psi \phi$ in 8 fb$^{-1}$ of $p \bar{p}$ collisions}",
      collaboration  = "D0",
      journal        = "\PRD",
      volume         = "85",
      pages          = "032006",
      doi            = "10.1103/PhysRevD.85.032006",
      year           = "2012",
      eprint         = "1109.3166",
      archivePrefix  = "arXiv",
      primaryClass   = "hep-ex",
      reportNumber   = "FERMILAB-PUB-11-463-E",
      SLACcitation   = "%%CITATION = ARXIV:1109.3166;%%",
}

@article{Atlfast,
      author         = "{ATLAS Collaboration}",
      title          = "{The ATLAS Simulation Infrastructure}",
      collaboration  = "ATLAS",
      journal        = "\EPJC",
      volume         = "70",
      pages          = "823-874",
      doi            = "10.1140/epjc/s10052-010-1429-9",
      year           = "2010",
      eprint         = "1005.4568",
      archivePrefix  = "arXiv",
      primaryClass   = "physics.ins-det",
      SLACcitation   = "%%CITATION = ARXIV:1005.4568;%%",
}

@article{Geant4,
      author         = "Agostinelli, S. and others",
      title          = "{\textsc{Geant4}---a simulation toolkit}",
      collaboration  = "GEANT4",
      journal        = "Nucl. Instrum. Meth. A",
      volume         = "506",
      year           = "2003",
      pages          = "250",
      doi            = "10.1016/S0168-9002(03)01368-8",
      reportNumber   = "SLAC-PUB-9350, FERMILAB-PUB-03-339",
      SLACcitation   = "%%CITATION = NUIMA,A506,250;%%"
}

@article{Aaij:2013Mix,
      author         = "{LHCb Collaboration}",
      title          = "{Precision measurement of the  $B^0_s - \bar{B}^0_s$ oscillation frequency with the decay 
                        $B^0_s \to D_s^- \pi^+$ }",
      collaboration  = "LHCb",
      journal        = "New J. Phys.",
      volume         = "15",
      pages          = "053021",
      doi = {10.1088/1367-2630/15/5/053021},
      year           = "2013",
      eprint         = "1304.4741",
      archivePrefix  = "arXiv",
      primaryClass   = "hep-ex",
      SLACcitation   = "%%CITATION = ARXIV:1304.4741;%%",
}

@Booklet{LHCb-CONF-2013-011,
    author         = "{LHCb\hspace{0.1cm}Collaboration}",
    title          = "{Updated average $f_s/f_d$ $b$-hadron production fraction ratio for 7 TeV $pp$ collisions}",
    howpublished   = "{LHCb-CONF-2013-011}",
    url            = "https://cds.cern.ch/record/1559262",
    year           = "2013",
}

@article{BABAR,
      author         = "{BaBar Collaboration}",
      title          = "{Search for the $Z(4430)^{-}$ at BABAR}",
      collaboration  = "BABAR",
      journal        = "\PRD",
      volume         = "79",
      pages          = "112001",
      doi            = "10.1103/PhysRevD.79.112001",
      year           = "2009",
      eprint         = "0811.0564",
      archivePrefix  = "arXiv",
      primaryClass   = "hep-ex",
      reportNumber   = "SLAC-PUB-13437",
      SLACcitation   = "%%CITATION = 0811.0564;%%",
}

@article{JINST.13.T05008,
      author         = "{B. Abbott et al.}",
      title          = "{Production and Integration of the ATLAS Insertable B-Layer}",
      journal        = "JINST",
      volume         = "13",
      number         = "05",
      year           = "2018",
      pages          = "T05008",
      doi            = "10.1088/1748-0221/13/05/T05008",
      eprint         = "1803.00844",
      archivePrefix  = "arXiv",
      primaryClass   = "physics.ins-det",
      SLACcitation   = "%%CITATION = ARXIV:1803.00844;%%"
}

@article{LHCbK0Star,
      author         = "{LHCb Collaboration}",
      title          = {Measurement of the polarization amplitudes in ${B}^{0} \rightarrow J/\psi K^{*}(892{)}^{0}$ decays},
      collaboration  = "LHCb",
      journal        = {Phys. Rev. D},
      volume         = {88},
      %issue          = {5},
      pages          = {052002},
      numpages       = {12},
      year           = {2013},
      publisher      = {American Physical Society},
      doi            = "10.1103/PhysRevD.88.052002",
      eprint         = "1307.2782",
      archivePrefix  = "arXiv",
      primaryClass   = "hep-ex",
      reportNumber   = "CERN-PH-EP-2013-104, LHCB-PAPER-2013-023",
      SLACcitation   = "%%CITATION = ARXIV:1307.2782;%%",
}

@article{LHCbLambdabProduction,
      author         = "{LHCb Collaboration}",
      title          = "{Study of the kinematic dependences of $\Lambda_{b}^{0}$
                        production in pp collisions and a measurement of the
                        $\Lambda_{b}^{0} \to \Lambda_{c}^{+}$ $\pi^{-}$ branching
                        fraction}",
      collaboration  = "LHCb",
      journal        = "JHEP",
      volume         = "08",
      year           = "2014",
      pages          = "143",
      doi            = "10.1007/JHEP08(2014)143",
      eprint         = "1405.6842",
      archivePrefix  = "arXiv",
      primaryClass   = "hep-ex",
      reportNumber   = "CERN-PH-EP-2014-106, LHCB-PAPER-2014-004",
      SLACcitation   = "%%CITATION = ARXIV:1405.6842;%%"
}

@article{BOTELLA20071,
title = "CP violation and limits on New Physics including recent $B_{s}$ measurements",
journal = "Nucl. Phys. B",
volume = "768",
number = "1",
pages = "1 - 20",
year = "2007",
%issn = "0550-3213",
doi = "10.1016/j.nuclphysb.2006.12.022",
%url = "http://www.sciencedirect.com/science/article/pii/S0550321306010182",
author = "Francisco J. Botella and Gustavo C. Branco and Miguel Nebot",
eprint         = "0608100",
archivePrefix  = "arXiv",
primaryClass   = "hep-ph",
}

@article{Pumplin:2002vw,
      author         = "Pumplin, J. and Stump, D. R. and Huston, J. and Lai, H.
                        L. and Nadolsky, Pavel M. and Tung, W. K.",
      title          = "{New generation of parton distributions with
                        uncertainties from global QCD analysis}",
      journal        = "JHEP",
      volume         = "07",
      year           = "2002",
      pages          = "012",
      doi            = "10.1088/1126-6708/2002/07/012",
      eprint         = "hep-ph/0201195",
      archivePrefix  = "arXiv",
      %primaryClass   = "hep-ph",
      reportNumber   = "MSU-HEP-011101",
      SLACcitation   = "%%CITATION = HEP-PH/0201195;%%"
}

\clearpage 
 
\begin{flushleft}
\hypersetup{urlcolor=black}
{\Large The ATLAS Collaboration}

\bigskip

\AtlasOrcid[0000-0002-6665-4934]{G.~Aad}$^\textrm{\scriptsize 102}$,    
\AtlasOrcid[0000-0002-5888-2734]{B.~Abbott}$^\textrm{\scriptsize 129}$,    
\AtlasOrcid[0000-0002-7248-3203]{D.C.~Abbott}$^\textrm{\scriptsize 103}$,    
\AtlasOrcid{O.~Abdinov}$^\textrm{\scriptsize 13,*}$,    
\AtlasOrcid[0000-0002-2788-3822]{A.~Abed~Abud}$^\textrm{\scriptsize 71a,71b}$,    
\AtlasOrcid[0000-0002-1002-1652]{K.~Abeling}$^\textrm{\scriptsize 53}$,    
\AtlasOrcid[0000-0002-2987-4006]{D.K.~Abhayasinghe}$^\textrm{\scriptsize 94}$,    
\AtlasOrcid[0000-0002-8496-9294]{S.H.~Abidi}$^\textrm{\scriptsize 168}$,    
\AtlasOrcid[0000-0002-8279-9324]{O.S.~AbouZeid}$^\textrm{\scriptsize 40}$,    
\AtlasOrcid{N.L.~Abraham}$^\textrm{\scriptsize 157}$
% ...
\bigskip
\\

% ...
$^{13}$Institute of Physics, Azerbaijan Academy of Sciences, Baku; Azerbaijan.\\
% ...
$^{40}$Niels Bohr Institute, University of Copenhagen, Copenhagen; Denmark.\\
% ...
$^{53}$II. Physikalisches Institut, Georg-August-Universit\"{a}t G\"ottingen, G\"ottingen; Germany.\\
% ...
$^{71}$$^{(a)}$INFN Sezione di Pavia;$^{(b)}$Dipartimento di Fisica, Universit\`a di Pavia, Pavia; Italy.\\
% ...
$^{94}$Department of Physics, Royal Holloway University of London, Egham; United Kingdom.\\
% ...
$^{102}$CPPM, Aix-Marseille Universit\'e, CNRS/IN2P3, Marseille; France.\\
$^{103}$Department of Physics, University of Massachusetts, Amherst MA; United States of America.\\
% ...
$^{129}$Homer L. Dodge Department of Physics and Astronomy, University of Oklahoma, Norman OK; United States of America.\\
% ...
$^{157}$Department of Physics and Astronomy, University of Sussex, Brighton; United Kingdom.\\
% ...
$^{168}$Department of Physics, University of Toronto, Toronto ON; Canada.\\
% ...
$^{a}$ Also at Borough of Manhattan Community College, City University of New York, New York NY; United States of America.\\
$^{b}$ Also at Centre for High Performance Computing, CSIR Campus, Rosebank, Cape Town; South Africa.\\
% ...
$^{*}$ Deceased

\end{flushleft}


\end{document}